\documentclass{osa-article}

\journal{osajournal}

\articletype{Research Article}
\usepackage{xcolor}

\usepackage{lineno}
\usepackage{siunitx}

\begin{document}

\title{\textcolor{black}{Optomechanics of optically-levitated particles: A tutorial and perspective}}

\author{George Winstone,\authormark{1}, Alexey Grinin,\authormark{1} Mishkat Bhattacharya,\authormark{2} Andrew A. Geraci,\authormark{1,*}  Tongcang Li,\authormark{3}  Peter J. Pauzauskie,\authormark{4,5} and Nick Vamivakas\authormark{6}} 

\address{\authormark{1}Center for Fundamental Physics and Center for Interdisciplinary Exploration and Research in Astrophysics, Department of Physics and Astronomy, Northwestern University, Evanston, IL, 60208\\
\authormark{2}School of Physics and Astronomy, Rochester Institute of Technology, 84 Lomb Memorial Drive, Rochester, 14623 NY\\
\authormark{3}Department of Physics and Astronomy and Elmore Family School of Electrical and Computer Engineering, Purdue University, West Lafayette, IN, USA\\
\authormark{4}Department of Materials Science and Engineering, University of Washington, Seattle, WA\\
\authormark{5}Physical and Computational Sciences Directorate, Pacific Northwest National Laboratory, Richland, WA\\
\authormark{6}Department of Physics, University of Rochester, Rochester, NY, 14627}

\email{\authormark{*}andrew.geraci@northwestern.edu} 



\begin{abstract}
Optomechanics, the study of the mechanical interaction of light with matter, has proven to be a fruitful area of research that has yielded many notable achievements, including the direct detection of gravitational waves in kilometer-scale optical interferometers. Light has been used to cool and demonstrate quantum control over the mechanical degrees of freedom of individual ions and atoms, and more recently has facilitated the observation of quantum ``mechanics'' in objects of larger mass, even at the kg-scale. \textcolor{black}{Optical levitation}, where an object can be suspended by radiation pressure and largely decoupled from its environment, has recently established itself as a rich field of study, with many notable results relevant for precision measurement, quantum information science, and foundational tests of quantum mechanics and fundamental physics.   This article provides a survey of several current activities in field along with a tutorial describing associated key concepts and methods, both from an experimental and theoretical approach. It is intended as a resource for junior researchers who are new to this growing field as well as beginning graduate students. The tutorial is concluded with a perspective on both promising emerging experimental platforms and anticipated future theoretical developments. 
\end{abstract}
\tableofcontents
\section{Introduction}
The field of optomechanics dates back at least to the early 1600s, when Johannes Kepler suggested the force imparted by solar radiation explained the direction of comet tails (for a historical account see Ref. \cite{NicholsandHull}). Radiation pressure was experimentally observed and confirmed in the laboratory by Nichols and Hull as well as Lebedev \cite{Nichols,Lebedew}, only after an erroneous demonstration decades earlier by Crookes due to radiometric forces \cite{Crookes}. Such radiometric forces, having to do with momentum imparted to an object by thermal currents of gas molecules surrounding it when the object is heated due light absorption, were later shown in some cases to exceed radiation pressure forces by five orders of magnitude \cite{Nichols}. With the invention of the laser \cite{Maiman1960}, the potential for the manipulation \textcolor{black}{and measurement} of particles and objects with radiation pressure became substantially enhanced \textcolor{black}{by taking advantage of coherence of the optical field. For example, the use of single wavelength sources enables avoiding unwanted optical absorption bands in materials, allows the implementation of optical standing wave trapping potentials, and facilitates the interferometric detection of particle displacement, e.g. homodyne detection.}   
\textcolor{black}{A celebrated modern day feat of optomechanics is the realization of quantum-limited precision measurement in macroscopic systems, in the context of laser-interferometer gravitational wave detectors \cite{GWreview89,Braginsky1978,ligofirst2016,aasi2013enhanced}.}

\begin{figure}[ht!]
\begin{center}
	\includegraphics[width=0.99\columnwidth]{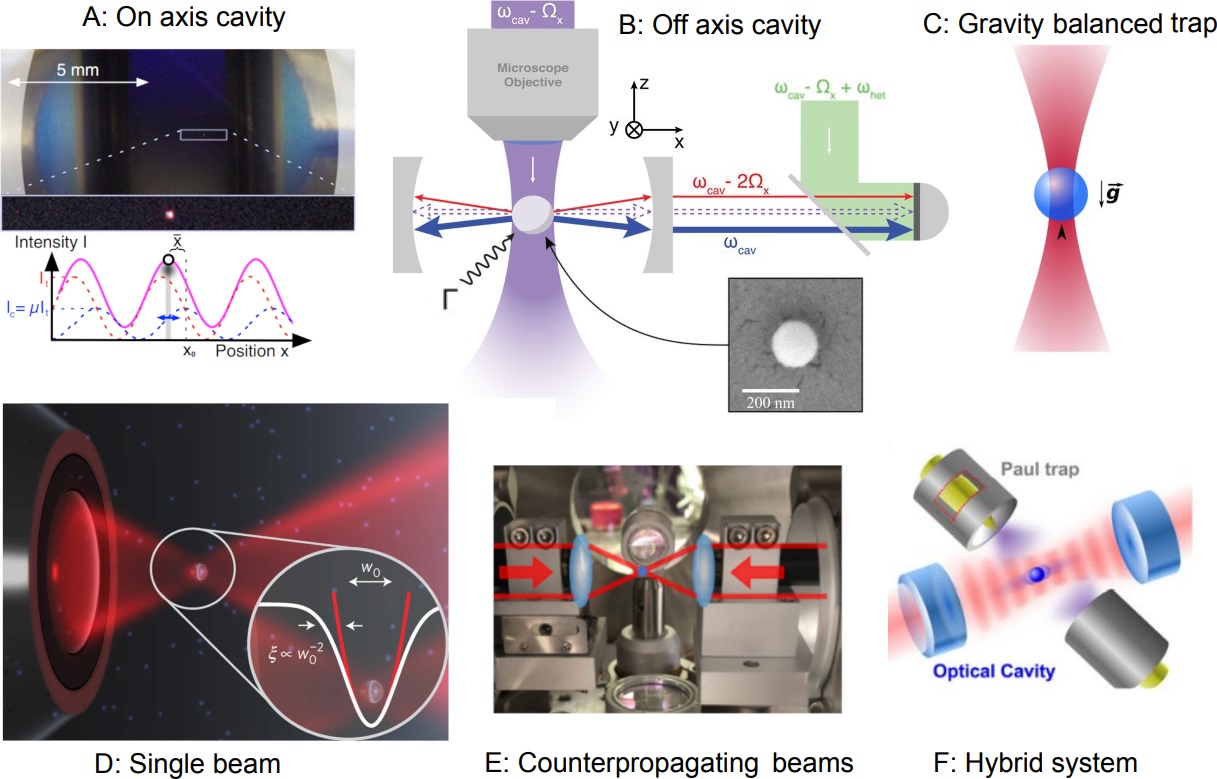}
	\caption{Optical trapping geometries employed in recent levitated optomechanics experiments (clockwise from upper left): A) cavity along trapping optical axis, reproduced from Ref. \cite{kiesel2013cavity}, B) cavity normal to trapping optical axis - reproduced from Ref. \cite{delic2020cooling}, C) scattering force - gravity balanced single beam trap, adapted from Ref. \cite{blakemore2019precision}, D) single objective focus trap with a single propagation direction, as in Ref. \cite{gieseler2013thermal}, E) two counterpropagating freespace beams in a non cavity configuration, adapated from Ref. \cite{winstone2022optical}, F) hybrid Optical and Paul trap (for the same trapped species) as in Ref. \cite{millen2015cavity}. 
}
	\label{trapping_configurations}
	\end{center}
\end{figure}

Pioneered by Arthur Ashkin and coworkers, early work on optical levitation and manipulation of dielectric particles led both to the development of optical tweezers (for a review see Ref. \cite{tweezerreview}), which has had a profound impact on biology and biophysics, and eventually to the laser cooling and trapping of atoms and molecules \cite{Nobel1997,ChuNobel}. 
Optical trapping of dielectric particles was first studied in a liquid environment \cite{Ashkin:1970}, and investigations of trapping in air and vacuum were carried out shortly afterwards \cite{Ashkin:1971,Ashkin:1976,Ashkin:1977}. In the late 2000s it was realized that optical trapping and manipulation of dielectric particles in vacuum is a promising route to observe quantum mechanical behavior, due to the extreme environmental isolation made possible by suspending a particle using radiation pressure  \textcolor{black}{\cite{Romero-Isartvirus2010,Chang:2010,Li:2011,Gieseler:2012,Barker:2010}. Alternatively ion trapping or hybrid ion-optical trapping has also been considered \cite{Chang:2010, Li:2011, Barker:2010}.} Experimental progress in levitated optomechanics in vacuum has been quite rapid  (see Ref. \cite{Millen:2020review} for a recent review), with quantum control including ground state cooling \cite{delic2020cooling,tebbenjohanns2021quantum} of optically trapped nanoparticles, and precision sensing including force sensing at the zeptonewton level \cite{ranjit2016zeptonewton},  acceleration sensing at the nano-$g$ level \cite{monteiro2020force}, along with torque sensing at the level \textcolor{black}{of $10^{-27}$ Nm} \cite{Li2018} and realization of GHz rotation rates of nano-objects \cite{Li2018,Novotny2018,Moore2018}. A variety of trapping architectures and trapped particles of a wide range of mass and size have been studied, with notable recent examples illustrated in Fig. \ref{trapping_configurations}. More broadly, levitodynamics, the levitation and control of microscopic objects in vacuum \cite{Levitodynamics}, has emerged as an active field, including not only schemes employing optical radiation pressure but instead relying on magnetic trapping or radio-frequency trapping, for example for charged levitated objects. \textcolor{black}{Although many exciting developments are in progress for schemes that do not rely on optical radiation pressure for suspending or trapping particles, in this tutorial we limit our focus to optical levitation.}

Intended primarily as an introductory resource for junior researchers, in this tutorial we describe the working principles behind controlling the dynamics of levitated optomechanical systems, and discuss recent developments in hybrid and coupled systems with multiple degrees of freedom.  We illustrate several examples of applications and science opportunities relying on optically levitated systems including precision sensing, fundamental tests of quantum mechanics and quantum coherence, searches for exotic new forces, dark matter, and gravitational waves, and tests of thermodynamics and radiative heat transfer in novel regimes. We conclude with perspective on future prospects in both theory and experiment, including biological and chemistry applications, tests and applications of quantum squeezing and entanglement, and sensing beyond the standard quantum limit.

\section{Controlling the dynamics of levitated optomechanical systems \label{sec:trapping}}

\textcolor{black}{
This chapter gives an overview of the most important theoretical concepts in the topic of optical levitation.} We describe the basic features of optical trapping that allow the levitation of \textcolor{black}{dielectric} particles and their subsequent manipulation. \textcolor{black}{We focus our discussion on a dielectric object levitated by a focused laser in vacuum. We discuss electromagnetic and thermal forces acting upon it and its equations of motion. We further discuss methods to measure and cool its center of mass (COM) motion. }\textcolor{black}{Our aim is to provide an intuitive understanding of the involved physics, followed by a presentation of the main relevant equations and their implications. When detailed rigorous derivations of these results is outside of the scope of this tutorial article, references are provided.}

\subsection{Principles of optical trapping}


\textcolor{black}{Levitation refers to the process or technique of making an object float in air or vacuum without any solid support. It can be achieved through various methods, such as utilizing acoustic waves to suspend objects in mid-air or using air currents for aerodynamic levitation. Electromagnetic forces, however, have the advantage of enabling the trapping of objects in a vacuum, thereby largely decoupling them from environmental interactions and disturbances. Moreover, electromagnetic fields can be controlled with unparalleled }precision, enabling accurate manipulation of trapped objects.

\textcolor{black}{Electromagnetic levitation has been demonstrated across a wide range of interaction types, particles, and configurations. To organize this diversity, one can consider the multipole expansion of an electric potential. An arbitrary particle's electromagnetic interaction can be characterized by its charge distribution $\rho(\mathbf{r}) = \rho_{\text{free}}(\mathbf{r})-\nabla\cdot \mathbf{P}(\mathbf{r})$ consisting of both free charge carriers (charges, dipoles, etc.) as well as induced dipoles $\nabla\cdot \mathbf{P}$. The following expansion gives the electric interaction potential with an external field.
\begin{equation}
    U_{\text{int}} = q \phi_{\text{ext}}(\mathbf{r}_0) - \mathbf{p} \cdot \mathbf{E}_{\text{ext}}(\mathbf{r}_0) +  \frac{1}{6}\sum_{i,j}Q_{i,j}\frac{\partial E_i(\mathbf{r}_0)}{\partial x_j} + \text{h.o.}
    \label{eq:multipole}
\end{equation}}
\textcolor{black}{Here $q$, $\mathbf{p}$, $Q_{i,j}$ are the charge (scalar), dipole moment (vector), and quadrupole moment (second rank tensor) of the charge distribution, respectively. $\phi_{\text{ext}}(\mathbf{r}_0)$, $\mathbf{E}_{\text{ext}}(\mathbf{r}_0)$, and $\frac{\partial E_i(\mathbf{r}_0)}{\partial x_j}$ denote the electric potential, field and field gradient at the center of gravity of the particle $\mathbf{r}_0$. In other words, charges couple to potentials, dipoles to fields, and quadrupoles to field gradients. The magnetic field expansion follows analogously, excluding the first term due to the absence of magnetic monopoles. Higher-order moments are seldom of interest, and for optical levitation, even quadrupole interactions are generally negligible.} 
\textcolor{black}{Charge moments can be either permanent (e.g., polar molecules) or induced (e.g., non-polar atoms, molecules, dielectrics). To understand the principal mechanism of optical trapping, we consider how an incident field with frequency $\omega$ can drive an induced dipole moment through the polarizability of the dielectric object.} 
\textcolor{black}{A trapping technique can be realized through an off-resonant, induced dipole interaction \footnote{Note this is in contrast to typical optical dipole trapping of atoms or molecules, where the laser wavelength is chosen near the frequency of a resonant atomic or molecular transition, although many considerations are common to both techniques. For a pedagogical introduction see e.g. Ref. \cite{Foot:book}}. Note that Eq. \ref{eq:multipole} encompasses only the conservative parts of the interaction responsible for trapping, while non-conservative forces will be addressed subsequently.}

\textcolor{black}{To begin with, we ignore non-conservative forces (e.g. due to light scattering), and see that for trapping, the potential energy $U_{\text{tr}}$ must necessarily exhibit a minimum, implying that its gradient or the corresponding force vanishes:
\begin{equation}
     \Vec{F}_{\text{tr}}(\Vec{r}_0) = -\vec{\nabla}U_{\text{tr}}(\vec{r}_0) = \nabla(\alpha \cdot \mathbf{E}_{\text{ext}}^{2}(\vec{r})) |_{\vec{r}=\vec{r}_0}  = \mathbf{0}
\end{equation}
Here, we assumed only induced dipole coupling and assumed a single polarizability constant $\alpha$ for the dipole moment of the object. 
For ordinary dielectric materials with positive polarizability, stable minima occur at the field maxima (i.e. the particles are high-field-seeking.) }
\begin{figure}[htbp]
\begin{center}
        \includegraphics[width=0.9\columnwidth]{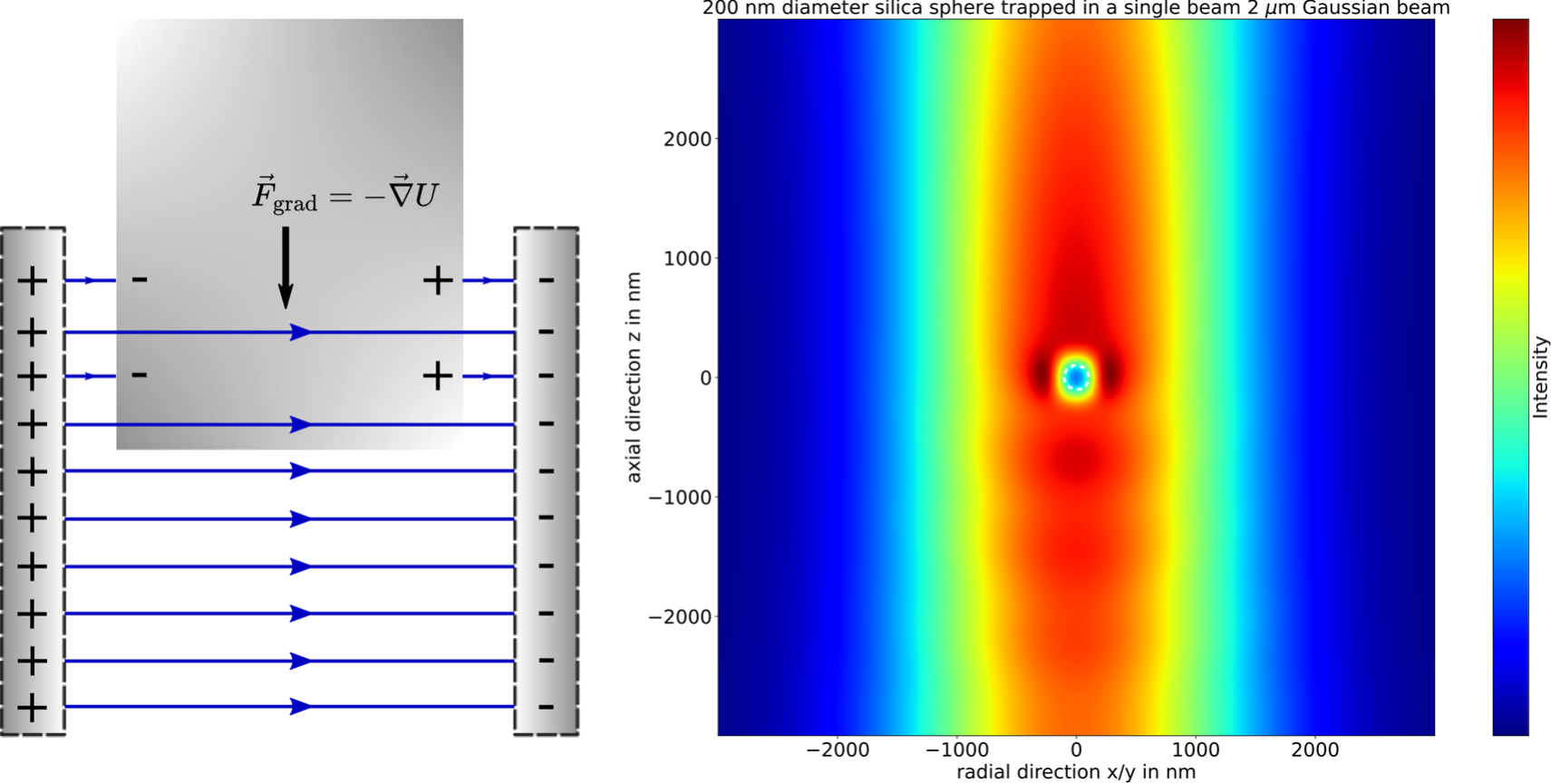}
        \caption{\textcolor{black}{(left) A dielectric slab when inserted from the side experiences an attractive force towards the interior of a parallel plate capacitor. Due to the presence of the external field, 
        a net dipole moment (polarization) is generated in the dielectric, which is attracted to the oppositely charged capacitor plates. The induced dipoles partially compensate for the external field, resulting in a lower energy configuration. The field is said to be displaced from within the dielectric. 
        (right) Intensity of a linearly-polarized Gaussian laser beam of 2\si{\micro\metre} waist and 1550\si{\nano\metre} wavelength trapping a 200\si{\nano\metre} diameter silica (n=1.45) sphere. Just as in the case of the plate capacitor, the external electric field due to the laser is displaced out of the sphere. The energy minimum is at the point of highest intensity (high-field seeker)}}
	\label{fig:tweezerpgdm}
\end{center}
\end{figure}

\textcolor{black}{This \textbf{gradient force} that provides the mechanism of confinement in optical tweezers is analogous to the phenomenon of a dielectric material being drawn towards the interior of a plate capacitor, as illustrated in Figure \ref{fig:tweezerpgdm}.}
\textcolor{black}{The energy density is decreased as the induced field of the dielectric is opposed to the original field of the place capacitor. We say the dielectric displaces the electric field from its interior.  The system tends towards an energy minimum, where the entire piece of dielectric is inside of the plate capacitor. 
Figure \ref{fig:tweezerpgdm} also shows the intensity profile of a tightly focused laser beam with a small glass sphere trapped in its focus. One can clearly see how the field has been displaced from the interior of the dielectric sphere in analogy to the plate capacitor case. }
\textcolor{black}{This behavior can be understood by considering the response of a dielectric sphere of permittivity $\epsilon$ in a uniform electric field $\vec{E_0}=E_0\hat{z}$, for example as described in Ref. \cite{Jackson}. The boundary conditions required by Maxwell's equations are that the tangential component of the electric field must be continuous at the surface of the sphere along with the normal component of the displacement $\vec{D}=\epsilon \vec{E} $. Solving the boundary value problem with a potential having appropriate azimuthal symmetry of the form 
\begin{eqnarray*}
    \Phi_{in}&=&\Sigma_{l=0}^{\infty}A_l r^l P_l(\cos{\theta})\\
    \Phi_{out}&=&\Sigma_{l=0}^{\infty}[B_l r^l+C_lr^{-(l+1)} ]P_l(\cos{\theta})
\end{eqnarray*}
with the added boundary condition that $\Phi_{out} \rightarrow -E_0 z=-E_0 r \cos{\theta}$ at infinity, the solution is
\begin{eqnarray*}
    \Phi_{in}&=& -\frac{3}{\epsilon/\epsilon_0+2}E_0 r \cos{\theta}\\
    \Phi_{out}&=&-E_0 r \cos{\theta}+\frac{\epsilon/\epsilon_0-1}{\epsilon/\epsilon_0+2}E_0 \frac{a^3}{r^2}\cos{\theta},
\end{eqnarray*}}
\textcolor{black}{where $a$ is the radius of the sphere. From this we see that the electric field in the interior of the dielectric sphere is reduced by a factor of $\frac{3}{\epsilon/\epsilon_0+2} < 1 $ relative to the applied field. Outside of the dielectric sphere the electric field corresponds to the superposition of the applied field along with the field of a point-dipole at the origin with total dipole moment 
\begin{equation}
\vec{p} = 4 \pi \epsilon_0 \frac{\epsilon/\epsilon_0-1}{\epsilon/\epsilon_0+2} a^3 \vec{E_0}. \label{eq:CM}
\end{equation}
where the polarizibility is $\alpha = 3 V \epsilon_0 (\frac{\epsilon/\epsilon_0-1}{\epsilon/\epsilon_0+2})$ and $V$ is the volume of the sphere.}

\textcolor{black}{In general, the polarizability can be regarded mathematically as a complex quantity with the real part accounting for the dispersion, while the imaginary part accounts for losses due to absorption or scattering. For larger objects, the polarizability will additionally reflect their shape and material properties.  The real part of the polarizability function $\alpha(\omega)$ determines the wavelength or frequency-dependent response strength of the material (dispersion) to electromagnetic fields. It determines the corresponding index of refraction. The dispersive and dissipative parts of the polarizability are causally connected. 
This important result is referred to as the \textbf{Kramers-Kronig} relation \cite{Jackson}.}

\textcolor{black}{For a time varying electric field due to a laser, the potential experienced by the dipole in the electric field $\vec{E}$ can be written as
\begin{equation}
    U_{tr}=-\vec{p}\cdot\vec{E}=-\alpha \vec{E}^2.\label{eq:dipolepot}
\end{equation}
The potential depends on the square of the electric field which is proportional to the intensity of the laser beam. For focused laser beam and a positive value of $\alpha$, this results in an attractive force towards a local maximum of laser intensity.}

\begin{figure}[ht!]
\begin{center}
	\includegraphics[width=0.5\columnwidth]{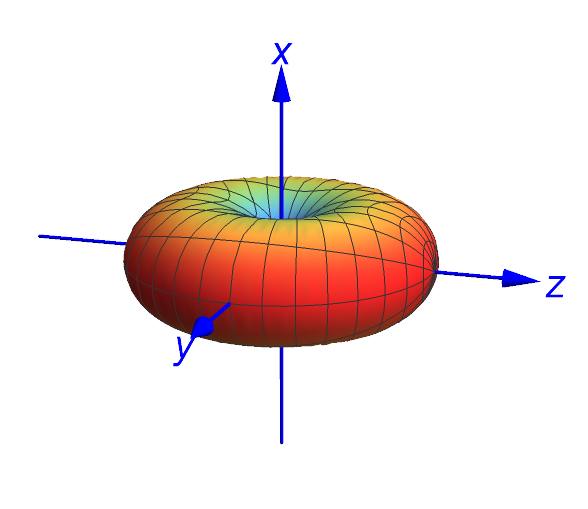}
	\caption{\textcolor{black}{Illustration of angular distribution of dipole radiation pattern described by Eq. \ref{dipolerad} assuming an incident electric field polarized along the $\hat{x}$ direction.  No radiation is emitted along the direction of the polarization of the incident beam and radiation is isotropic in the plane normal to the polarization of the incident radiation.}}. 
	\label{fig:dipolescat}
	\end{center}
\end{figure}

\textcolor{black}{Even for a lossless particle, scattering of the incident radiation is also present. This is a consequence of Maxwell's equations, which dictate that an accelerated charge always radiates. 
An oscillating dipole driven by the electric field of an incident laser will emit radiation. Assuming a sinusoidally oscillating current $\vec{J}=\vec{J}\exp{(-i\omega t)}$ corresponding to the oscillating induced dipole moment, in the Lorentz (or radiation) gauge, the vector potential can be written \begin{equation*}
    \vec{A}(\vec{r})=\frac{\mu_0}{4\pi} \int \vec{J}(\vec{r}') \frac{e^{ik|\vec{r}-\vec{r}'|}}{|\vec{r}-\vec{r}'|}d^3r',
\end{equation*}
where in the far-field we can consider the approximation $|\vec{r}-\vec{r}'| \approx r - \hat{r} \cdot \vec{r}'$, so that at leading order
\begin{equation*}
    \vec{A}(\vec{r})=\frac{\mu_0}{4\pi} \frac{e^{ikr}}{r} \int \vec{J}(\vec{r}') d^3r'.
\end{equation*}
Integration by parts yields
\begin{equation*}
    \int \vec{J}(\vec{r}') d^3r' = -\int \vec{r}' (\vec{\nabla}'\cdot \vec{J}(\vec{r}')) d^3r'= -i \omega \int \vec{r}' \rho(\vec{r}') d^3r'
\end{equation*}
where the integral in the last term we recognize as the total dipole moment $\vec{p}$,
yeilding
\begin{equation*}
      \vec{A}(\vec{r})=\frac{i \omega \mu_0}{4\pi} \frac{e^{ikr}}{r} \vec{p}.
\end{equation*}
Then using Maxwell's equations for the source-free region away from the sphere we have $\vec{H}=\frac{1}{\mu_0} \vec{\nabla} \times \vec{A}$ and $\vec{E} = ic/\omega \sqrt{\mu_0/\epsilon_0} \vec{\nabla} \times \vec{H}$.  The solutions are given in Ref. \cite{Jackson} as \begin{eqnarray*}
    \vec{H}&=&\frac{ck^2}{4\pi} (\vec{n} \times \vec{p})\frac{e^{ikr}}{r}(1-\frac{1}{ikr})\\
    \vec{E}&=&\frac{1}{4\pi\epsilon_0} \frac{e^{ikr}}{r}\left[ k^2(\vec{n} \times \vec{p})\times \vec{n} +(3\vec{n}(\vec{n} \cdot \vec{p})-\vec{p})(\frac{1}{r^2}-\frac{ik}{r})  \right].
\end{eqnarray*}
In the near field for $kr<<1$, i.e. for small $r$ or very long wavelength radiation we see the electric field reduces to the form from a static dipole modulated by the time dependence, and the magnetic field times $\sqrt{\mu_0/\epsilon_0}$ is suppressed by a factor of $kr<<1$, and vanishes in the limit of $k \rightarrow 0.$ 
In the far field, we find the time averaged power radiated per unit angle by the oscillating dipole is given by
\begin{equation}
    \frac{dP}{d\Omega}= \frac{1}{2} Re [ r^2 \vec{n} \cdot \vec{E} \times \vec{H}^*]=\frac{c^2}{32\pi^2}\sqrt{\frac{\mu_0}{\epsilon_0}} k^4 |\vec{p}|^2\sin^2{\theta},\label{dipolerad}
\end{equation}
    which is a typical dipole radiation pattern with $\theta$ measured relative to the direction of $\vec{p}$.  The pattern is illustrated in Fig. \ref{fig:dipolescat}.}

\textcolor{black}{Whether the direction of a photon's travel is changed due to refraction or scattering, this results in a total momentum change of the electromagnetic field. This momentum change must necessarily be compensated by a force on the particle, which in the case of scattering is commonly called the \textbf{scattering force}. 
Photons carry momentum $\hbar \vec{k}$ where $|k|=2\pi/\lambda$ is their wavenumber, and photons reflecting from a perfectly reflecting mirror of area $A$ will impart a total force $F= \frac{2IA}{c}$ on the mirror, where $I$ is the intensity of the laser. A perfectly absorbing mirror, e.g. which converts the incident energy into heating the material. will experience a force $F= \frac{2IA}{c}$ due to conservation of momentum.  The scattering force experienced by a dipolar scatterer can be expressed according to its scattering cross section $C_{\mathrm{scat}}$ as \begin{equation}
\vec{F}_{\mathrm{scat}}(r)=\hat{k}\frac{1}{c}C_{\mathrm{scat}}I(r).
\end{equation}
This force is directed along the propagation axis of the laser.  After a scattering event, described by Eq. \ref{dipolerad}, the scattered photon exits in a random direction with its probability of being emitted at a particular angle given by the differential scattering cross section. Due to conservation of momentum, there must be a corresponding recoil of the center of mass of the dielectric sphere.  While on the average the net force from a series of such recoils will average to zero for the isotropic scattering of a dipole, the repeated scattering process leads to a momentum diffusion process known as \textbf{photon recoil heating} \cite{Gordon1980,Chang:2010,jain2016direct}. We note that even for the dipole force, which being well described by a time averaged potential as in Eq. \ref{eq:dipolepot}, the discrete quantum nature of photons in the laser trap yeilds a fluctuating force on a trapped dielectric particle, analaogous to that studied for atoms in early work by Gordon and Ashkin \cite{Gordon1980}.}

\subsubsection{\textcolor{black}{Light-matter interaction and optical forces within size regimes}}

As with many fields of physics, \textcolor{black}{the behavior of a system depends greatly on its characteristic length scales.} In the case of \textcolor{black}{optical levitation}, the ratio between the wavelength of the trapping light and the size of the levitated object is critical. 
As such, for spherical objects, the optical size parameter: 

\begin{equation}
\xi=\frac{2 \pi R}{\lambda_0}
\end{equation}

is commonly defined, in which $\lambda_0$ is the wavelength of the \textcolor{black}{trapping} light and $R$ is the radius.
The case where the trapped object's circumference is much smaller than the wavelength is commonly referred to as the Rayleigh regime. \textcolor{black}{In this case, or when all of the dimensions of the particle are much smaller than the wavelength, the exact shape of the object (i.e. spherical, cylindrical, cubic, or irregular etc.) has less impact on the scattering properties, as the object effectively scatters light as if it is a dipole with a given dipole moment per unit volume.} The case in which \textcolor{black}{the dimensions of object} are approximately equal to the wavelength (i.e. $\xi \sim 1$) is known as the Mie-Lorentz regime. \textcolor{black}{In this regime the differential scattering cross section is fairly complicated and depends on resonances associated with detailed constructive and destructive interference within the dielectric object. For a textbook reference on Mie scattering \cite{Mieorig} for spheres see for example Ref. \cite{Miebook}.} The case in which the object is much larger than the wavelength is the geometric optics regime. \textcolor{black}{In this regime, one can effectively consider tracing rays of light through refraction and reflection from the dielectric-vacuum or dielectric-air interfaces at the surface of the levitated object. The associated bending of light rays at the surface of the dielectric object is associated with a gradient force which tends to attract particles towards the focus of a laser beam. While many optical levitation experiments operate in the Rayleigh or Mie-Lorentz regimes, recent work has also occurred in the geometric optics regime, with spherical particles of size ranging from $5$ to $30$ $\mu$m \cite{Moore2017}.  }
\textcolor{black}{When light either from the optical trap or from another light source is used for particle detection, the light scattering profile and dependence on the particle size and geometry with respect to the laser wavelength also plays a key role, as we discuss further in Sec. \ref{sec:detection}. }

\begin{figure}[htbp]
\begin{center}
	\includegraphics[width=0.9\columnwidth]{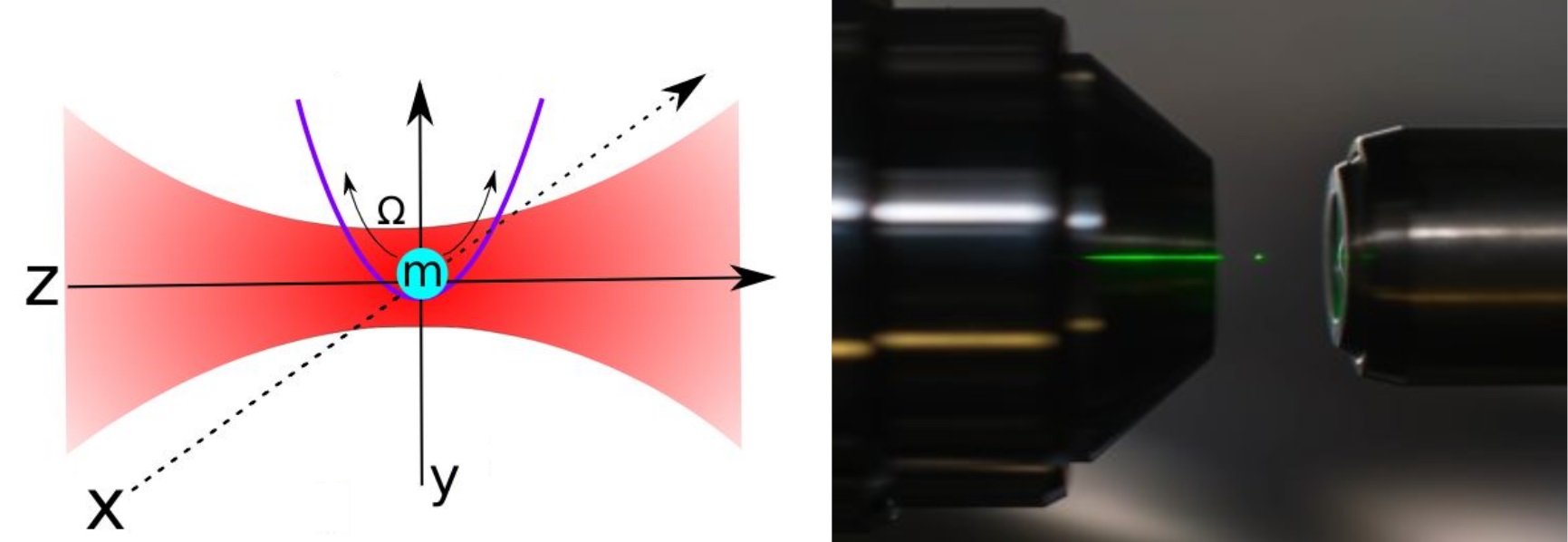}
	\caption{A commonly used architecture for a levitated optomechanical system employing a horizontally-oriented, \textcolor{black}{linearly-polarized}, single-beam optical tweezer trap. A spherical dielectric nanoparticle of mass $m$ is levitated at the center of an optical potential. \textcolor{black}{The particle oscillates about the potential minimum in three dimensions, with center frequencies }(eigenfrequencies) $\Omega_q$, where $q=(x,y,z)$. \textcolor{black}{For a single-beam tweezer trap, the normal mode frequency for oscillations in the z-direction is typically lower than for the transverse directions, since the Rayleigh range $z_R = \pi w_0^2 /\lambda_0$ of the beam determines the length scale for the intensity gradient, which is typically larger than the beam waist $w_0$. The degeneracy between the normal mode frequencies in the x- and y- directions is generally broken due to the direction of linear polarization for a tightly focused tweezer beam \cite{hecht} or for a beam with an elliptical profile.} Image adapted from Ref. \cite{Rodenburg:16}.
	\label{canonical}}
	\end{center}
\end{figure}





\textcolor{black}{More generally, the radiation pressure forces acting on the levitated object} can be obtained from the self consistent EM field configuration in and around the particle, in which scattered photons represent transient fluctuations.
\textcolor{black}{We focus our attention first on the Rayleigh regime, as several notable recent optical-levitation experiments in high vacuum have trapped sub-wavelength particles in single-beam optical tweezer traps and have  achieved many interesting results, including cooling to the motional quantum ground state of the trapping potential \cite{delic2020cooling,Windey_2019}. }
Within the Rayleigh regime, \textcolor{black}{the nanoparticle can be regarded as a point-dipole, and} the gradient and scattering forces can be considered to be separable, with the gradient force \textcolor{black}{being} proportional to the derivative of the intensity of the EM field and the scattering force \textcolor{black}{proportional to} the product of the intensity and the particle's scattering cross section. It is important to note that the gradient force, \textcolor{black}{which manifests itself as a restoring force about the focus,} typically overpowers the scattering force along the axis of beam propagation within the Rayleigh regime for focused beams. \textcolor{black}{ This is apparent since the gradient force depends linearly on the volume of the particle (and total dipole moment) while the scattering force is proportional to the square of the volume of the particle. Thus as the size of the particle is reduced eventually the dipole force dominates.} The scattering and gradient forces in the Rayleigh limit for a laser beam of intensity $I(\vec{r})$ are given
by \cite{li2012fundamental,harada1996radiation}:

\begin{equation}
\vec{F}_{\mathrm{scat}}(\vec{r})=\hat{z}\frac{n_{\mathrm{med}}}{c}C_{\mathrm{scat}}I(\vec{r})=\hat{z}\frac{128 \pi^5 R^6}{3 c \lambda_{0}^4}\Big(\frac{m_{\mathrm{obj}}^2 -1}{m_{\mathrm{obj}}^2 +2}\Big)^2 n_{\mathrm{med}}^5 I(\vec{r}),\label{eq:fscat}
\end{equation}
and
\begin{eqnarray}
\vec{F}_{\mathrm{grad}}(\vec{r})&=& \vec{\nabla} [\vec{p}(\vec{r}) \cdot \vec{E}(\vec{r})]\\
&=& \frac{4 \pi n_{\mathrm{med}}^2 \epsilon_0 R^3}{c} (\frac{m_{\mathrm{obj}}^2 -1}{m_{\mathrm{obj}}^2 +2}) \cdot \frac{1}{2}\vec{\nabla} E^2(\vec{r}) = \frac{2 \pi n_{\mathrm{med}} R^3}{c} (\frac{m_{\mathrm{obj}}^2 -1}{m_{\mathrm{obj}}^2 +2}) \vec{\nabla} I(\vec{r}) \label{eq:fgrad}
\end{eqnarray}
where $\hat{z}$ is the displacement vector along the optical axis, \textcolor{black}{i.e. the direction of beam propagation, and $\vec{r}=(x,y,z)$.} $n_{\mathrm{med}}$ is the refractive index of the medium surrounding the object while $m_{\mathrm{obj}} \equiv n/n_{\mathrm{med}}$ is the refractive index
of the object relative to the surrounding medium. $p(r,t)$ is the dipole induced on the object from the electric field
$E(r,t)$ of the incident light. \textcolor{black}{A connection can be made with Eq. \ref{eq:CM} using the relation that $m_{\textrm{obj}}^2 = \epsilon/\epsilon_0.$ }$C_{\mathrm{scat}}$ is the scattering cross section of the object. 
Note that both the dipole response and forces are time averaged. The characteristic size scalings (with respect to the levitated object) of the two forces are $F_{\mathrm{scat}}(r) \propto R^6$ and $F_{\mathrm{grad}}(r) \propto R^3$ in the Rayleigh limit. 
\textcolor{black}{While this scaling enables trapping of very sub-wavelength particles in a tightly-focused single-beam tweezer, as the particle size is increased, this scaling also illustrates the requirement for the scattering force to be counterbalanced by another field (optical, gravitational etc) in the direction of beam propagation, i.e. when the particle radius is a large fraction of the wavelength and we approach the Mie regime.} A commonly employed trapping configuration for a Rayleigh-size particle is a linearly-polarized single beam optical tweezer and is depicted in Fig. \ref{canonical}. \textcolor{black} {Here the laser beam propagates along the $\hat{z}$-direction and is linearly polarized along $\hat{x}$.} The \textcolor{black}{optical confinement} of the particle can be described by an asymmetric three-dimensional harmonic potential in the limit of small motional amplitudes. 

We also note that the gradient force described by Eq. (\ref{eq:fgrad}) is not universally applicable. In particular, dielectric particles sufficiently large compared to the trap wavelength and with large enough refractive index (e.g. silicon) can also exhibit strong magnetic dipole responses, with different force coefficients \cite{opticaltrapmag1,opticaltrapmag2}. Further, the force induced by the imaginary component of the Poynting vector generally needs to be considered \cite{Poynting1,Poynting2}. However for low-optical-loss \textcolor{black}{sub-wavelength silica nanoparticles} typically used in levitated optomechanics experiments, we expect Eqs. (\ref{eq:fscat}) and (\ref{eq:fgrad}) to be a reasonable approximation \cite{opticaltrapmag1} \textcolor{black}{since the electric dipole response is dominant}. 

For most setups, the beam profile is typically Gaussian, or a superposition of counter-propagating Gaussian beams, with an electric field \begin{equation*}
    \vec{E}(\vec{r},t) = E_0 \hat{x} \frac{w_0}{w(z)} \exp{[-r^2/w^2(z)}] \exp{[i(kz+2kr^2/(R(z))-\psi(z)]}\exp{(-i\omega t)}
\end{equation*}
for linearly polarized light in the $\hat{x}$ propagating in the $\hat{z}$ direction for example. 
where the intensity profile is given by

\begin{equation}
I(x,y,z) = \frac{ I_{0}}{ \left(\frac{z^{2}}{z_{R}^{2}} + 1\right)} e^{\frac{- 2 x^{2} - 2 y^{2}}{w_{0}^{2} \left(\frac{z^{2}}{z_{R}^{2}} + 1\right)}}\label{eq:Gauss},
\end{equation}
with beam waist $w_0$, the beam radius of curvature $R(z) = z (1+z^2_R/z^2)$, $w(z)=w_0\sqrt{1+z^2/z^2_R}$ where the Rayleigh range is defined as $z_R \equiv \pi w_0^2 /\lambda_0$.

For the case of the gradient force, taking the derivative of such a function leads to a series of high order terms for each axis, providing coupling between the degrees of freedom for even a perfect sphere. In the limit of small displacement, these terms are negligible, and a Gaussian potential \textcolor{black}{is well approximated by a quadratic ($x^2$) potential, i.e. corresponding to a harmonic oscillator.} 
This results in a linear restoring force as a function of displacement along all three spatial directions. 
\textcolor{black}{For small particles in the Rayleigh limit, the restoring force can be calculated analytically using the expression in Eq. \ref{eq:fgrad} and the intensity profile given in Eq. \ref{eq:Gauss}, and for larger particles or more complex geometries this can generally be done numerically, as we discuss further below.}

\begin{figure}[htbp]
\begin{center}
	\includegraphics[width=1.0\columnwidth]{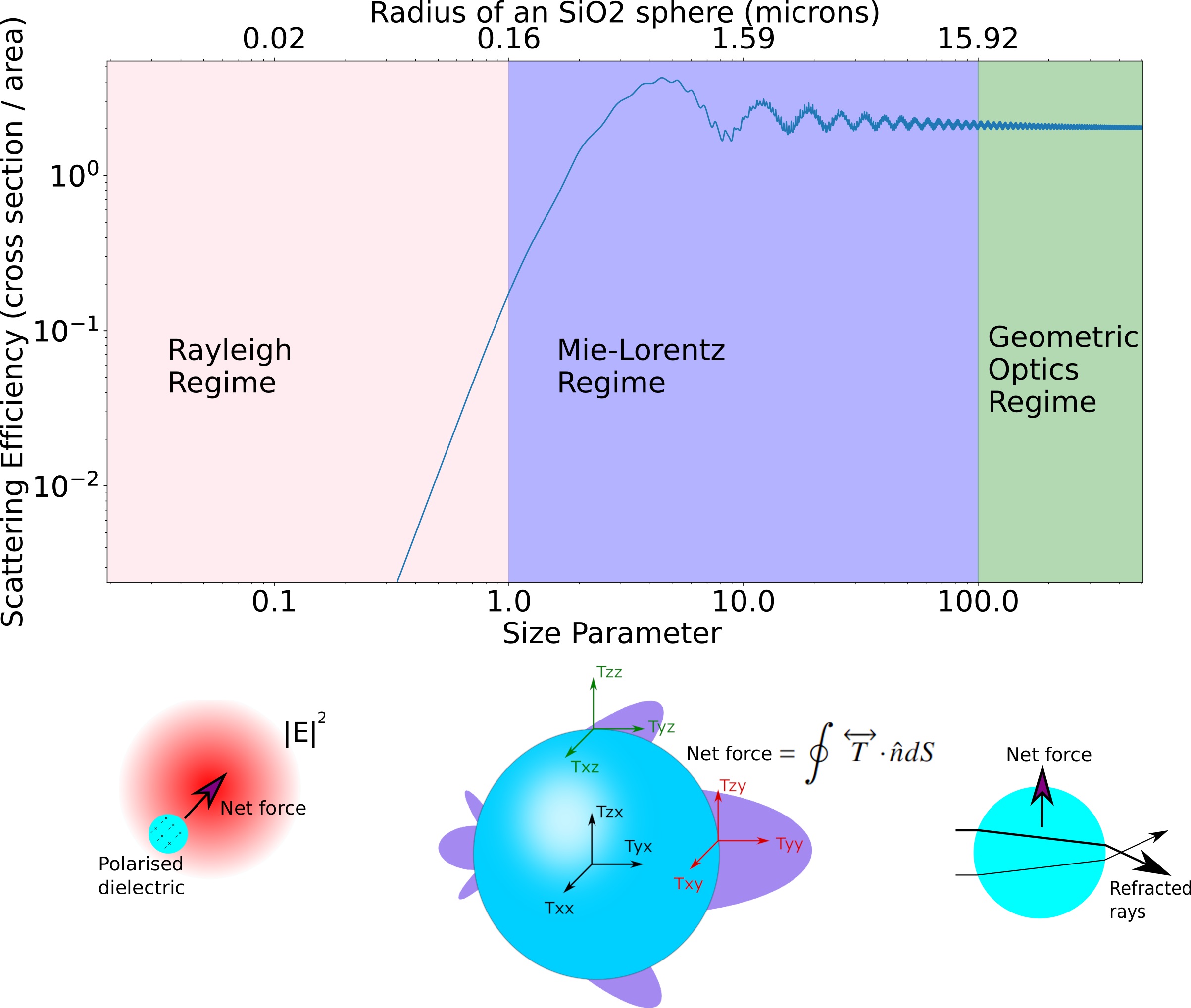}
	\caption{Scattering cross section of a dielectric sphere  as a function of its size parameter, with regimes labelled. \textcolor{black}{In the Rayleigh regime an analytic form for the cross section is apparent from Eq. \ref{eq:fscat}.} Upper horizontal axis shows the corresponding radius of a SiO$_2$ sphere when scattering light of wavelength $1$ $\mu$m. Lower: \textcolor{black}{Intuitive simplified description of physical origin of forces within each regime.} (right) Geometric optics regime in which the levitated object is much larger than the interacting wavelength, the forces and thus motional dynamics are dominated by conservation of momentum of the refracted rays. (left) Rayleigh regime in which the levitated object is much smaller than the interacting wavelength, the dominating force is gradient force, and to a good approximation the nano-object acts like a dipole. (middle) \textcolor{black}{A diagram of a Lorentz-Mie sized sphere in a conceptually depicted typical Mie scatter pattern. The vectors components of the Maxwell stress tensor integral are illustrated overlaid on the $(x,y,z)$ Cartesian normal facing surfaces of  the sphere.  }The optical force \textcolor{black}{on a particle and corresponding cross section} in the Mie-Lorentz regime is obtained by integrating the maxwell stress tensor (as indicated in Eq. \ref{maxwell_stress_tensor}) over a surface enclosing the object.
	}
	\label{all_regimes}
	\end{center}
\end{figure}

The single beam tweezer is a commonly employed trapping method for dielectric nanoparticles. 
The tweezer intensity profile is typically Gaussian and for this the gradient force is linear in nanoparticle displacement (for amplitude of motion smaller than the beam waist), resulting in a harmonic trap to lowest order in nanoparticle displacement \cite{PhysRevLett.109.103603}. 

For low-optical-loss silica nanoparticles, with refractive index $n=1.45$, we expect Eqs. (\ref{eq:fscat}) and (\ref{eq:fgrad}) to be a reasonable approximation, and expanding about the \textcolor{black}{equilibrium} position for a focused Gaussian beam, it can be seen that a nanosphere will be confined in an approximately harmonic potential, and attracted to the point of maximal intensity with a nearly linear restoring force \cite{tweezerreview}. In vacuum, the motion of the trapped particle largely corresponds to a 3-dimensional simple harmonic oscillator in the under-damped regime, damped by background \textcolor{black}{gas} collisions and driven by thermal noise. 
    The trap frequency for oscillations transverse to the direction of propagation of the laser and in the axial directions is approximately given by \begin{equation} \omega_r=\sqrt[]{\frac{6 I_0}{c \rho w_0^2} \left( \frac{m_{\mathrm{obj}}^2-1}{m_{\mathrm{obj}}^2+2} \right) } \end{equation}
and \begin{equation}\omega_z=\sqrt[]{\frac{3 I_0 \lambda_0^2}{c \pi^2 \rho w^4_0} \left( \frac{m_{\mathrm{obj}}^2-1}{m_{\mathrm{obj}}^2+2} \right), }
\end{equation} respectively, where $\rho$ is the density of the sphere, assuming a trap in vacuum ($n_{\mathrm{med}}=1$).

For small amplitudes of motion and a spherical particle, the oscillation along each Cartesian direction is independent. For large amplitudes of oscillation, the tweezer displays anharmonicities, resulting in Duffing-type nonlinearities in the next-to-harmonic order \cite{Gieseler2013}. These nonlinearities can not only make \textcolor{black}{the oscillations in each direction} anharmonic (due to self-Duffing nonlinearities), \textcolor{black}{but} they can \textcolor{black}{also} couple the various directions (due to cross-Duffing nonlinearities).

In the Mie-Lorentz regime, in which the optical size parameter is roughly equal to one, \textcolor{black}{the computation of optical forces is achieved by a variety of other means}. Fig. \ref{all_regimes} illustrates how the scattering cross section of a dielectric sphere varies with size parameter across the Rayleigh, Mie-Lorentz, and geometric optics regimes. 

Exact analytic solutions for the EM field configuration within the particle and therefore the force exerted by the light field only exist for certain symmetric particle shapes, materials and beam profiles. For example, the Mie equations describe the forces on spheres \cite{tzarouchis2018light}, spheroids\cite{asano1979light} and infinite cylinders \cite{ren1997scattering} for a limited set of optical potentials. Outside of these limits approximations exist to extend these solutions for objects of refractive indices similar to the environment \cite{barber1978rayleigh} or of limited difference in optical size parameter to the Rayleigh regime. Furthermore, semi-analytic solutions, such as the T-matrix\cite{mishchenko1996t} method may also be used for objects similar to those with an exact Mie solution - finite cylinders as opposed to infinite cylinders. Beyond these limited cases, fully numerical methods are required to compute the optical forces in this regime. 

In the general case, the \textcolor{black}{stress} \textcolor{black}{acting} upon the particle may be obtained by solving for the Maxwell Stress Tensor :



\begin{equation}
\label{maxwell_stress_tensor}
T_{ij}  = \epsilon_0 [E_iE_j + c^2B_iB_j  - \frac{1}{2}(|\overrightarrow{E}^2| + c^2 |\overrightarrow{B}|^2 \delta_{ij}].    
\end{equation}

The force and torque exerted on the object by the optical field can then be calculated \textcolor{black}{by integrating over the stress tensor over a surface enclosing the particle}:

\begin{equation}
\overrightarrow{F} = \oint \overleftrightarrow{T} \cdot \hat{n} dS
\end{equation}

\begin{equation}
\overrightarrow{\tau} = \oint \overleftrightarrow{M} \cdot \hat{n} dS
\end{equation}
where the tensor $M$ is given by $\overrightarrow{M} = - \overleftrightarrow{T} \times  \overrightarrow{r}$ and $\overrightarrow{r}$ is the displacement vector \cite{seberson2020stability,jackson2012classical}.

This naturally requires the user to know the configuration of optical fields within the object, for complex objects not matching the cases above this can be obtained with an numerical solver. The Discrete Dipole Approximation (DDA)\cite{draine1994discrete} or Green's Dyad Method (GDM)\cite{wiecha2018pygdm} \textcolor{black}{are} two popular examples. Several other popular solvers exist in including MEEP (a Finite-Difference Time-Domain (FDTD) method)\cite{oskooi2010meep} and COMSOL\cite{comsol}.

While the Maxwell stress tensor is often quoted as the canonical textbook method, equivalent integrals - for example over volume rather than area are often used\cite{wiecha2022pygdm,chaumet2000coupled}, these have different computational resource scalings and numerical convergences for each method.


It should be noted that while the optical potential in the Mie-Lorentz regime is still defined by the gradient force, momentum carried in the propagation direction of the light field in the Mie-Lorentz regime is significant and creating a stable trap usually requires this force to be balanced against another optical beam, field, or gravity.


In the geometric optics limit, in which the optical size parameter is much larger than one, \textcolor{black}{forces are generally given by can be computed by the ray optics model, or extensions thereof \cite{sindhu2022optical,shao2019calculation}}. \textcolor{black}{ In the context of an optical cavity, modal analysis can be used to derive the relevant interactions with optical forces \cite{sidles2006optical}. For example, this is the method used to derive effects such as tilt induced instabilities within advanced LIGO \cite{savov2006estimate}.}
\textcolor{black}{In the context of large optically levitated test masses, there are often practical limitations in transferring adequate power to counteract gravity. Some recent levitated optomechanics experiments have taken place within this regime\cite{PhysRevA.101.053857,PhysRevLett.105.213602}.}


\subsubsection{Trapping configurations}

A variety of trapping architectures have been presented
in the literature. A selection of these is presented within figure \ref{trapping_configurations}. These include single \cite{gieseler2013thermal}, dual \cite{winstone2022optical} and multi-\cite{rondin2017direct} beam tweezers. For single beam dipole traps, the required focusing of the tweezing beam has been obtained by using high-numerical aperture (NA) lenses \cite{PhysRevLett.109.103603} as well as parabolic reflectors \cite{PhysRevLett.121.253601}. Counter propagating beams do not require high NA optics due to the self cancellation of the scattering force - which allows larger particles to be trapped than single beam configurations. Configurations with \cite{Kiesel2013} and without \cite{Novotny2012}  optical cavities have been realized. Optical tweezers have also been combined with other confinement devices such as Paul traps \cite{PhysRevLett.117.173602}.

\subsection{\textcolor{black}{Equations of motion and dynamics of an optically trapped particle}}

\textcolor{black}{In addition to the optical confinement forces and force due to photon scattering and photon recoil-heating, several other forces generally act on the trapped particle, including the force due to Earth's gravity, forces due to collisions with background gas molecules, as well as any other applied external forces.} 
\textcolor{black}{In general, the motion of an optically confined dielectric nanoparticle is subject to thermal noise and associated damping due background gas collisions, as well as laser noise. Laser noise includes both technical noise, e.g., vibration, pointing noise, and relative intensity noise as well as fundamental noise, e.g., photon shot-noise and its associated fluctuations in radiation pressure which impart forces due to photon recoil). }

\textcolor{black}{The motion of an optically trapped particle is well described by 
the classical equations of motion of an oscillator as long as its energy is large compared to the corresponding ground state energy. 
Although the center of mass motion temperature is strictly speaking only defined for a large enough ensemble of particles, we can still formally equate the average kinetic energy to $k_{B}T_{\mathrm{cm}}$. Further, using the equipartition theorem the average potential energy of each degree of freedom is equal to its kinetic energy $E_{\mathrm{pot}}=E_{\mathrm{kin}}=1/2k_{B}T_{\mathrm{cm}}$. If we set the potential energy to be the ground state energy of a harmonic oscillator $\hbar\omega$, we obtain a temperature at which the classical description breaks down. For a typical oscillation frequency of $\omega = 2\pi\times 100$\si{\kilo\hertz}, the classical center of mass motion temperature must be much higher than:
\begin{equation}
    T_{\mathrm{cm}} \gg \frac{\hbar \omega}{k_B} \approx 5\si{\micro\kelvin} 
\end{equation}}

\textcolor{black}{In the vicinity of the potential minimum, we can approximate any binding potential with the harmonic potential and obtain the following equations of motion for each degree of freedom ($q = x, y, z$)
\begin{equation}
    \ddot{q}+\gamma_g\dot{q}+\Omega_{q}^2q=\frac{F_{\mathrm{th}}}{M}\eta(t)+\frac{F_{\mathrm{qba,q}}}{M}\eta_{\mathrm{qba,q}}(t)+\frac{F_{\mathrm{ext,q}}(t)}{M}.
\end{equation}
Here the last term includes the constant gravitational force and time-averaged value of the scattering force, as they only lead to a shift in the particle's equilibrium position but do not otherwise affect the dynamics, as well as any other external perturbing forces, e.g. due to technical laser intensity noise. Here $\gamma_g$ is the gas damping constant, $\Omega_q$ is the mechanical frequency, and $\eta$ is a random variable. The random acceleration $\frac{F_{\mathrm{qba}}}{M}\eta_{\mathrm{qba}}$ describes the effect of fluctuating force due to radiation pressure shot noise, i.e. quantum back action, which includes photon recoil heating.}

\begin{figure}
    \centering
    \includegraphics[width=0.95\linewidth]{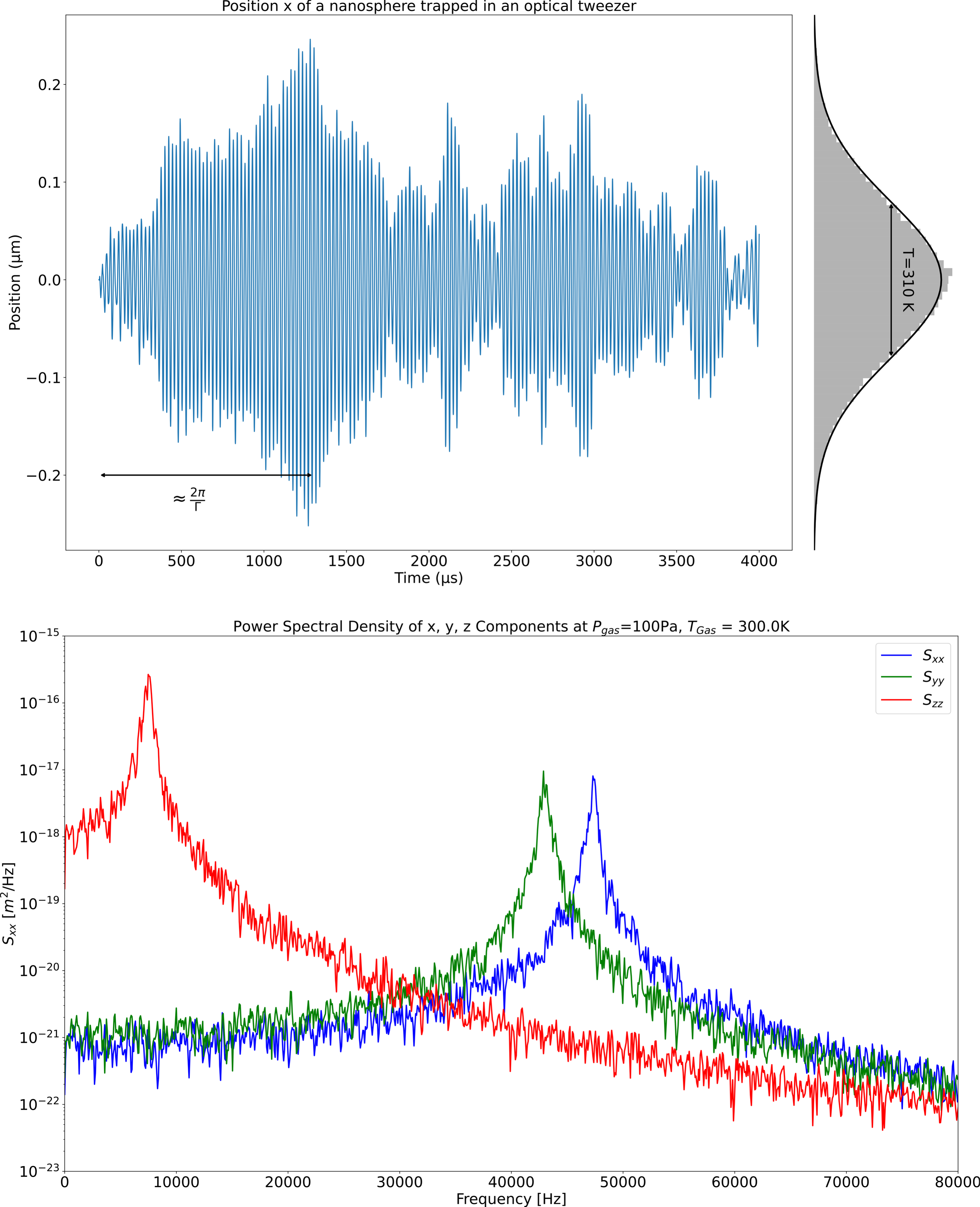}
    \caption{The upper plot shows a sample of the simulated motion (x-axis, 4\si{\milli\second}) of an optically trapped nanosphere (silica,  200 \si{\nano\metre} diameter, 1 W power, 2 \si{\micro\metre} waist, $\omega_x = 47.328$ \si{\kilo\hertz}, $\omega_y = 43.025$ \si{\kilo\hertz}, $\omega_z = 7.505$ \si{\kilo\hertz}, gas pressure is $1$ mbar, temperature 300 \si{\kelvin}). The histogram on the right side shows the position distribution fitted with the Maxwell distributions. The characteristic mechanical oscillation frequency and the damping rate $\Gamma = \gamma_g$ are clearly visible. The lower plot shows the corresponding power spectral densities for all three coordinated.}
    \label{fig:time_trace}
\end{figure}

\textcolor{black}{As we further motivate and discuss in Sec. \ref{sec:force}, the thermal fluctuating force satisfies $ \langle F_{\mathrm{th},q} (t) F_{\mathrm{th},q'} (t') \rangle =2 M \gamma_g  k_B T_{\mathrm{gas}} \delta_{qq'} \delta (t-t')$. The Dirac delta function signifies the Markov approximation, i.e. the lack of memory of the random collisional process of collisions with the background gas molecules. This corresponds to a power spectral density \begin{equation*}
    S_{FF}(\omega) = \int_{-\infty}^{\infty} e^{i \omega \tau}  \langle F_{\mathrm{th}} (0) F_{\mathrm{th}} (\tau) \rangle 
\end{equation*}of the thermal force, which for the case of gas damping is given by 
\begin{equation}
    S_{FF} = 2Mk_BT_{\mathrm{gas}}\gamma_g.
\end{equation}
The presence of the damping constant $\gamma_g$ in the power spectral density of the thermal force that arises from random gas collisions with the particle has an  intuitive explanation. Random collisions of a moving particle cause damping by converting its kinetic energy into heat (coupling to the bath).
This fundamental relation between damping and thermal noise is called \textit{fluctuation-dissipation theorem} and can be applied to various physical situations \cite{CallenWelton1951}. 
If the steady state energy e.g. in thermal equilibrium is $E_\infty$, 
the rate of the change of the average energy is \begin{equation*}
   \frac{d}{dt} \bar{E} = -\gamma_g (\bar{E}- E_{\infty}).
\end{equation*}
Describing the energy in terms of phonons with an average occupation $\bar{n}_q=k_B T_{\mathrm{gas}} / (\hbar \Omega_q)$, we can normalize the heating power $\gamma_g E_{\infty}$ in terms of the number of phonons to arrive at the thermalization rate
\begin{equation}
    \Gamma_{\mathrm{th}} \equiv \gamma_g \frac{k_B T_{\mathrm{gas}}}{\hbar \Omega_q}
\end{equation}
which has units of Hz. The term $\gamma_g/\Omega_q$ we also recognize as the inverse of the mechanical quality factor $Q$, describing 
the ratio of the total energy to the energy dissipated per cycle of oscillation of the harmonic oscillator. In the high vacuum regime, mechanical $Q$ factors of optically trapped nanoparticles in excess of $10^8$ have been observed \cite{Gieseler2013}, and in a Paul trap recent work demonstrated a $Q$ of order $10^{10}$ \cite{Northup2024}.
Fig. \ref{fig:time_trace} shows a simulation of a typical displacement time-series and corresponding Fourier spectrum of a nanoparticle undergoing Brownian motion in an optical tweezer at an intermediate vacuum of 1 mbar. }

\textcolor{black}{The force due to radiation pressure fluctuations depends on the scattered power $P_{\mathrm{scat}}$ \cite{jain2016direct} as the total force acting in on the particle from scattering $F^{q}(\omega) = P^{q}_{\mathrm{scat}}(\omega)/c$ for photons scattered in the $q$-th direction. For a dipole scatterer with $\hat{x}$-polarized radiation incident on it, the the power spectral density of force fluctuations is given by 
$S^{y}_{FF,qba} (\omega) = \frac{2}{5} \frac{\hbar \omega_0}{2 \pi c^2} P_{\mathrm{scat}}(\omega)$ assuming shot noise is the dominant source of fluctuations \cite{jain2016direct}. Here $\omega_0$ is the trap laser frequency, and the numerical factor of $2/5$ accounts for the angular distribution from a dipole polarized in the $\hat{x}$ direction. An identical expression is given for the $\hat{z}$-direction, while the expression for the $\hat{x}$ direction is a factor of two smaller. This is since for the dipole radiation pattern, radiation emission is suppressed along the direction of polarization, i.e. ($\hat{x}$) in this case.  The corresponding phonon heating rate due to photon recoil heating is given by
\begin{equation}
    \Gamma_{\mathrm{sc},q}=\frac{1}{5} \frac{P_{\mathrm{scat}}}{Mc^2} \frac{\omega_0}{\Omega_q}
\end{equation}
for either $q=y$ or $q=z$, whereas in the $\hat{x}-$ direction this rate is reduced by half.}
\subsubsection{Gas damping and radiometric heating}


\textcolor{black}{For gas collisions, two relevant length scales are the mean free path of the gas molecules and the size of the levitated object.  For pressures where the mean free path exceeds the size of the particle, gas collisions can be regarded as being in the ballistic regime. Diffusive behavior occurs in the opposite regime where the mean free path is very short compared to the size of the trapped particle. } In particular, at higher pressures (>~50mbar depending on object size and geometry) the motion is completely dominated by random momentum interactions with the background gases and undergoes a random walk around the optical potential. At lower pressures (<10mbar) - where the damping is linear and no longer the leading term in the equation of motion, the levitated object behaves like a driven damped harmonic oscillator, manifesting as a Lorentzian peak in frequency space for each degree of freedom, with a width defined by the sum of its damping processes.
A more detailed discussion of gas damping and Brownian motion is also included in Sec. \ref{sec:thermo}.

In the intermediate regime, where the mean free path of the gas molecules is of order the size of the dielectric particle to be trapped, additional instabilities are possible due to radiometric forces, which can dominate over the confining forces due to radiation pressure, as was first studied by Ashkin and coworkers in the late 1970s \cite{Ashkin:1977}. Thus one of the challenges of reaching high vacuum \textcolor{black}{when a particle is initially trapped at high pressure} is to overcome radiometric forces. Of prime importance is to minimize heating of the particle surface from optical absorption, as temperature gradients across the surface of the levitated object can drive currents of gas molecules which are de-stabilizing over a range of pressure spanning orders of magnitude. A detailed study of these effects was performed in Ref. \cite{Atherton:2015} for trapped silica microspheres. Heating from optical absorption can be mitigated by choice of material as well as trapping laser wavelength. As described later in Sec. \ref{sec:thermo}, solid-state laser cooling may enable negligible heating of the particle surface, or cooling substantially below room temperature.
To compensate for radiometric forces during pump-down, laser feedback cooling at a rate \textcolor{black}{$\gamma_{\mathrm{fb}}$} can be utilized to restore damping that is no longer provided by the background gas being evacuated from the trapping chamber. \textcolor{black}{Alternatively, directly trapping particles at high vacuum conditions by use of laser cooling, a slowing laser beam, or time-dependent potentials is a possible promising route towards avoiding radiometric forces. direct "trapping" at UHV has also been demonstrated by trapping in a second chamber at intermediate pressure and then passing the nanoparticle through via a hollow core fiber link \cite{lindner2023hollow} }

\subsubsection{\textcolor{black}{Feedback cooling of the center-of-mass}}

\textcolor{black}{Feedback cooling of the center of mass motion of the trapped nanoparticle is often used to faciliate stable trapping at high vacuum condiitions, provided that optical absorption does not prohibit this due to internal heating. Methods of feedback cooling are discussed further in Sec. \ref{subsec:optomech}.}

For each degree of freedom, the temperature or \textcolor{black}{mean steady state phonon occupation number $n_\infty$} is defined by the ratio of the heating and cooling processes in the system \cite{jain2016direct}:

\begin{equation}
n_{\infty} = \frac{\Gamma_{\mathrm{th}} + \Gamma_{\mathrm{fb}}+\Gamma_{\mathrm{sc}} +\Gamma_{\mathrm{other}} }{\gamma_{g} + \gamma_{\mathrm{fb}} + \gamma_{\mathrm{photon}}}
\end{equation}

where $\Gamma_{\mathrm{th}}$ describes the heating corresponding to interactions with background gases, $\Gamma_{\mathrm{sc}}$ accounts for photon recoil scattering events, $\Gamma_{\mathrm{fb}}$ heating events corresponding to phase imprecision in the feedback loop, and $\Gamma_{\mathrm{other}}$ accounts for other noise sources e.g. technical noise. The      lower case gammas $\gamma$ correspond to cooling events from the same physical interactions, i.e., $\gamma_g$ is the gas damping rate, $\gamma_{\mathrm{fb}}$ is the feedback cooling damping rate, $\gamma_{\mathrm{photon}}$ is due to damping from the velocity dependent doppler shift of scattered photons (typically a small effect) \cite{jain2016direct}. Each of these pairs can be thought \textcolor{black}{of} as the levitated object's connection to a heat bath. At higher pressures the dominating term is the collisions with background gases, and at \textcolor{black}{lower pressures photon recoil heating or technical laser noise dominates.}

In the limit of negligible other technical heating and noise sources, i.e. $\Gamma_{\mathrm{other}}=0$, as well as negligible heating from feedback imprecision $\Gamma_{\mathrm{fb}} \approx 0$ and  $\gamma_{\mathrm{photon}} \approx 0$, 
we arrive at the theoretical mean steady state phonon number for this \textcolor{black}{limiting} case
\begin{equation}
n_{\infty} \approx \frac{\Gamma_{\mathrm{th}}+\Gamma_{\mathrm{sc}}}{\gamma_{g}+\gamma_{\mathrm{fb}}} \approx \frac{n_i \gamma_g + \Gamma_{\mathrm{sc}}}{\gamma_g+\gamma_{\mathrm{fb}}}
\end{equation}
where in the last equation we specialize to the case of a nanosphere optically levitated with trap frequency $\omega_0/2\pi$ in contact with a thermal gas at temperature $T$ where $n_i=k_B T / \hbar \omega_0$.

The time required to enter equilibrium (thermalize) with any given heat bath is related to the heating rate of the process (in Hz) and the frequency of the oscillator, and can be on the order of milliseconds to seconds depending on the process and frequency, providing an accessible  platform for thermodynamics, and studies of levitated interactions with non Markovian system noise\cite{ren2022event}, as we discuss futher in Sec. \ref{sec:thermo}.  




\subsection{Particle sources and materials for optical levitation}
\subsubsection{Sources}
A variety of methods for loading nanometer to micron sized objects into optical and hybrid traps exist, such as PZT driven launchers\cite{li2012fundamental}, electrosprays, nebulisers\cite{summers2008trapping} and Laser-Induced Acoustic Desorption (LIAD) methods \cite{nikkhou2021direct}.

Particle sources are typically matched to the size, shape and material of the desired levitated species, as well as the type of trap being used. For smaller objects in the 10's or 100's of nm regime, nebulisers have historically been used \cite{summers2008trapping} which disperse a colloidal solution of suspended nanoparticles into the trapping region. These require operation at atmospheric pressures and typically contaminate the chamber with whatever liquid is used for the colloidal suspension. This problem has been overcome to some degree by separating the trapping and science chambers.


Larger objects in the 100's of nm to 10's of $\mu m$ regime have often been trapped using PZT driven mechanical launchers, in which the particle's are physically detached from a surface by vibration.  Ref. \cite{weisman2022} describes an apparatus which uses a glass substrate clamped to a piezoelectric transducer for in-vacuum loading of dielectric nanoparticles into an optical trap. The device has been shown to generate accelerations of order $10^7$ $g$, which is sufficient to overcome stiction forces between glass nanoparticles and a glass substrate for particles as small as $170$ nm diameter, consistent with theoretical expectations. Using this device, single spherical particles with sizes ranging from $170$ nm to $3$ $\mu$m \cite{Atherton:2015,ranjit2016zeptonewton,montoya2022scanning}, as well as clumps of such particles \cite{Atherthesis}, have been successfully loaded into optical traps at pressures ranging from 1 bar to 0.6 mbar. 
Similar work was also recently reported and described in Ref. \cite{aspelmeyerlaunch}. For illustrative purposes, Visualization 1 features a short video clip showing particles being launched downwards towards a vertically oriented optical levitation trap, resulting in the capture of a silica microsphere of diameter $3$ $\mu$m.

Pulsed laser launchers directly transfer momentum from a light field to an area of nanofabricated objects on substrate, kinetically directing them towards a trapping area. Like the PZT sources, this has a distinct advantage of being able to to launch at vacuum, however its high launch velocities does make it better suited to a deep trapping potential. This source also has the advantage of not requiring an intermediate solution of material to be either loaded into a nebuliser or prepared on a microscope slide - as is the case for the PZT method. For the PZT method, the velocity distribution for a variety of particles launched from the substrate was studied in Refs. \cite{weisman2022,aspelmeyerlaunch} and results indicate promise for direct loading into ultra-high-vacuum with sufficient laser feedback cooling.   In contrast to  LIAD, where particles tend to be launched with speeds well in excess of $1$ m/s \cite{nikkhou2021direct}, the PZT method produces sufficient flux of particles with initial launch velocity less than $1$ m/s to in principle allow direct ultra-high-vacuum trapping with the assistance of a slowing laser beam or other arresting field. 


\subsubsection{Materials and morphologies}

\begin{figure}[htbp]
\begin{center}
	\includegraphics[width=1.0\columnwidth]{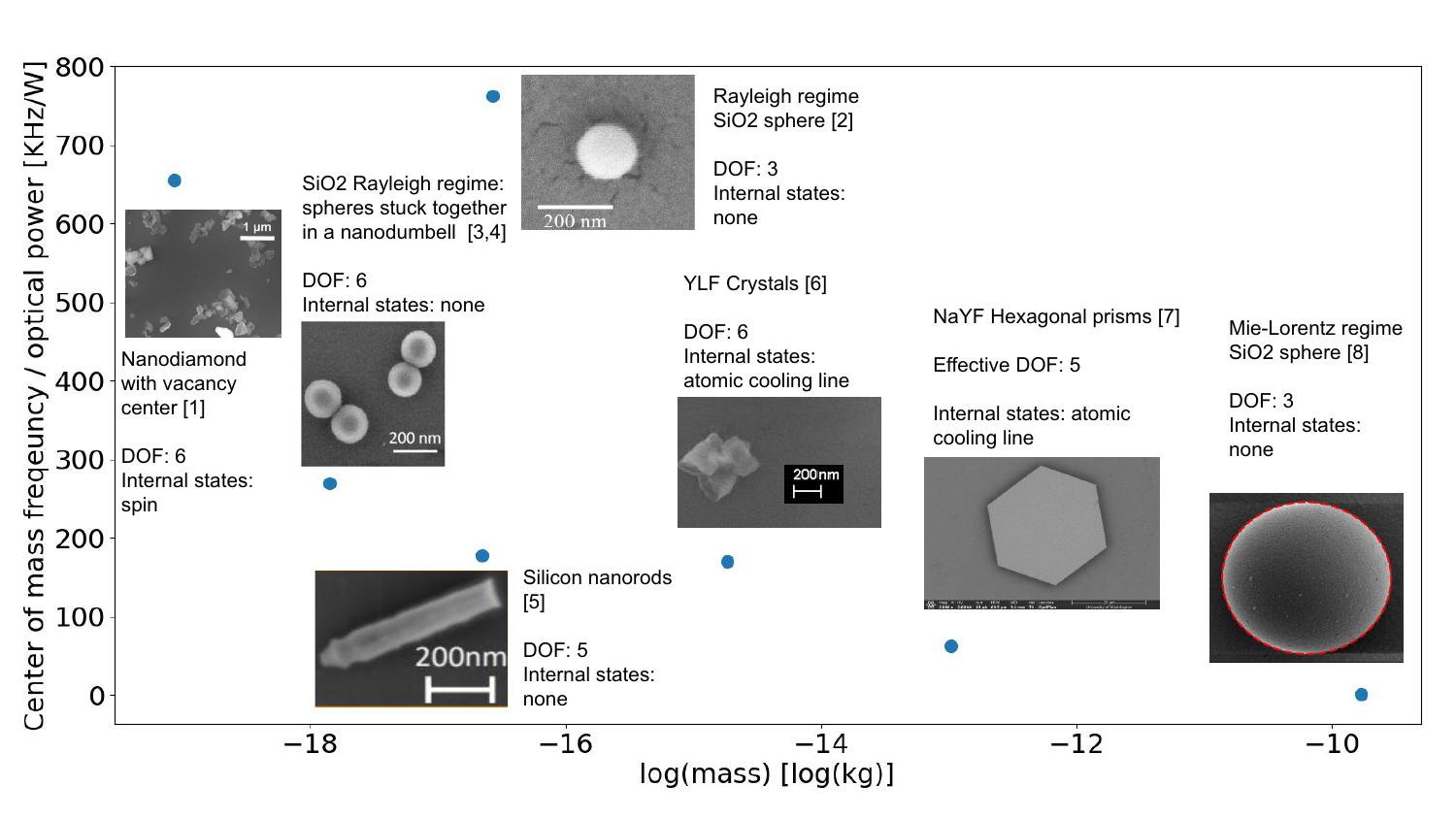}
	\caption{A sampled zoology of optically trapped species, different geometries and materials give different optical couplings, internal states and degrees of freedom. 1: Nanodiamonds\cite{frangeskou2018pure}, 2: Rayleigh regime spheres \cite{delic2020cooling},3,4: Nanodumbells, combined spheres and agglometerates \cite{ahn2018optically},5: Silicon nanorods \cite{kuhn2017full},6: YLF crystals \cite{rahman2017laser},7: NaYF high-aspect-ratio hexagonal plate \cite{winstone2022optical},8: Mie-Lorentz Regime spheres \cite{blakemore2019precision}. The ratio of the COM frequency and the optical power can be used as a limited metric of the optomechanical coupling. 
	\label{zoology}}
	\end{center}
\end{figure}

Many different shapes and materials of the levitated object have been investigated in optical trapping experiments. Each combination and permutation providing different tunable properties to match the desired experiment, such as optical coupling, dispersion relation, a controllable distribution of the scattered light and additional degrees of freedom. Example of different investigated geometries include; SiO2 spheres from 10's of nm \cite{delic2020cooling} \cite{jain2016direct} to 10's of $\mu$m \cite{monteiro2020force}, NaYF disc like objects \cite{winstone2022optical}, prolate spheroids \cite{rashid2018precession}, nanorods\cite{kuhn2017full}. A sampled parameter space of these is given in Figure \ref{zoology}.

Control over the specific material properties is especially and increasingly important given that such attributes of objects at this scale frequently deviate from those of the bulk, such as the (lower) density of silicon dioxide spheres made from the stober process for example \cite{blakemore2019precision}. This has spurred efforts in generating/creating/engineering new levitatable nanomaterials with bespoke desirable properties. Experiments to date include vaderite spheres \cite{ghosh2019fabrication}, and species with potentially useful internally accessible states: nanodiamonds\cite{rahman2016burning} for a spin degree of freedom, 
 NaYF\cite{Luntz-Martin:21,winstone2022optical}\cite{laplane2024inert} and YLF\cite{rahman2017laser} crystals for optically accessible internal state cooling. Composite test masses composed of multiple materials may be advantageous for certain science applications, for example, core-shell particles to enhance internal state cooling efficiency \cite{laplane2024inert}. \textcolor{black}{Further discussion on these possibilities is provided in Sec. \ref{sec:hybrid} and \ref{sec:thermo}.}

\subsection{Detection \label{sec:detection}} 
\begin{figure}[hb]
\centering
	\includegraphics[scale=0.99]{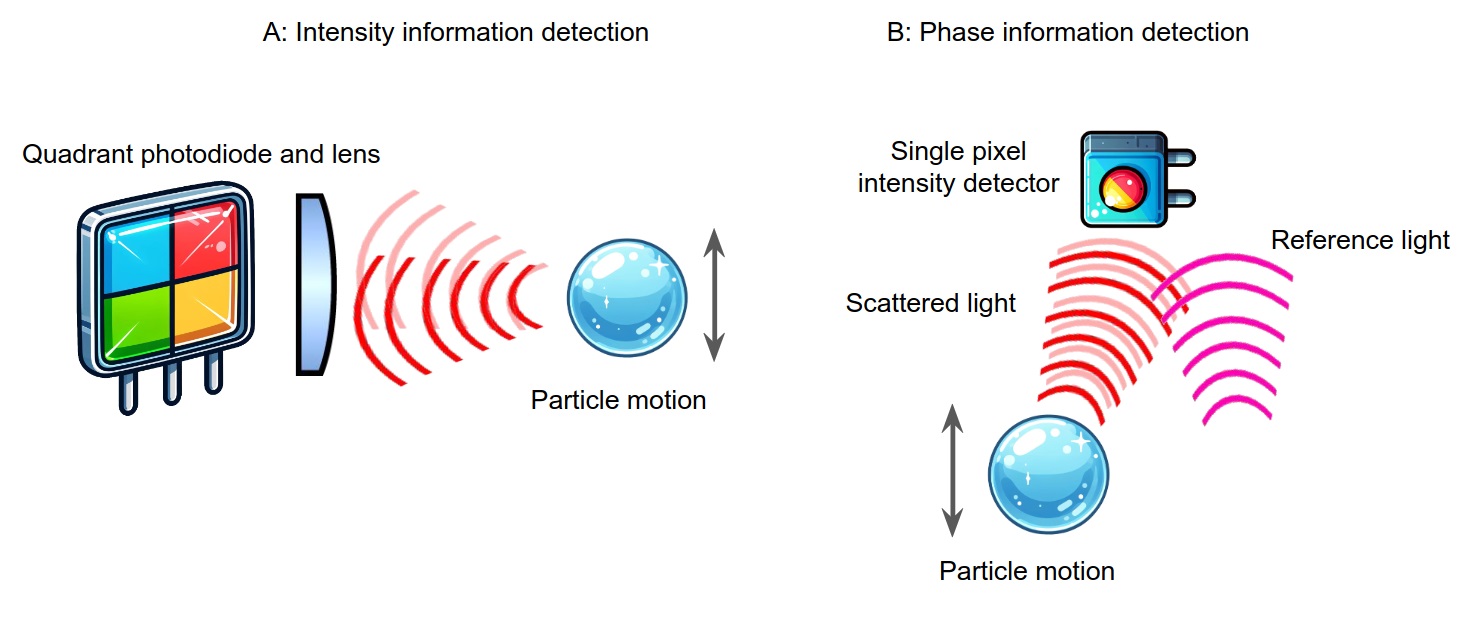}
	\caption{\textcolor{black}{Information carried by different quadratures of a light field scattered from a levitated Rayleigh scale nanosphere and their respective detection mechanisms. A: Information from light intensity varying over time recorded on a detector - often a position sensitive detector such as a quadrant photodiode (QPD). This mechanism is  similar to a microscope and is generally not sensitive to information encoded in the phase of the scattered light. B: Scattered light is interfered with a local oscillator reference light, sensitive to information encoded in the phase of the scattered light. In many ``homodyne like'' schemes, the reference light is light that traversed through the trapping zone of the levitated object and did not interact with it - however, it should be pointed out that this approach does not allow for the phase of the local oscillator reference field to be varied for homodyne state tomography (such as in Ref. \cite{militaru2022ponderomotive})}.
}
	\label{detection_mechanisms_figure}
\end{figure}

\textcolor{black}{
Detecting a particle's position and momentum is necessary both to cool its center of mass motion and to precisely sense external forces. Two optical methods are commonly used to infer the particle's position: (1) interferometry and (2) imaging the particle using scattered light.} 
In experiments with cavities, the mechanical displacement information can be found from the transmission through or reflection from the optical cavity \cite{RevModPhys.86.1391}, where the suspended object acts as a dispersive element modifying the refractive index of the cavity depending on its position \cite{Chang:2010}. In experiments without cavities, oscillator displacement may be obtained by interfering the light scattered by the nanoparticle with a local reference field \cite{PhysRevLett.109.103603}, or by directly observing the scattered light of the particle in a spatially sensitive detection system \cite{ranjit2016zeptonewton}, as illustrated in Fig.  \ref{detection_mechanisms_figure}.

\textcolor{black} {For interferometric detection using a reference laser, the amplitude of the scattered light, also herein referred to as the signal light, will be modulated by the varying intensity of the beam at different positions, while the phase of the emitted light with respect to a fixed detector is proportional to the optical path difference between them, which also depends on the particle's position. 
While amplitude modulation can be detected by just measuring the scattered light with a photosensitive detector (photodiode, photo-multiplying tube, etc.), any phase measurement requires a reference phase. 
The frequency of the reference laser can either be the same as the scattered light (e.g. as in homodyne detection) or different - referred to as heterodyne detection (where the difference typically is chosen in the MHz range).  Consider two interfering fields - the reference $\mathbf{E}_{\mathrm{ref}} = \text{Re}(\mathbf{E_{r}}e^{i\phi_r})$ and the signal $\mathbf{E}_{\mathrm{sig}} = \text{Re}(\mathbf{E_{s}}e^{i\phi_s})$. The signal at a photosensitive detector will be proportional to the power of the incident light.
\begin{equation}
P_{\mathrm{PD}} \propto |\mathbf{E}_{r}|^2+2\mathbf{E}_{r}\cdot\mathbf{E}_{s}\cos{(\phi_s-\phi_r)}+|\mathbf{E}_{s}|^2
    \label{eq:interference}
\end{equation}
The last term can be neglected since the power of the reference beam is typically chosen several orders of magnitude larger than that of the scattered light (e.g., a few milliwatts vs. nanowatts to microwatts). This helps increase the photodetector signal above the detector noise. It also reduces the uncertainty of the phase measurement due to the quantum shot noise of the light, which is equal to the square root of the photon number $\sigma_{\Delta\phi}=1/(2\sqrt{N_{\text{ph}}})$. By using a balanced (differential) photodetector and a split-off of the reference laser, we can get rid of the constant offset $|\mathbf{E}_{r}|^2$ and the associated laser noise and are left only with the interference term. By choosing the phase of the reference laser to match the signal phase $\phi_r=\phi_s$, the cosine term becomes unity, and we are left with pure amplitude modulation. On the contrary, by choosing the reference phase to lag behind the average signal phase by ninety degrees, we obtain a sine of the signal phase and by dividing out the previously obtained amplitude quadrature, we end up with a linear pure phase modulation
\begin{equation}
P_{\phi} \propto \sin{(\phi_s) \approx \phi_s}
    \label{eq:phasequadrature}
\end{equation}
Here, we assume that the signal phase is small and use the small angle approximation. }

\subsubsection{Optimal detector placement}

At each point in space at which we might place a detector, the light field carries information in its intensity and phase quadratures. 
Solutions for optimal placement of the detector with respect to the trapping location can be found in Ref.  \cite{tebbenjohanns2019optimal} for Rayleigh sized particles, and in Ref.  \cite{maurer2022quantum} for Mie-Lorentz regime sized spheres. 
\begin{figure}[htbp]
\begin{center}
	\includegraphics[width=1.0\columnwidth]{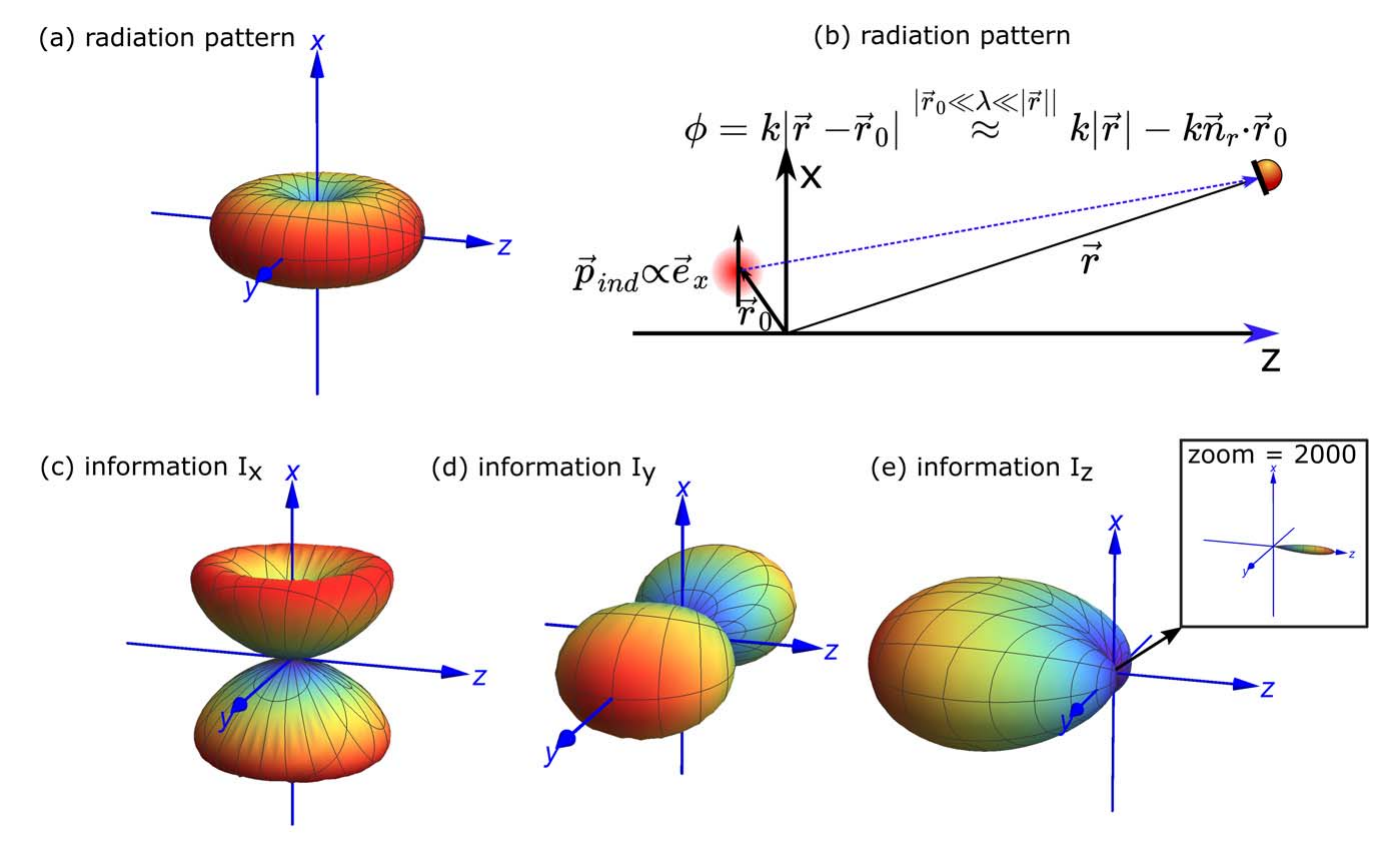}
	\caption{\textcolor{black}{(a) Dipole emission pattern. The dipole oscillates along the x-axis. (c,d,e) Dipole radiation information patterns  $I_{x/y/z}$  \cite{tebbenjohanns2019optimal} for the x, y, and z coordinates of the center of mass motion of a nanosphere trapped in a focused single beam optical tweezers. The information radiation pattern $I_{i}(\theta,\phi)$ is defined as the inverse of the power spectral density of the imprecision noise (for more details see \cite{tebbenjohanns2019optimal}). It characterized the lowest possible position power spectral density noise floor as given by the photon shot noise and obeys the imprecision-back action uncertainty principle. (b) The emitted electric field of an induced dipole (e.g. dielectric nanosphere) in a focused laser beam as measured by a detector in far-field at $\boldsymbol{r}$ is given by $\boldsymbol{E}_{\mathrm{sc}}(\boldsymbol{r})=\boldsymbol{E}_{\mathrm{dip}}(\boldsymbol{r})\exp[-ik(\boldsymbol{r}_{0}\cdot \boldsymbol{n}_{r}-Az_{0}]$. Here $\boldsymbol{E}_{\mathrm{dip}}(\boldsymbol{r})$ is the strength of the dipole field in the direction of the detector $\boldsymbol{n}_{r}$,  $\boldsymbol{r}_{0}$ is the position vector of the particle, and $k$ is the wave number and $Az_{0}$  is a term related to the Gouy phase of the beam. Although the dipole emits light equally along the forward and backwards direction along $\hat{z}$ according to (a), 
 the information measurement in positive the z-direction (beam propagation direction) is suppressed due to the Gouy phase term $Az_{0}$ which cancels the particles position phase term to a large extent. The small 2000-times magnified term is shown in the inset of the plot (e). 
	\label{fig:irp}}}
\end{center}
\end{figure}
In terms of practical experiment construction, phase bound information is recovered by interfering the scattered light from the levitated object with a local reference field - conditioned for maximum interference and spatial mode overlap, while in the case of intensity modulated detection the objects motion is usually literally imaged on a spatially sensitive detector. 
A review of the ultimate sensitivity limits of both methods (for Rayleigh regime sized spheres) may be found in Ref. \cite{tebbenjohanns2019optimal}.
For example, when collecting scattered light from a spherical particle small enough to be within the Rayleigh regime, the backwards scattered radiation carries far more information about the axial displacement of the particle along the laser beam than the forward scattered light \cite{tebbenjohanns2019optimal}. This phenomenon is illustrated in Fig. \ref{fig:irp}. 

\subsubsection{Interpreting data - conversion and calibration of displacement signal, e.g. volts to meters}

To determine the physical displacement of a trapped nanoparticle e.g. in meters, it is necessary to convert the signal from the photosensitive detector(s) used to monitor the scattered light from the particle from volts to meters.  
Given any systematic errors in the conversion of photons to volts to nanometers, it is important to calibrate the sensor under its final operating conditions, ideally with a reference to a known length scale, or known force, for example from an electrostatic calibration \cite{ranjit2016zeptonewton}. For example depending the placement of the detector upon on the tabletop, drifts in path lengths or fluctuations in beam pointing due to thermal drifts could affect the measurement. 
For example, when trapping non-spherical particles, the location of the detectors can be chosen to either suppress or enhance sensitivity to rotational information. 
\textcolor{black}{Nonlinearities present in the motional spectrum of the particle e.g. due to Duffing nonlinearities can make it challenging to calibrate the motional spectrum assuming it is a linear oscillator coupled to a thermal bath in equilibrium. It has been demonstrated that this is a deterministically nonlinear function \cite{rondin2017direct}. Also absorbed laser power can increase the surface temperature of the trapped particles and result in larger motional amplitudes for the particle than would be present in thermal equilibrium at the ambient temperature of the vacuum chamber. Thus the experimental vacuum pressure, which determines the thermalization rate from gas collisions, along with the degree of nonlinearities can significantly affect the volts-meters conversion.} While a complete discussion is beyond the scope of this tutorial article, references for the interpretation of detector signals and transforming then into physical displacement units can be found in Refs. \cite{hebestreit2018calibration, dawson2019spectral}.
Finally, we note that in the limit of the oscillators occupation number being small enough such that quantum effects become relevant, the center of mass temperature assigned to the particle's external degrees of freedom may also be found by sideband thermometry \cite{PhysRevLett.124.013603,PhysRevLett.123.153601}.

\subsection{Quantum optomechanics \label{subsec:optomech}}
The dynamics of the light-matter interaction between the tweezer (or other) beams and the nanoparticle motion are rich and have been investigated in two main types of configurations. The first of these involves optical cavities which contain levitated nanoparticles \cite{Chang:2010}, see Fig.. Here the coupling between the nanoparticle and optical cavity mode is dispersive, meaning the motion of the nanoparticle only shifts the cavity mode resonance frequency. This configuration can be described using the standard cavity optomechanics Hamiltonian \cite{RevModPhys.86.1391} 

\begin{equation}\label{eq:HLin}
    H=\hbar \omega_{c}a^{\dagger}a+\hbar \omega_{m}b^{\dagger}b+\hbar ga^{\dagger}a(b^{\dagger}+b)+H_{o}+H_{m},
\end{equation}
where $\omega_{c}(\omega_{m})$ is the cavity mode (mechanical) frequency, $a(a^{\dagger})$ and $b(b^{\dagger})$ are the creation and annihilation operators for the optical mode and nanoparticle motion, respectively, obeying $[a,a^{\dagger}]=1, [b,b^{\dagger}=1]$, and $g$ is the optomechanical coupling. It should be mentioned that the to realize the given kind of optomechanical coupling, linear in the particle position $\sim a^{\dagger}a( b^{\dagger}+b)$, the nanoparticle has to be positioned between a node and an anti-node of the standing wave cavity mode (i.e. at an intensity gradient). A quadratic coupling, proportional to the square of the particle displacement, $\sim a^{\dagger}a( b^{\dagger}+b)^{2}$, can also be engineered, by positioning the nanoparticle at the antinode of the cavity mode (i.e. at an intensity maximum). The applications of the linear and quadratic couplings, which may be implemented simultaneously using two cavity modes, will be mentioned below. The additional terms $H_{o}(H_{m})$ describe the coupling to the environment which is responsible for optical (mechanical) damping at the rate of the cavity (mechanical) linewidth $\gamma_{o}(\gamma_{m})$
and optical as well as thermal fluctuations. The Hamiltonian of Eq.~(\ref{eq:HLin}) applies to experiments where shot noise, blackbody radiation and particle asphericity are negligible and the nanosphere's internal degrees of freedom are decoupled from its center of mass motion.

In principle, the tweezer-in-a-cavity principle can be used to realize standard optomechanical effects such as cooling the nanoparticle to its vibrational ground state, ponderomotive (mechanics-induced) quadrature squeezing of the optical mode, and quantum state transfer between the mechanics and optics \cite{RevModPhys.86.1391}.
Ground state cooling, for example, can be implemented by trapping the nanoparticle using a quadratic optomechanical coupling and cooling its motion by using a linearly-coupled optical mode (quadratic-coupling-based cooling has also been shown \cite{PhysRevResearch.3.L032022}). However, in practice, there are undesirable limitations. For example, cooling requires large optical intensities, and is limited by co-trapping due to the linearly-coupled mode, cooling rates fall off with power, and laser phase noise degrades performance at the nanoparticle oscillation frequencies \cite{aspelmeyercavity}.

These problems can be overcome by a modification of the cavity-based setup, using the technique of coherent scattering, see Fig. In the standard setup described above, an external laser is responsible both for driving the cavity as well as scattering from the nanoparticle. In contrast, the coherent scattering method is based on a tweezer positionally transverse to the cavity axis, and close in frequency to a cavity mode.
The Hamiltonian for the configuration can be written as \begin{equation}
\label{eq:HamCoh}
H_{\mathrm{coh}}=-\frac{\alpha}{2}|\vec{E}_{tw}|^{2}-\frac{\alpha}{2}|\vec{E}_{cav}|^{2}
-\alpha \mathcal{R}\left(\vec{E}_{tw} \cdot \vec{E}_{cav}^{*}\right),
\end{equation}
where $\alpha$ is the nanoparticle polarizability, $\vec{E}_{tw(cav)}$ is the tweezer(cavity) field and $\mathcal{R}$ denotes the real part. 

The third term in Eq.~(\ref{eq:HamCoh}) is responsible for the interference between the tweezer and cavity fields and thus coherent scattering. Due to this term, the photons scattered by the nanoparticle are preferentially enter the mode of the initially \textit{empty} cavity. The optimum position for the nanoparticle in this method turns out to be the node of the cavity mode. When located at that node, the scattering of photons from the particle into the cavity mode is prohibited by the interference between the mode and the optical tweezer beam. However, as the nanoparticle vibrates, it is displaced from the node and scatters photons, which after Doppler-shifted in frequency due to the particle motion. The Stokes (anti-Stokes) photons are responsible for heating (cooling) the particle motion. Preferential cooling of the particle can be arranged by choosing the frequency of the tweezer beam appropriately near a cavity mode. Although we do not go into the details of the mechanism involved, it should be noted that the polarization of the tweezer plays an important role in the dynamics \cite{aspelmeyercavity}. Coherent scattering techniques have been proposed \cite{PhysRevResearch.3.023071} and demonstrated to eventually cool all six (vibrational and rotational) degrees of freedom of an anisotropic nanoparticle optically levitated in a cavity \cite{Pontin_2023}.

In the configuration described by Eq.~(\ref{eq:HLin}), the cavity provides passive feedback to the nanoparticle motion by recirculating the photons in the optical mode,
while in the coherent scattering technique modeled by Eq.~(\ref{eq:HamCoh}), the cavity serves to stimulate optical scattering at a desired frequency. While cavity-based levitated optomechanics has achieved many milestones \cite{RevModPhys.86.1391}, there are a number of limitations posed by such configurations. For example, the nanoparticle is limited to interact only with those optical wavelengths which are resonant with the cavity, the cavity poses bandwidth limitations, and, perhaps most importantly, it is not easy to freely engineer the light-matter interaction
beyond what is provided by the optical resonator.

\begin{figure}[hb]
\centering
	\includegraphics[scale=0.30]{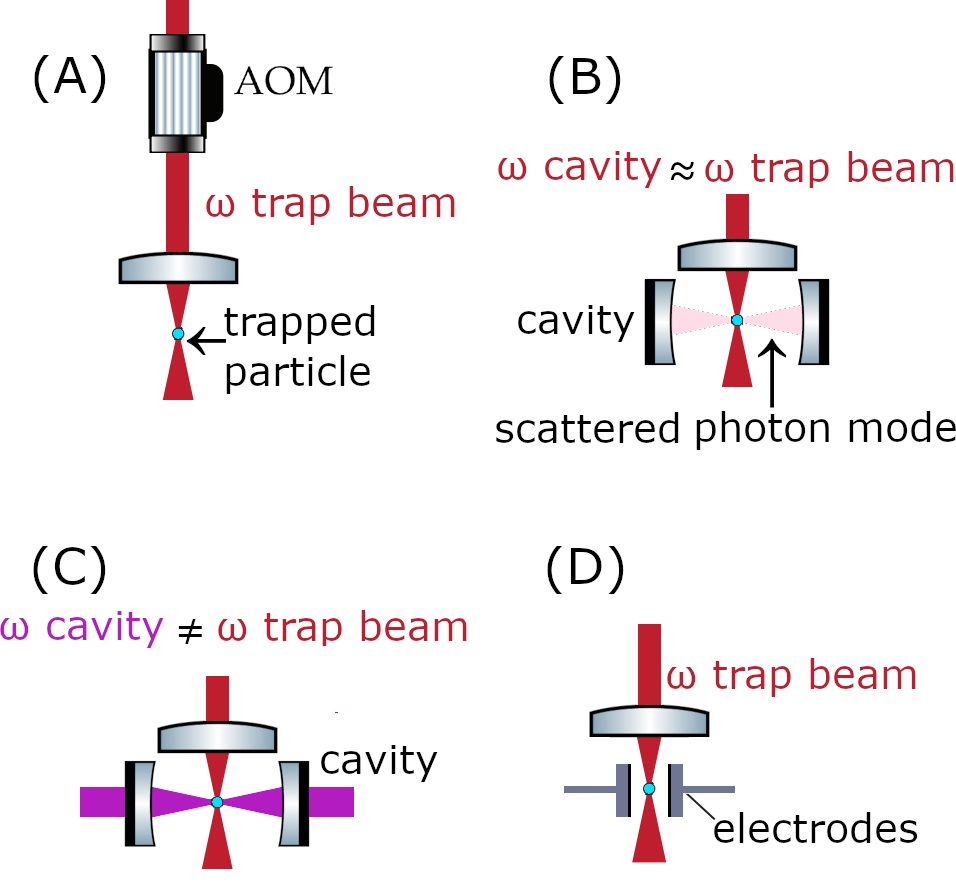}
	\caption{Different types of feedback demonstrated in optically levitated experiments. A) Direct time domain modulation of light intensity via AOM \cite{jain2016direct} - compatible with feedback schemes such as parametric. B) feedback/cooling via an initially empty off axis cavity populated only with photons emitted from levitated species, often referred to as coherent scattering\cite{delic2019cavity}. C) Dissapative optical cavity with a pre-populated cavity mode from a second laser\cite{delic2019cavity}, the wavelength of which is tuned to make the equation of motion of the levitated species dissapative. D) Electrostatic feedback \cite{tebbenjohanns2021quantum} - suitable for linear feedback schemes such as cold damping. Adapted from ref. \cite{delic2019cavity}.
}
	\label{feedback_types}
\end{figure}

These limitations can be overcome by replacing the cavity with active feedback. This kind of feedback works typically by using position detection of the nanoparticle, followed by supplying this information as active feedback through a transfer function implemented by a device such as an electro-optic modulator which can appropriately modulate the optical beam trapping the tweezer, see Fig. \ref{feedback_types}. The optomechanical theory of the active feedback scheme is somewhat more involved than  standard cavity optomechanics. Initial experiments were carried out in the classical regime of nanoparticle motion and can be described using the dynamical equation for a damped oscillator with feedback \cite{Millen:2020review}
\begin{equation}
\ddot{q}(t)+\gamma_{m}\dot{q}(t)
+\omega_{q}^{2}q(t)=A\eta(t)+u_{\mathrm{fb}}(t),    
\end{equation}
where $q(t),\gamma_{m},$ and $\omega_{q}$ are the position, center-of-mass damping, and oscillation frequency of the nanoparticle, respectively. The stochastic noise due to sources such as Brownian fluctuations from the background gas, optical scattering and feedback backaction is represented by the random variable $\eta(t)$ of strength $A$. The feedback applied to the nanoparticle is accounted for by $u_{\mathrm{fb}}(t)$. Various kinds of feedback have been used. Linear damping proportional to the velocity, with $u_{fb}(t)=C_{l}\dot{q}(t)$, has been used to cool the particle motion \cite{Li:2011}, as well as nonlinear damping \cite{gieseler2012subkelvin}, with $u_{fb}(t)=C_{nl}q(t)\dot{q}(t)$. The terms linear and nonlinear are used here with the significance that the particle damping rate is independent of the energy for linear damping and depends on energy for nonlinear damping. A comparison of the efficacy of the different types of feedback for cooling has been carried out \cite{PhysRevA.104.023502}. State base Bayesian models from control theory (such as Kalman filters), have also been demonstrated experimentally for feedback \cite{magrini2021real}.

The cavityless situations considered above regard only a single vibrational mode of the levitated nanoparticle. Both experiments and theory have considered multi-mode optomechanics. These include various combinations of vibration \cite{Kuang2023} as well as torsional \cite{PhysRevLett.117.123604,PhysRevLett.121.253601,PhysRevLett.127.123605} and rotational \cite{PhysRevLett.110.143604} modes \cite{PhysRevLett.126.163603}. Some of these configurations involve composite nanoparticles, like dumbbells \cite{PhysRevLett.110.143604} or nanoplatelets \cite{PhysRevB.96.035402}. Even untrapped nanoparticles have been modified by their transient interaction with a cavity \cite{Asenbaum2013}. For these cases, the mathematical descriptions provided above have to be suitably modified or substituted.

\subsection{Quantum regime and applications}
For levitated optomechanics, the quantum regime is of interest coming from two directions. The first is the 'top-down' direction where the 
system may consist of a mesoscopic or nanoscopic mechanical object and the aim of the investigation is to introduce the object into the quantum domain. This is the motivation behind the preparation of levitated nanoparticles in the ground state \cite{delic2020cooling,tebbenjohanns2021quantum} or squeezed states \cite{rashid2016experimental,Ge_2016,Chauhan_2020}. The second is the 'bottom-up' direction, where the system may consist of a sample of cold \cite{PhysRevLett.105.133602} or degenerate atomic gas \cite{Brennecke_2008,konishi2021universal}. These gases are already at or close to the ground states of their confining potentials and the object is to use them to establish optomechanical effects cleanly.  

For cavity-based systems, strong-coupling effects have been observed \cite{delosRiosSommer2021}, optically levitated nanoparticles have been prepared in their ground state of motion using coherent scattering into cavity modes \cite{delic2020cooling}. More recent work has demonstrated simultaneous ground state cooling of two translational modes of a levitated nanosphere in a cavity \cite{Piotrowski2023}.

For systems without cavities, the photon recoil-dominated regime has been reached \cite{PhysRevLett.116.243601}, ground state preparation has been achieved via active negative feedback in cryogenic environments \cite{tebbenjohanns2021quantum}; light scattered from such particles has been squeezed ponderomotively \cite{PhysRevLett.129.053601, PhysRevLett.129.053602}; the mechanical analog of a laser has been realized with a combination of positive linear feedback and negative nonlinear feedback \cite{Pettit2019}. 

Both experiments \cite{yan2023demand} and theory \cite{Chauhan_2020} are now moving towards considering multiple simultaneously levitating nanoparticles. 

\subsection{Other levitation schemes}
Although it is beyond the scope of this tutorial to describe in detail the many alternative platforms for levitodynamics, we briefly mention several systems which have recently been the subject of research. 
Magnetically levitated liquid helium droplets \cite{PhysRevA.96.063842}, superconductors, magnets \cite{PhysRevApplied.11.044041, PhysRevLett.128.013602,PhysRevLett.126.193602}, degenerate gases \cite{PhysRevLett.127.113601} (Esslinger, Brantut), gravimagnetic trap (d'Urso), dynamic electric traps (\cite{PhysRevLett.129.013601}).

One useful metric of comparison between systems is the mass-frequency-Q factor parameter space, as well 
as the fundamental limiting noise source and floor. 

In general at time of publication, Magnetic and ion trap systems allow for higher levitated masses and
non photon limited noise sources at the expense of having a significantly lower mechanical frequency. Notably, nanospheres in Paul traps were recently demonstrated to have quality factors in excess of $10^{10}$ \cite{dania2023ultra}.

Liquid helium systems have also demonstrated extremely high Q factors with substantial tunability. 

\section{Hybrid and coupled systems with multiple degrees of freedom\label{sec:hybrid}}

This section examines advances that have been made with levitated optomechanical systems consisting of multiple degrees of freedom. Such systems present opportunities to explore novel optomechanical dynamics unique to levitated systems, new paradigms for sensing as well as opportunities for exploring fundamental physics. What are some examples of these modalities? One example is an isolated levitated particle that engage multiple modes of oscillation, torsion and/or rotation.  A second example is levitated diamond nanocrystals containing nitrogen vacancy center defects. The embedded defects internal degrees of freedom, for example spin, can couple with the particles center of mass degrees of freedom.  Another system described is an atomic gas that envelopes a levitated nanoparticle and the coupling of these two hybrid systems. Finally, work with cold and degenerate atomic gases coupled with optical cavities will be described.

Progress in cooling of levitated nano-objects into the quantum regime can lead to the development of new hybrid quantum systems where the motion of levitated optomechanical systems can be coupled with other quantum systems such as these.
In these setups, mechanical oscillators can act as transducers, providing coupling between photons, spins, and charges via phonons (see for example Refs. \cite{Kurizki2015-yv,Kumar2023}). Such transducers could play an important role in quantum networks, by allowing coupling between different types of quantum systems, each with differing advantages.

\subsection{Levitated multi-degree of freedom optomechanical   systems}
\begin{figure}[htbp]
	\includegraphics[width=1.0\columnwidth]{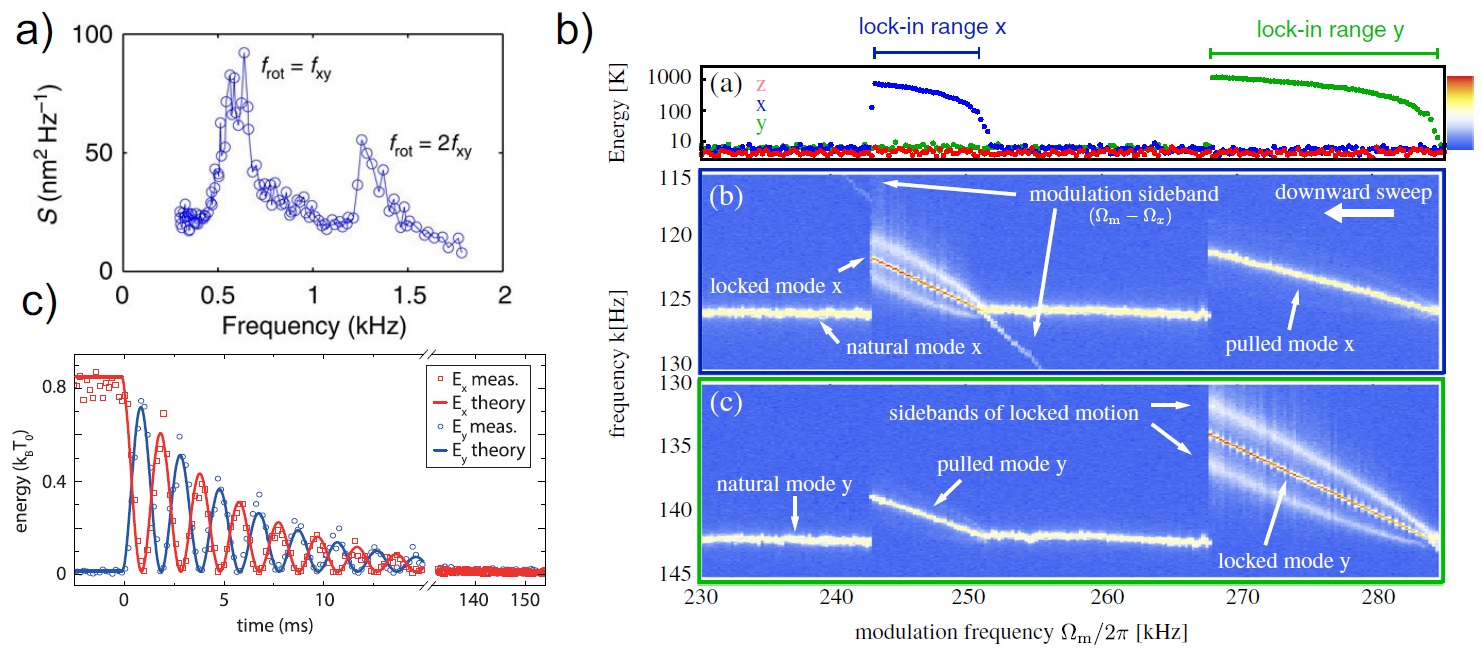}
	\caption{a) Position spectral density data as the frequency of rotation is tuned through the fundamental and second harmonic frequency of the translational motion. b) Top panel: The parametric modulation frequency is swept from 285 to 230 kHz to excite both the $y$ and $x$ resonances. Notice when the y-mode is phase locked with the modulation the nonlinearly coupled x-mode frequency also has its frequency pulled to lower frequencies. c) Sympathetic cooling of  the $x$ mode via coherent coupling to the $y$ mode. Energy transferred from $x$ to $y$ is subsequently actively damped out of the system lowering the $x$ mode's effective temperature.
	}
	\label{Nickfig1}
\end{figure}

A unique feature of levitated optomechanical systems is they possess multiple degrees of freedom - translation, rotation and torsion - that can be coupled. In this section we review research  that has explored the controlled/intentional coupling of these degrees of freedom. The coupling provides a new opportunity for controlling the optomechanical system’s dynamics.
One of the earliest works was by Arita \textit{et. al.} \cite{arita2013laser} studying how rotation of a levitated birefringent vaterite microcrystal could couple with the crystal’s translational motion.  The translational motion in the plane transverse to the optical axis was harmonic and had a fundamental resonance on the order of $600$ Hz ($f_{\rm{xy}}$).  Rotation was accomplished via spin angular momentum transfer  from the trapping light to the birefringent microcrystal. The circularly polarized light induces a rotation and a rotation frequency ($f_{\rm{rot}}$) of $5$ MHz was observed at $0.1$ Pa. The control came  through changing the microcrystal’s rotation frequency by varying the background gas pressure. Figure \ref{Nickfig1}a presents an exemplary data set where resonant enhancement is observed when the rotation frequency is equal to the oscillation frequency ($f_{\rm{rot}} \sim$ $f_{\rm{xy}}$) or twice the oscillation frequency ($f_{\rm{rot}} \sim$ $2f_{\rm{xy}}$ ).  The former enhancement is the result of a driven resonance whereas the  latter is a parametric resonance.  Exciting is when the rotation frequency is commensurate with  the translation frequency. When the previous is achieved, a passive cooling of the microcrystal’s harmonic translational motion was possible achieving minimum effective temperatures of 40K.

Parametric coupling was also used by Gieseler \textit{et. al.} \cite{Gieseler2014} as a means to promote nonlinear mode coupling between the transverse modes of translational motion. In their experiments, an additional frequency and amplitude modulation was imparted to the trapping laser with a frequency in close vicinity to the oscillation frequency. First, the external modulation and one direction of oscillation can become phase synchronized. The motional degree of freedom’s linewidth can narrow, exhibit amplification and have its frequency pulled toward the external modulation. The coupling to the other translational degree of freedom is mediated by the optical trap’s potential well’s intrinsic Duffing nonlinearity. For large enough oscillation amplitude the transverse degrees of translational motion our coupled via the nonlinearity. Here, via the Duffing nonlinearity, one mode of particle oscillation is controlled by the external modulation through the mode the external modulation is directly actuating. As shown in Fig. (\ref{Nickfig1}b), the mode locked to the external drive (y-mode) has its frequency pulled.  The nonlinearity leads to the controlled frequency shift of the x-mode. Similar effects are observed for the y-mode when the x-mode is phase locked to  the external modulation.

Coherent control of a levitated nanoparticle’s motion was implemented by Frimmer \textit{et. al.} \cite{Frimmer2016} to sympathetically cool one translational mode of motion that was coherently coupled to a mode    that was being actively cooled.  Much of the theoretical work describing the experiments can be found in an excellent article in the American Journal of Physics \cite{Frimmer2014}.  The key to this work       was to couple two orthogonal transverse modes of oscillation. Such coupling was achieved by rotating the linear orientation of the trapping laser via an electro-optic modulator sinusoidally:    $\phi(t) = \phi_0 \cos (\omega t)$. The small angle harmonic polarization rotation creates an effective coupling between the $x$ and $y$ directions in the original, unrotated, frame.

Sympathetic cooling was achieved by using Rabi oscillations in the coupled motion to pump energy out of the uncooled mode via coherent control and then dissipate the removed energy by active feedback cooling. The measurement is presented in Fig. (\ref{Nickfig1}c). The feedback cooled mode
(y) is initialized to $0.01$ $k_BT_0$ and the mode to be cooled (x) begins at $0.8$ $k_B T_0$ (a fluctuation away from $k_B T_0$). When the polarization coupling is introduced Rabi oscillations are evident and the energy in both mode’s decay’s exponentially at a timescale determined by the feedback cooling gain. The long-time steady energy in each mode $0.01$ $k_B T_0$ and sympathetic cooling is achieved. Such a technique would also enable the cooling of dark modes that might be invisible to the specific measurement technique.


\subsection{Spin systems}
\begin{figure}[htbp]
	\includegraphics[width=1.0\columnwidth]{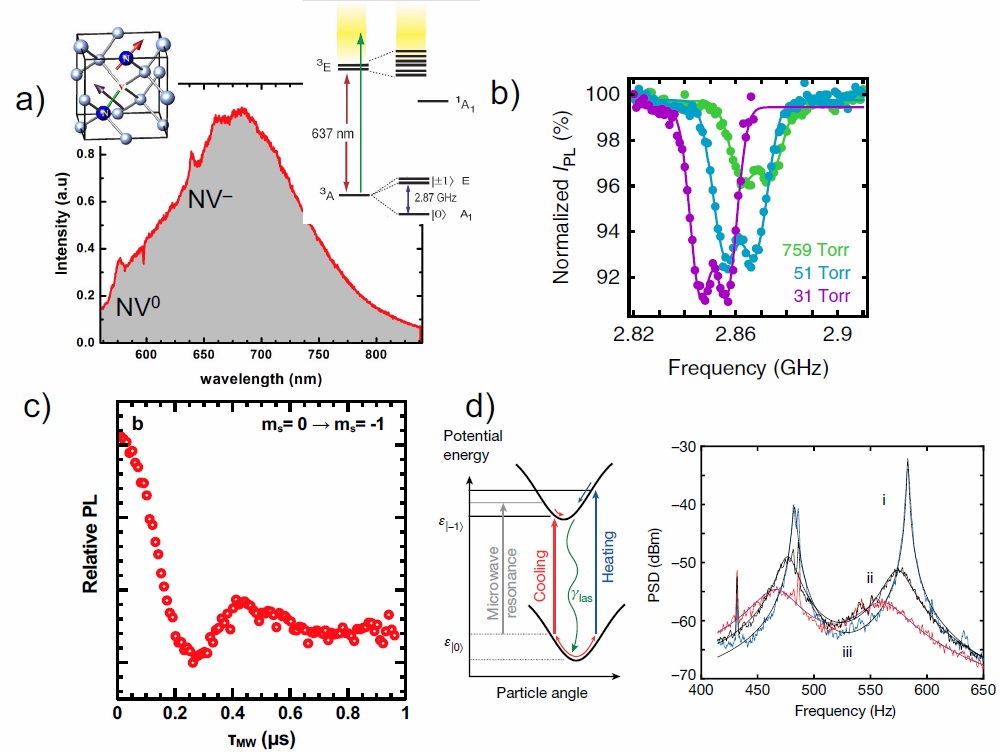}
	\caption{a) The nitrogen-vacancy center photoluminescence spectra. Left inset: The NV center crystal structure. Right inset: The NV center electronic structure. b) The NV center electronic structure illustrating the zero-field split paramagnetic ground state. c) Exemplary NV center en- semble photoluminescence spectrum. Visible are both the NV$^{-1}$ and NV$^0$.
	}
	\label{Nickfig2}
\end{figure}

Levitated optomechanical particles that contain intrinsic defects provide a unique platform to study the coupling of a quantum degree of freedom, such as a spin, to the particle motion. An exemplary system in this regard are diamond nano/microcrystals that host the nitrogen vacancy (NV) defect \cite{Childress2007}. The left inset of Fig. \ref{Nickfig2}a) illustrates the NV$^-$ defect’s crystal structure, a substitutional nitrogen adjacent to a vacancy in the diamond lattice. Figure \ref{Nickfig2}a) shows the NV-center photoluminescence spectrum.  Notable features in the spectrum are the zero phonon line at $\sim 637$ nm and the series of phonon-assisted transitions extending out to $\sim 800$ nm.  The neutral NV center has a zero phonon line at $\sim 575$ nm.  All features are evident in the ensemble NV  center PL spectra presented in Fig.  \ref{Nickfig2}a).  The electronic structure also consists of a paramagnetic ground state that in bulk diamond has a splitting between the $m_s = 0$ and $m_s  = \pm 1$ of
$2.87$ GHz. See the right inset of Fig. \ref{Nickfig2}a). Important for experiments is the the NV$^{-1}$ spin active ground state can be optically prepared, controlled and detected. A number of theoretical proposals \cite{Yin2013,Kumar2017} and experiments \cite{Neukirch2013,Neukirch2015,Hoang2016,Pettit2017, Delord2020} have explored this hybrid levitated optomechanical system. 
Early experiments demonstrated that it was possible to optically trap diamond nanocrystals con- taining NV centers and observe the NV center photoluminescence \cite{Neukirch2013}. The previous was a first step towards using the PL as a channel to monitor the NV center spin dynamics. Subsequent studies explored the possibility of engaging the NV$^{-1}$’s spin. One study focused on how the center of mass dynamics for the nanodiamond influenced the photophysics and observed optically detected magnetic resonance \cite{Neukirch2015}. A second work explored how robust the spin properties are to the gaseous nanodiamond environment.  Figure \ref{Nickfig2}b) shows how the NV$^{-1}$’s ODMR resonance is influenced by the background gas pressure \cite{Hoang2016}. The possibility for coherent spin control was evidenced by imprinting Rabi oscillations into the time resolved ODMR \cite{Pettit2017}, see Fig. \ref{Nickfig2}c). Spin coherence times of $\sim 100$ ns for the NV center on the levitated nanodiamond were estimated from these measurements.

More recently, instead of using an optical trap, Delord \textit{et. al.} \cite{Delord2020} levitated a microdiamond in a Paul trap taking advantage of the native diamond surface charge.  Although this review has focused on optical levitation, the experiments in the Paul trap are exciting in that motional cooling is observed dependent on how the spin resonance is actuated. The left panel of Fig. \ref{Nickfig2}d) illustrates how torsional cooling is realized. First, as a result of diamond rotation, the microwave frequency detunes from the NV spin resonance. The dynamic detuning leads to a torque that stiffens or relaxes the Paul trap confinement dependent on whether the microwave frequency is blue or red detuned. As a result of the previous and the fact the spin lifetime is commensurate with torsional oscillation, the diamond torsional motion is cooled when the microwave frequency is red detuned from the spin resonance. The right panel of Fig. \ref{Nickfig2}d) presents data , trace (iii), where the torsional motion is cooled from 300K to 80K \cite{Delord2020}. In 2023, optical detection of the electron spin resonance of a nanodiamond levitated in a surface ion trap in high vacuum was demonstrated \cite{jin2023quantum}. 

\subsection{Atomic gases and nanoparticles}

It has been shown theoretically that cooling of an optically trapped nanosphere is achievable by sympathetically cooling a cold atomic gas optically coupled to the nanoparticle. 
In addition to being a potential method to cool the particle to its quantum ground state of motion in one dimension, it allows the dynamics of the hybrid system to be explored with cooling turned off, in principle enabling observation of strongly-coupled dynamics between the atoms and nanosphere.  In addition, unlike other methods capable of cooling the center of mass motion of levitated particles to the quantum ground state such as cavity cooling by coherent scattering \cite{delic2020cooling}, such a setup would require only a modest cavity finesse \cite{sympcool}.

Along this direction, recent experimental results have demonstrated sympathetic cooling of a membrane in a cavity using atoms in a separate vacuum chamber \cite{treutlein2010,treutlein2014,Christoph_2018,treutlein2022}. By placing a cryogenic membrane in a medium-finesse optical cavity, the strong coupling regime and ground state cooling of the membrane is predicted to be achievable via sympathetic cooling of the atoms \cite{Bennett_2014,treutlein2014}. Sympathetic cooling can significantly enhance optomechanical cooling of mechanical resonators even outside of the resolved sideband regime \cite{mukund}.  

\begin{figure}[htbp]
	\centering
	\includegraphics[width=0.7\columnwidth]{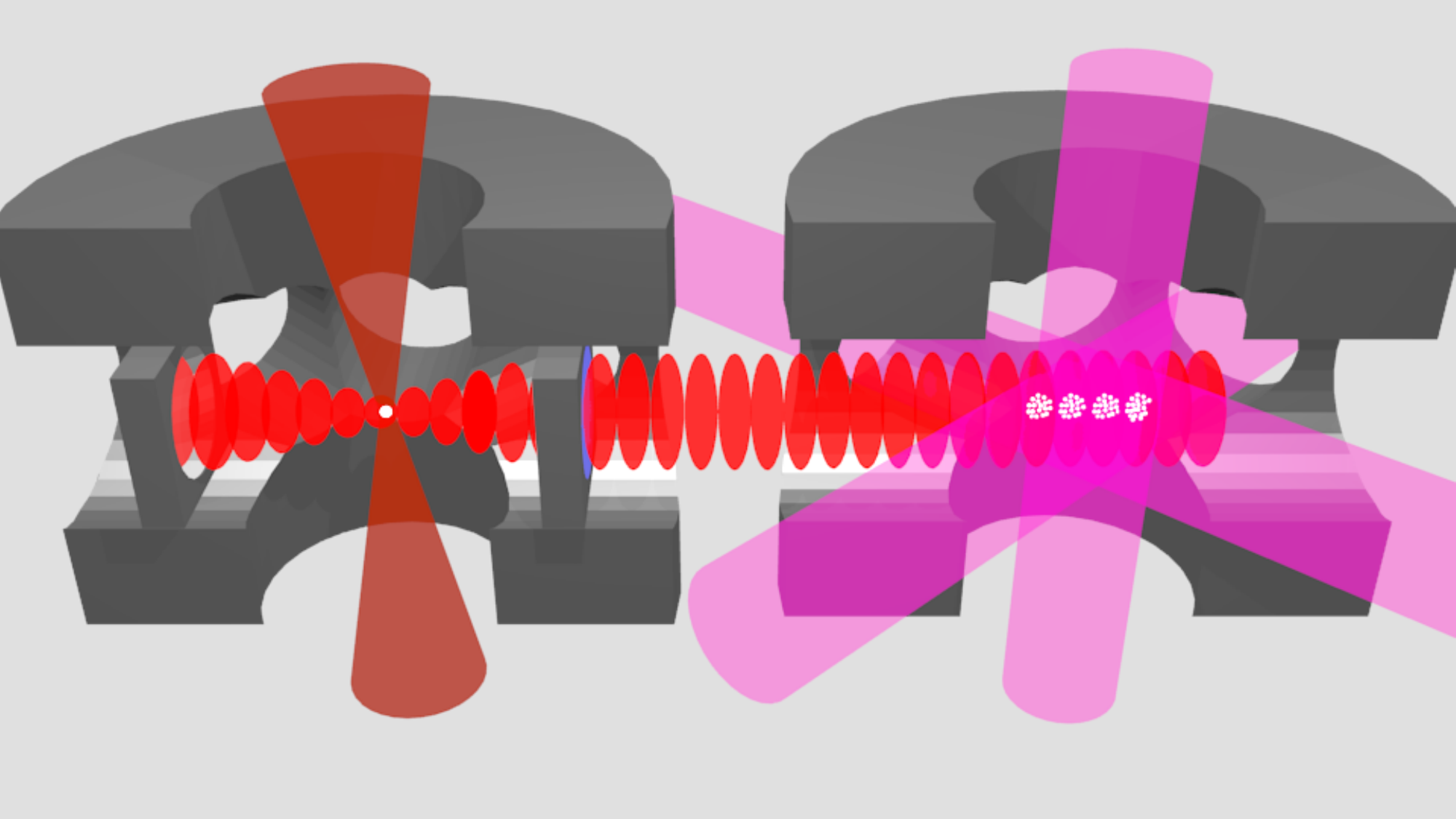}
	\caption{An ensemble of laser-cooled atoms trapped in an optical lattice (right chamber) could be used to sympathetically cool the center of mass motion of a trapped nanobead in a separate optical cavity (left chamber). The left-most mirror of the cavity containing the bead has higher reflectivity to provide increased coupling of the reflected field to the cold atoms. A dual-beam optical tweezer traps the sphere at a position of maximal linear opto-mechanical coupling between the sphere and lattice beam, as discussed in the text. Adapted from Ref. \cite{sympcool}}.
	
	\label{fig:sympcool}
\end{figure}

For either a (tethered) membrane oscillator or a trapped dielectric particle in an optical cavity, a two-way mechanical link is established between the two systems. The finesse of the optical cavity acts as a lever to enhance the coupling between the systems and helps to compensate for the impedance mismatch that results from the lower mass of the atomic ensemble relative to the levitated or tethered macroscopic mechanical system.  The setup is illustrated in Fig. \ref{fig:sympcool}. Cold atoms are trapped in an optical lattice in one vacuum chamber and the same lattice beam also drives an optical cavity in a second vacuum chamber containing the optomechanical element. The optomechanical coupling between the oscillator and the cavity results in a phase shift of the light reflected by the cavity if the nanoparticle moves. This phase shift results in a shift of the position of the optical standing wave potential which results in a force imparted on the atoms. Correspondingly, for atoms confined in the one-dimensional standing wave optical lattice potential formed by the beam reflecting from the optical cavity, atoms feel a restoring force confining them to the antinodes of the laser field for a laser that is red-detuned with respect to the atomic resonance.  If the atom moves away from this equilibrium position at the maximial intensity anti-node, a force is established through absorption and stimulated emission of trap laser photons which impart momentum on the atoms to bring them back the trap minimum.  Due to conservation of momentum, to establish this however the intensity balance between the right-moving and left-moving components of the standing wave must be disturbed. This results in an intensity fluctuation on the light incident on the optical cavity where the nanoparticle is trapped. since the nanoparticle is trapped at a location where there is a linear optomechanical coupling, i.e. at a location where there is a slope of the intensity of the standing wave inside the cavity, this results in a force on the bead due to the motion of the atoms.  Given this mechanical link, it is thus possible to reduce the kinetic energy of the bead by performing laser cooling of the moition of the atoms in their optical trap. 

For an atom-membrane coupled system such as that realized in Ref. \cite{treutlein2014,Christoph_2018}, ground state cooling can be achieved if the coherent dynamics occur at a sufficiently fast time scale compared with dissipative effects. The fast thermalization rate \begin{equation} \Gamma_{\rm{th}}=\frac{k_B T}{\hbar Q}, \label{gth} \end{equation} where $T$ and $Q$ are the initial temperature and mechanical quality factor of the oscillator, requires cryogenic pre-cooling of the membrane \cite{treutlein2014,Christoph_2018}.  On the other hand, because their coupling to the thermal environment is very weak under ultra-high vacuum conditions, the thermalization rate in Eq. (\ref{gth}) for levitated mechanical systems \cite{levreview,Millen:2020review} can be exceedingly low even at room temperature, as the mechanical quality factors are predicted to exceed $Q>10^{11}$. 
In Ref. \cite{sympcool} it was shown that optically trapped nanospheres in vacuum can theoretically be cooled from room temperature to the ground-state by sympathetic cooling with atoms in an optical lattice. With $Q \sim 10^{11}-10^{12}$ at room temperature the thermalization rate in Eq. (\ref{gth}) becomes $\sim 10-100$ Hz.  For an ensemble of $5 \times 10^7$ Rb atoms in an optical lattice, the opto-mechanical coupling rate can exceed several kHz, making this system able to attain the strong coupling regime. With a suitable sympathetic cooling rate of order $\sim 10$ kHz, ground state cooling of a $300$ nm or smaller diameter sphere is possible. 

Recent work has demonstrated a light-mediated coupling between the spins of cold atoms and the mechanical motion of a membrane in a cavity \cite{treutlein2022}. By utilizing a cryogenically pre-cooled membrane,this method is predicted to allow cooling of the mechanical oscillator close to its quantum mechanical ground state and permit the preparation of non-classical states. Similar experiments using the spins degrees of freedom of cold atoms could also be applied to levitated optomechanical systems.

\subsection{Cold and degenerate atomic gases and optical cavities}
\begin{figure}[htbp]
	\centering
	\includegraphics[width=0.9\columnwidth]{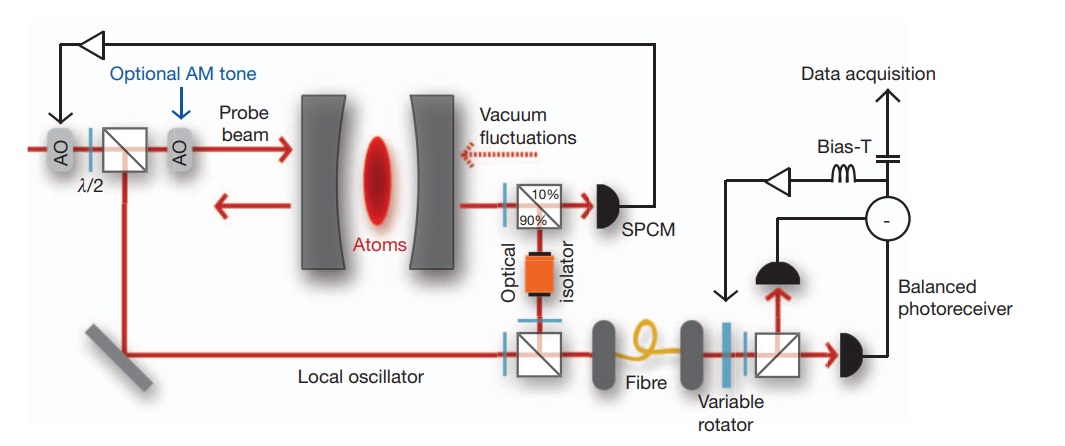}
	\caption{An atom cloud coupled to an optomechanical system, reproduced from Ref. \cite{brooks2012non}}
	
	\label{fig:atom_opto}
\end{figure}

The ready availability of samples of cold as well as degenerate samples of atoms have led to levitated optomechanical studies of such gases coupled to optical modes of electromagnetic cavities. \textcolor{black}{Fig. \ref{fig:atom_opto} depicts the arrangement recently studied in Ref. \cite{brooks2012non}.}

Typically, the atoms are levitated using the dipole force from the optical lattice formed by the standing wave optical mode of the cavity. The attraction for using atoms for cavity optomechanics is based on the strong light matter interaction strengths achievable using dispersive coupling, that is, by detuning the cavity optical mode far enough away from the atomic transition that spontaneous emission has a very low probability of occurring.  This interaction strength is much larger than the dissipation - due to finite cavity lifetimes and atom loss from traps - in the system, leading to a large optomechanical cooperativity (which is the ratio of the coherent light-matter interaction to the product of the optical and mechanical mode linewidths), up to $\sim 10^{4}$ as can be seen in Fig.12 of \cite{aspelmeyercavity}. These are about the largest cooperativities displayed by any optomechanical system, and make effects such as ground state cooling and radiation pressure shot noise easier to observe compared to other platforms.

Cold - but not degenerate - atoms were introduced into optical cavities in an optomechanical context in \cite{gupta2007cavity}. In these systems, about $10^5$ atoms at $\sim 0.8 \mu$K are localized at the wells formed by the optical wave and oscillate harmonically about those points, forming a collection of oscillators. With such setups optomechanical bistability was observed with intracavity photon numbers far less than unity ($\sim 0.05$), pointing to the strong light-matter coupling in the system. Signatures of collective atomic motion were also observed in the cavity transmission in this work. The same group also reported observation of measurement quantum backaction in the system \cite{murch2008observation}, followed by the demonstration of both linear and quadratic optomechanical coupling \cite{purdy2010tunable} and ponderomotive squeezing \cite{brooks2012non}. The ability of this configuration to detect forces near the standard quantum limit was then shown in Ref. \cite{Schreppler2014}.

Degenerate bosonic atoms - in the form of a Bose-Einstein condensate - were first coupled experimentally to an optical cavity in \cite{brennecke2008cavity}. This configuration coupled the optical lattice formed by a cavity mode to density oscillations in the condensate created by matter-wave diffraction from the optical field. In effect, the 
condensate provided a Kerr-like medium for the cavity optical field. The density oscillations modulated the cavity transmission, which in turn provided information about the mode dynamics in the condensate. It is worth noting that this information is obtained with small backaction to the condensate; this minimally destructive measurement may be contrasted with the usual completely destructive measurements made usign absorption imaging on condensates. Bistability in the system was also observed, for photon numbers below unity. A large number of works treating the system, both experimental [\cite{baumann2010dicke}, \cite{schmidt2014dynamical}, \cite{schuster2020supersolid}] as well as theoretical \cite{chen2010classical},\cite{strack2011dicke},  \cite{kroeze2019dynamical}] have been published, addressing density waves characterized by their linear \cite{cola2004robust}, pseudo-spin [\cite{kroeze2018spinor}] or angular \cite{kumar2021cavity}] momentum .

Degenerate fermionic atoms - in the form of a quantum degenerate Fermi gas - were first coupled experimentally to an optical cavity in \cite{konishi2021universal}. The configuration had earlier been proposed in [\cite{kanamoto2010optomechanics}]. Optomechanical response of a Fermi gas was investigated in [\cite{helson2022optomechanical}], across the so called BCS-BEC transition (where BCS refers to a gas of weakly bound Fermi pairs and a BEC to a Bose-Einstein Condensate). A superradiant transition was demonstrated in [\cite{zhang2021observation}]. A unitary Fermi gas - a strongly correlated quantum fluid - was strongly coupled to cavity light, demonstrating the formation of polaritons in \cite{roux2020strongly}.

The high degree of control that the experimentalists have been able to establish over cold and degenerate samples of gases; the versatility that such platforms display in offering selection over spin, atomic species, interaction strengths; the minimally destructive readout offered by the dispersive interactions; and the amenability of these systems to microscopic yet tractable theoretical modeling are very great attractions for continuing the study of such optomechanical systems.    

\section{Thermodynamics\label{sec:thermo}}

\subsection{Non-equilibrium thermodynamics of small systems}
Non-equilibrium processes of small systems are widespread in the fields of biology, chemistry, and physics, yet they can be difficult to understand. For instance, how does a molecular motor work in noisy environments where fluctuations are prevalent? Unlike macroscopic systems, thermal fluctuations have a significant impact on microscopic systems. An optically levitated particle serves as an ideal model for investigating stochastic thermodynamics. By lowering the air pressure, the levitated particle can transition into an under-damped regime, as opposed to the over-damped regime typically seen in liquid-based particles. The same micro or nanoparticle can be stably trapped in air or vacuum for many days, enabling in-depth examination of its dynamic behavior.

\begin{figure}[htbp]
	\centering
	\includegraphics[width=0.75\columnwidth]{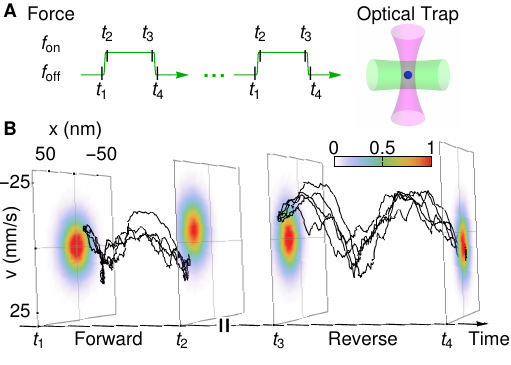}
	\caption{ (A) When a silica nanosphere is trapped in an optical tweezer (magenta), a series of laser pulses (green) applies an optical force on the nanosphere to drive the system out of equilibrium. (B) Examples of measured trajectories of the nanosphere in the phase space. Figure reproduced from \cite{Hoang2018Experimental}.
	}
	\label{figure_nonequilibrium}
\end{figure}

A levitated particle will do Brownian motion due to the collisions between surrounding air molecules and the particle. For small displacements, we can approximate the optical tweezers as a harmonic trap with a trapping frequency of $\omega_{m}$. Due to the energy equipartition theorem, the rms amplitude of the Brownian motion displacement of a levitated particle along one  axis is $x_{rms}=\sqrt{k_B T/(m\omega^2_m)}$, where $m$ is the mass of the particle. At short time scales, the mean square displacement (MSD) of the Brownian motion scales as $\left< [\Delta x(t)]^2 \right>=(k_B T/m) t^2$, which is very different from Einstein's law for diffussive Brownian motion of a free particle: $\left< [\Delta x(t)]^2 \right>=2Dt$, where $D$ is the diffusion constant. With sufficient temporal and spatial resolution, we can measure the instantaneous velocity of the Brownian motion of a levitated particle. In 2010, with a levitated glass microsphere in air and in vacuum, Li \emph{et al.} measured the instantaneous velocity of the Brownina motion of a particle for the first time \cite{li2010measurement}. The measured velocity of the Brownian motion of a levitated particle follows the Maxwell-Boltzmann distribution. The rms velocity of the particle along one direction will be $v_{rms}=\sqrt{k_B T/m}$. 

With a measured position trajectory, we can calculate its Fourier transform and the power spectral density (PSD). The expected value of the PSD of the Brownian motion of a levitated particle in a harmonic trap is:
\begin{equation}
	S(\omega)=\frac{2k_B T}{m \omega^2_m} \frac{\omega^2_m \gamma_{g}}{(\omega^2_m-\omega^2)^2+\omega^2 \gamma_{g}^2},
\end{equation} 
where $\gamma_{g}$ is the damping rate of the levitated particle due to air molecules. This equation is often used to calibrate the detection system in levitated optomechanical experiments. For a spherical particle, the damping rate at thermal equilibrium is \cite{Li:2011}
\begin{equation}
\gamma_{g}=\frac{6\pi \eta R}{m} \frac{0.619}{0.619+Kn}(1+c_k),
\end{equation} 
where $\eta$ is the air viscosity coefficient, $R$ is the radius of the levitated spherical particle, $Kn=l/R$ is the Knudsen number, $l$ is the mean free path of air molecules. $c_k=(0.31 Kn)/(0.785+1.152 Kn + Kn^2)$. This equation enables us to estimate the radius $R$ of a levitated particle.

Over the past three decades, several thermodynamic relationships related to nonequilibrium processes, known as fluctuation theorems \cite{jarzynski2011equalities,jarzynski1997nonequilibrium,crooks1999entropy,liphardt2002equilibrium,collin2005verification}, have been discovered. These fluctuation theorems generalize the second law of thermodynamics and can be unified into a differential fluctuation theorem (DFT) \cite{maragakis2008differential,jarzynski2000hamiltonian}. To test DFT and deepen our understanding of nonequilibrium physics including dissipation and irreversibility, we need large statistics and the ability to measure the instantaneous velocities of Brownian motion, which is doable with a levitated particle.
Recently, Hoang \emph{et al.} conducted the experimental test of the DFT using an optically levitated nanosphere in both underdamped and overdamped scenarios (Fig. \ref{figure_nonequilibrium}) \cite{Hoang2018Experimental}. In the experiment, a 1550-nm optical tweezer levitated the nanosphere in a vacuum chamber. The nonequilibrium processes were controlled by a 532-nm laser that applied a time-dependent optical force from the side. With a large set of experimental data, DFT and several theorems derived directly from DFT, including a generalized Jarzynski equality applicable to any initial state, were  tested \cite{Hoang2018Experimental}.  This work demonstrated the power of using a levitated particle for studying stochastic energetics.

Recently, Debiossac \emph{et al.} investigated the thermodynamics of an optically levitated microparticle under continuous, time-delayed feedback control \cite{debiossac2020thermodynamics}. With a time-delay in the feedback loop, the dynamics becomes non-Markovian. This work tested the generalized second law under non-Markovian feedback control. Rademacher \emph{et al.} utilized feedback cooling to achieve rapid and controlled temperature variations in a levitated particle, and investigated a fluctuation theorem that included the effects of both mechanical and thermal control \cite{rademacher2022nonequilibrium}. When a thermalization process is interrupted by a sudden change of the bath temperature, the Kovacs memory effect will lead to a nonmonotonic evolution of its energy. It is challenging to control the real temperature of the environment suddenly. Instead, the thermal bath can be simulated by an external white Gaussian noise source. By changing the effective thermal bath temperature much faster than thermal relaxation time fo the system, the Kovas memory effect was observed with an optically levitated nanoparticle \cite{militaru2021kovacs}. 

Many microscopic processes, such as chemical reactions and protein folding, involve multiple metastable states. An optically levitated particle can also be used to study thermodynamics involving two or more metastable states. The required multistable potential profiles can be generated by two or more optical tweezers \cite{Volpe_2023,shen2021chip,yan2023demand}. 
Recently, Rondin \emph{et al.}  studied stochastic processes of a levitated nanoparticle in a bistable optical potential generated by two optical tweezers \cite{rondin2017direct}. They measured the thermally activated transition rate between two traps for different damping levels and  observed the Kramers turnover, which predicted that the escape rate reached a maximum at intermediate damping. Militaru \emph{et al.} added an engineered stochastic force that mimicked self-propulsion to study active escape dynamics of a levitated nanoparticle  in a bistable potential \cite{militaru2021escape}. They observed an additional activity-related turnover in the escape rate.

\subsection{Laser Refrigeration}

During experiments in high vacuum a levitated particle no longer thermalizes with the surrounding gas environment due to an increase in the mean-free-path between particle collisions.   Consequently, the power from optical absorption no longer can be carried away by the reduced number of molecular collisions, leading to an increase in the particle’s internal absolute temperature (T$_{int}$).  The molecules that do collide with a trapped particle leave the particle’s surface with a temperature (T$_{em}$) greater than the surrounding gas temperature (T$_{gas}$), leading to a new damping term $\Gamma_{em}/2\pi = 1/16 \cdot \sqrt{(T_{em}/T_{gas})} \cdot \gamma_{g}$.   If optical absorption is high enough, additional effects become significant including photophoretic forces \cite{RN296}, changes in the trapped particle’s physical properties (density, refractive index), particle melting, or even complete particle destruction through evaporation \cite{Chang:2010} or burning \cite{Rahman2016}.

One potential approach to mitigate the detrimental consequences of internal particle heating is through solid-state laser refrigeration (SSLR)\cite{seletskiy2010,xia2020}.  SSLR is a process that has been demonstrated to cool the internal temperature of materials with near-unit external radiative quantum efficiencies, $\eta_{ext}$, through the efficient emission of anti-Stokes (upconverted) luminescence.  The process is based on the emission of luminescence with an average photon energy ($h\nu_{ave}$) that is greater than the average photon energy that is absorbed ($h\nu_{abs}$).  In many ways the process is analogous to operating a continuous-wave laser in reverse.  First, coherent, low-entropy photons are used to create an electronic excited state in a solid. 
Second, the excited state extracts phonons from the host matrix with phonon scattering time scales on the order of picoseconds.  Third, the excited state relaxes, emitting a photon with both a higher energy and a higher entropy relative to the photon absorbed originally from the excitation laser.  The cooling cycle is consistent with the second law of thermodynamics when considering that the net entropy of the emitted photons \cite{landau1946} (summing over their phase, polarization, direction, and energy) is greater than the net entropy of the photons from the excitation laser.  The theoretical thermodynamic efficiency of the process $\eta = (h\nu_{ave}-h\nu_{abs}) / h\nu_{abs}$ is typically on the order of a few percent.

In 1995 a team led by Richard Epstein at the Los Alamos National Laboratory reported a seminal experimental observation of SSLR \cite{epstein1995} using an amorphous fluoride (ZBLAN) glass material doped with trivalent ytterbium ions.  A measurement based on photothermal deflection was used to quantify cooling in vacuum on the order of 1K below room temperature.  An excellent recent review article\cite{seletskiy2010} discusses subsequent advances by Mansoor Sheik-Bahae's group at the University of New Mexico which demonstrated that using crystalline hosts for Yb(III) ions leads to dramatic reductions in the minimum achievable temperature that can be reached when using a continuous-wave laser to induce cooling.  In comparison with amorphous ZBLAN glass materials, using a stoichiometric crystalline material such as yttrium lithium tetrafluoride (YLiF$_{4}$) significantly increases the magnitude of laser cooling due to a combination of the near-unit external radiative quantum efficiency of Yb(III) ions in this crystal, the crystal field splitting of the ytterbium ion's 4f orbitals by the crystal's S$_{4}$ point group site symmetry, and a corresponding increase in the optical absorption coefficient for crystal field transitions between $^{2}$F$_{7/2}$ and $^{2}$F$_{5/2}$ manifolds.  A helpful review article has been published\cite{hehlen2013} on the use of Judd-Ofelt theory to interpret experimental spectra arising from interconfigurational 4f <-> 4f transitions within lanthanide ions that are frequently used for solid state laser refrigeration.  It is possible to reach cryogenic temperatures ($\sim$90K) using laser gain crystals of yttrium lithium fluoride (YLF) doped with ytterbium (III) ions \cite{seletskiy2010} which could have significant advantages in all-optical cooling of high precison silicon reference cavities\cite{kedar2023}.  Recent reports have demonstrated the use of SSLR in the optical cooling of focal plane arrays\cite{hehlen2018} for vibration-free imaging.

One challenge in the field of SSLR has been the time required to grow bulk crystals with low background absorption levels using either the classical Czochralski or Bridgemann growth methods.  These processing approaches normally require large amounts of purified reagents and time scales on the order of weeks to grow, cut, and polish a bulk crystal. Recently in 2015 it was shown that a low-cost hydrothermal processing approach can be used to synthesize crystalline host materials that are capable of laser cooling\cite{roder2015}.  These materials were used to report the first experimental observation of cold Brownian motion (CBM) \cite{roder2015} using optically-levitated microcrystals of Yb:YLF in water.   This was the first experimental demonstration of CBM since Einstein's seminal paper\cite{einstien1905} on Brownian motion in 1905. Before this work it was not clear whether it would be possible to observe SSLR in liquid water due to the absorption of the pumping laser.  It was also not obvious that materials grown through hydrothermal processing could be laser cooled due to the potential for background absorption from hydroxyl ions in the resulting ceramic fluoride microcrystals.  The theoretical modeling of CBM is identical to what has been described recently for hot Brownian motion\cite{rings2010} except the diffusing particle’s temperature is less than that of its surrounding medium. 

CBM has also been reported for YLiF$_{4}$ (YLF) nanocrystals levitated in vacuum\cite{rahman2017laser}.  The particles of YLF were observed to have irregular shapes following their preparation through top-down milling of bulk Czochralski single crystals.  The irregular morphology of milled YLF crystals was reported to create significant shape induced birefringence that caused particle rotation during levitation, and motivates the synthesis of materials for SSLR with spherical morphology.

The ability to control the size, shape, composition, and phase of particles for SSLR via bottom-up, chemical synthetic methods has opened up novel experimental possibilities using a wide range of nano- to micro- meter size materials\cite{xia2020,xia2021}.  For example, hydrothermal processing has been used to investigate non-stoichiometric crystalline materials for SSLR, including the hexagonal ($\beta$) and cubic ($\alpha$) phases of sodium yttrium fluoride (NaYF)\cite{zhou2016}.  The $\beta$ phase of NaYF is of great interest after being reported to be an efficient emitter of upconverted photons\cite{kraemer2004}.  However, the growth of large bulk crystals of $\beta$-NaYF has not been reported to date using Czochralski or Bridgmann methods due to large anisotropic thermal expansion stresses related to the crystal's hexagonal Bravais lattice. 

The hydrothermal synthesis of crystals with well-defined shapes also has enabled the use of van der Waals bonding to attach Yb-doped microcrystals for the refrigeration optomechanical semiconductor sensors\cite{pant2020}.  Hydrothermal SSLR materials have also been used to demonstrate the laser cooling of nanodiamonds containing the negatively charged nitrogen-vacancy center\cite{pant2020.2} that are used for quantum sensing through optically detected magnetic resonance spectroscopy\cite{bala2008}.  The synthesis of spherical nanocrystals from cubic $\alpha$-phase of NaYF has helped mitigate challenges in controlling both shape birefringence\cite{rahman2017laser} and optical birefringence during optical levitation in vacuum\cite{Luntz-Martin:21}.  Recent groundbreaking reports of SSLR using amorphous silicon dioxide glass doped with ytterbium ions\cite{mobini2019,mobini2020} has opened up promising future directions towards the formation of spherical materials for SSLR based on the processing of high purity optical fibers\cite{knall2020}.  Most recently, the synthesis of two dimensional hexagonal $\beta$-NaYF discs have been reported\cite{felsted2022} that could be used for in future experimental searches for high-frequency gravitational waves\cite{winstone2022optical} based on how their two-dimensional morphology minimizes photon recoil heating due to light scattering\cite{aggarwal_2022}.  The design and synthesis of new materials that demonstrate SSLR is a promising future research direction, where the particle's shape, size, density, mass, composition, and birefringence can be selected to achieve a number of experimental objectives in precision sensing and quantum information science, discussed in more detail below. 

\section{Sensing \label{sec:force}}
Optically levitated dielectric objects in ultra-high vacuum exhibit an excellent decoupling from their environment, making them highly promising systems for precision sensing. In particular, the center of mass modes of optically-trapped silica nanospheres have exhibited high mechanical quality factors in excess of $10^7$ \cite{Gieseler:2012} and zeptonewton ($10^{-21}$ N) force sensing capabilities \cite{ranjit2016zeptonewton}. Such devices make promising candidates for sensors of extremely feeble forces \cite{geraci2010}, accelerations \cite{andyhart2015, Moore2017,novotnydrop}, torques \cite{Li2016}, and rotations \cite{Li2018,Novotny2018,Moore2018}, testing the foundations of quantum mechanics \cite{oriol2011}, observing quantum behavior in the vibrational of modes of mechanical systems \cite{chang2009,delic2020cooling, aspelmeyercavity}.
\subsection{Force}

\begin{figure}[htbp]
	\centering
	\includegraphics[width=0.6\columnwidth]{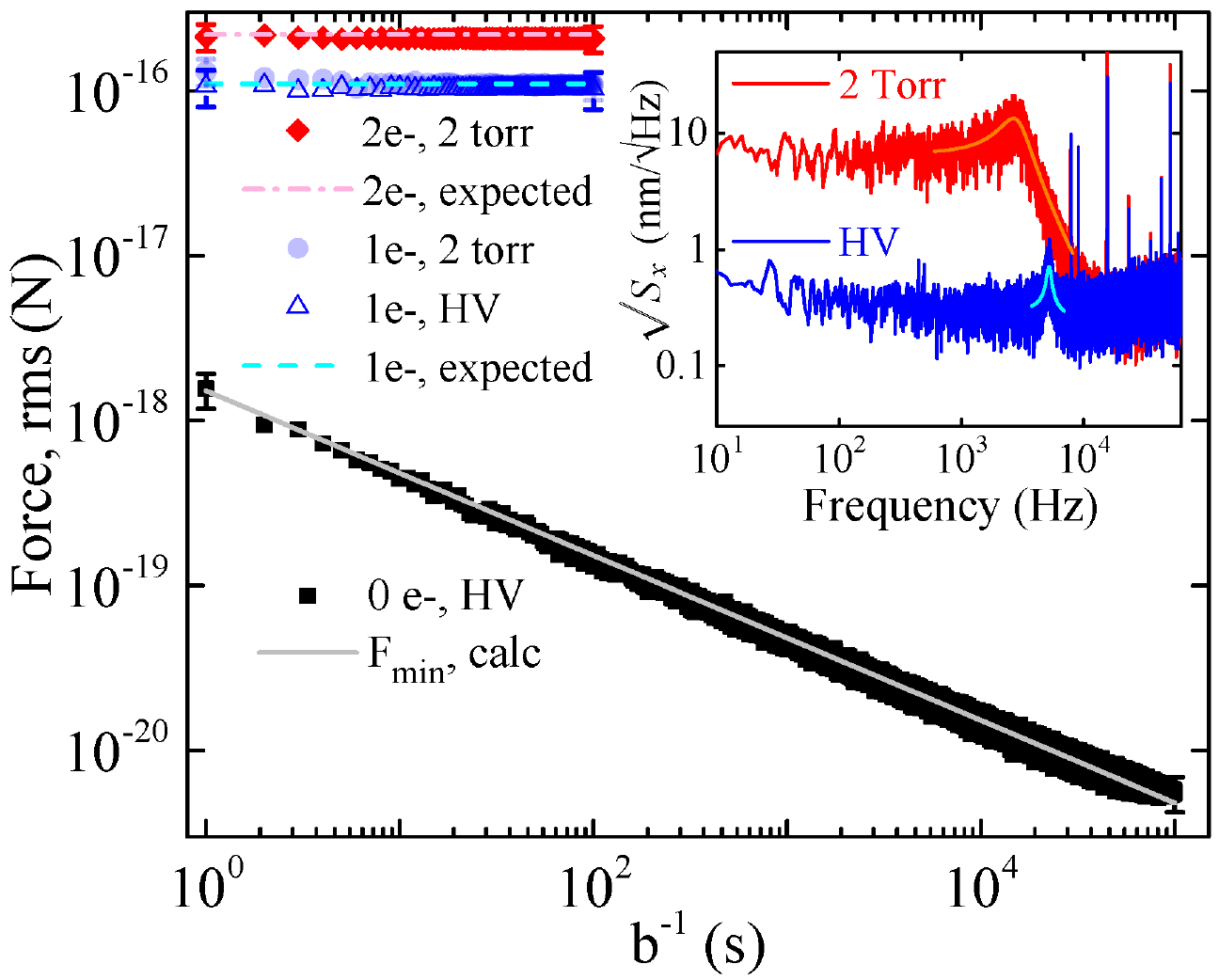}
	\caption{ Adapted from Ref. \cite{ranjit2016zeptonewton}. Force sensing with 300~nm nanospheres using charged and uncharged beads (0,1, or 2 excess electrons). A known applied electric field permits a calibration of the measured force which is deduced from the displacement spectral density of the particle's center of mass motion (inset). Under high vacuum (HV) of $5 \times 10^{-6}$ Torr, background-free force measurement is demonstrated to the $\sim 6$~zN level for long averaging times.}.
	
	\label{fig:forcefig}
\end{figure}

Force sensing has a variety of applications, ranging from force microscopy at the nanoscale to electromagnetic field sensing to gravitational wave detection. High force sensitivity resonant sensors have typically consisted of solid-state micro-fabricated structures \cite{Geraci2008,Rugar2004,Doolin_2014,Sankey2016,Moser2013,Norte2016}, for example cantilever beams or membranes.   In these systems, the internal materials losses and losses from clamping mechanisms are responsible for limiting the $Q$ factor of the oscillator in high-vacuum environments, to typically below $Q \sim 10^8$.  In contrast, in ultra-high vacuum (UHV),  the center of mass motion of micrometer sized dielectric spheres is well decoupled from the environment, and could exhibit $Q$ factors approaching $10^{12}$, leading to force sensitivity well below 1 aN/Hz$^{1/2}$.  Force sensing with levitated particles is typically accomplished by measurement of the displacement of the trapped particle, given knowledge of the mechanical susceptibility of the particle, which depends on the spring constant and dissipation. Gradients in force are also able to be deduced by measuring a change in the mechanical resonance frequency of the oscillator, where a linear force gradient manifests as a modification of the effective spring constant. 
Our discussion will assume that the response of the trapped particle is linear. Non-linearities e.g. Duffing terms, have been shown to occur for nanoparticles trapped in optical traps in vacuum due to the non-harmonic nature of the focused trap for sufficiently large particle amplitude \cite{Gieseler2013}. By cooling the motion of the particles, these nonlinear effects can be minimized in practice. Fig. \ref{fig:forcefig} illustrates a prior demonstration of calibrated zeptonewton level force sensing with levitated nanoparticles, using electrostatic calibration \cite{ranjit2016zeptonewton}. 

Feedback cooling and dynamical backaction cooling are possible to implement simultaneously with force sensing, and generally do not improve the force sensitivity for example from the thermal-noise limit. The cooling does allow stabilization of the particle under high vacuum conditions as well as control of the bandwidth of the sensor. For example in ultrahigh vacuum the mechanical quality factor of trapped particles can exceed $10^7$, requiring 100s or more for perturbations to the system to ring-down. By applying active or passive feedback cooling, the ratio of the temperature to the damping rate remains approximately constant, thus maintaining the same force sensitivity, however the sensing bandwidth can be significantly increased by lowering the $Q$ factor. The displacement detection of the particle however becomes more challenging the more the damping is applied, particularly for on-resonance detection schemes (see also Sec. \ref{sec:trapping} for further discussion of displacement sensing). 

\textit{Minimum detectable force} -- For force detection, it is desirable to have minimal dissipation, leading to a high quality factor $Q$ for the center of mass motion of the object.  This can be seen as the minimum detectable force due to thermal noise scales as $Q^{-1/2}$, and can be expressed as
\begin{equation}
F_{\rm{min}}  = \sqrt{\frac{4kk_BTb}{\omega_0 Q}}, 				
\label{fmin}
\end{equation}
where $b$ is the bandwidth of the measurement, $T$ is the temperature of the center of mass motion, $\omega_0$ is its resonance frequency, and $k$ is the spring constant.  This result can be derived by considering the equipartition theorem: for decoupled degrees of freedom and harmonic motion, the root-mean-square displacement $x_{\mathrm{rms}}$ for each translational mechanical degree of freedom is related to the thermal energy according to $k_B T = k x_{\mathrm{rms}}^2$. Assuming the particle is driven with a white-noise spectrum due to random thermal motion, writing the frequency-independent power spectral density of force noise as $S_{\mathrm{FF}}$ we find the displacement power spectral density as $S_{\mathrm{XX}}(\omega)=S_{\mathrm{FF}} |\chi(\omega)|^2$ where $\chi(\omega)$ is the mechanical susceptibility of the oscillator. Using the Wiener-Khinchin theorem \cite{RevModPhys.86.1391} for a stationary random process, we can solve  \begin{equation} <x^2>=\frac{1}{k^2}\int_0^{\infty} S_{\mathrm{FF}} d\omega \frac{\omega_0^2}{\sqrt{(\omega_0^2-\omega^2)^2+\frac{\omega_0^2 \omega^2}{Q^2}}}\end{equation} for $x_{\mathrm{rms}}$, and we find $x_{\mathrm{rms}}=\frac{S_{FF}^{1/2}}{k}(\frac{\omega Q}{4})^{1/2}$. After applying the equipartition theorem, we arrive at Eq. \ref{fmin}, as $F_{\mathrm{min}}=S_{FF}^{1/2}b^{1/2}$.

\textit{Photon Recoil heating.} 
For optically levitated particles, photon recoil heating represents a second limitation to force sensitivity.  The recoil heating rate for a nanosphere is $\Gamma_{sc}  = \frac{2}{5} \frac{\pi^2 \omega_0 V}{\lambda^3} \frac{(\epsilon-1)}{(\epsilon+2)}$ where $V$ is the volume of the nanosphere and $\lambda$ is the trap laser wavelength. 
Including the contributuion from recoil heating, the minimum detectable force $F_{\rm{min}}$ for a particle with center-of-mass temperature $T_\mathrm{CM}$ is approximately \cite{GWprl}
\begin{equation}
F_{\mathrm{lim}}=\sqrt{4k_BT_{\mathrm{CM}}\gamma_gb M \left[1+\frac{\Gamma_{\mathrm{sc}}}{n_i\gamma_g}\right]}, \label{eq:noise}
\end{equation}
where $n_i=k_BT_\mathrm{CM}/\hbar\omega_0$ is the mean initial phonon occupation number of the center-of-mass motion. In the high-vacuum regime, $\gamma_g = \frac{16P}{\pi \bar{v} \rho r}$ is the gas damping rate at pressure $P$ with mean gas speed $\bar{v}$ for a nanosphere of radius $r$ and density $\rho$. 

\textit{Minimum detectable frequency shift -- gradient detection.}
Another modality of detection relies on observing the gradient of an applied force, which manifests as a frequency shift of the oscilator. Such techniques have been previously employed in solid-state mechanical sensors such as cantilever beams \cite{rugarfreqshift}. In this case by driving the oscillator with a well defined frequency and phase into a amplitude $z_{\mathrm{rms}}$ which is typically larger than the thermal motion amplitude, the minimum detectable frequency shift improves. The minimum detectable frequency shift due to thermal
noise is given by  
\begin{equation}
|\delta \omega_0 /
\omega_0|_{\rm{min}}=\sqrt{\frac{k_BT_{\rm{CM}}
b}{k\omega_0Q_{\rm{eff}}z_{\rm{rms}}^2}}, 
\end{equation}
where $T_{\rm{CM}}$ is the effective temperature of the center of mass motion. Photon recoil heating and technical noise depending on the setup also represent limitations in principle. 
\subsection{Acceleration}
Depending on the application, the acceleration or force sensitivity of a levitated sensor, rather than its displacement sensitivity, is typically of interest, particularly when operated in a regime where the displacement detection, limited by shot noise, or electronic noise or other technical noise, is not the limiting factor. Optically levitated particles can serve as accelerometers both in a trapped configuration \cite{Moore2017,Moore2018} and in a free-fall configuration \cite{novotnydrop}.  For trapped nanoparticles acting as accelerometers, the minimal detectable acceleration due to thermal noise and photon recoil heating can be deduced from Eq. (\ref{eq:noise}) by simply dividing by the mass of the trapped particle.  Thus while force sensitivity tends to improve for smaller size particles, acceleration sensitivity tends to be better for larger objects, although the bandwidth of the accelerometer may be limited for larger particles due to correspondingly lower trapping frequencies.   Acceleration sensitivities for trapped $\sim 10$ $\mu$m size microspheres in vacuum have been demonstrated at the $\sim$100~n$g$/$\mathrm{\sqrt{Hz}}$ level in Ref \cite{acceleration2020}, with a resonance frequency of order 100 Hz.  

For free-fall accelerometers, a sensitivity to static forces of $10$ aN  was shown in Ref. \cite{novotnydrop}. In this case a particle was prepared in an optical trap and then released from the trapping potential. During free fall a static electrostatic force was applied via a known electic field, and the resulting displacement of the particle was observed upon its recapture up to approximately $100$ $\mu$s afterwards. Similar cooling and free-fall schemes may also be possible with matter-wave interferometer based accelerometers, as described in Sec. \ref{sec:future}. 
\subsection{Rotation}
\begin{figure}[htbp]
	\includegraphics[scale=0.4]{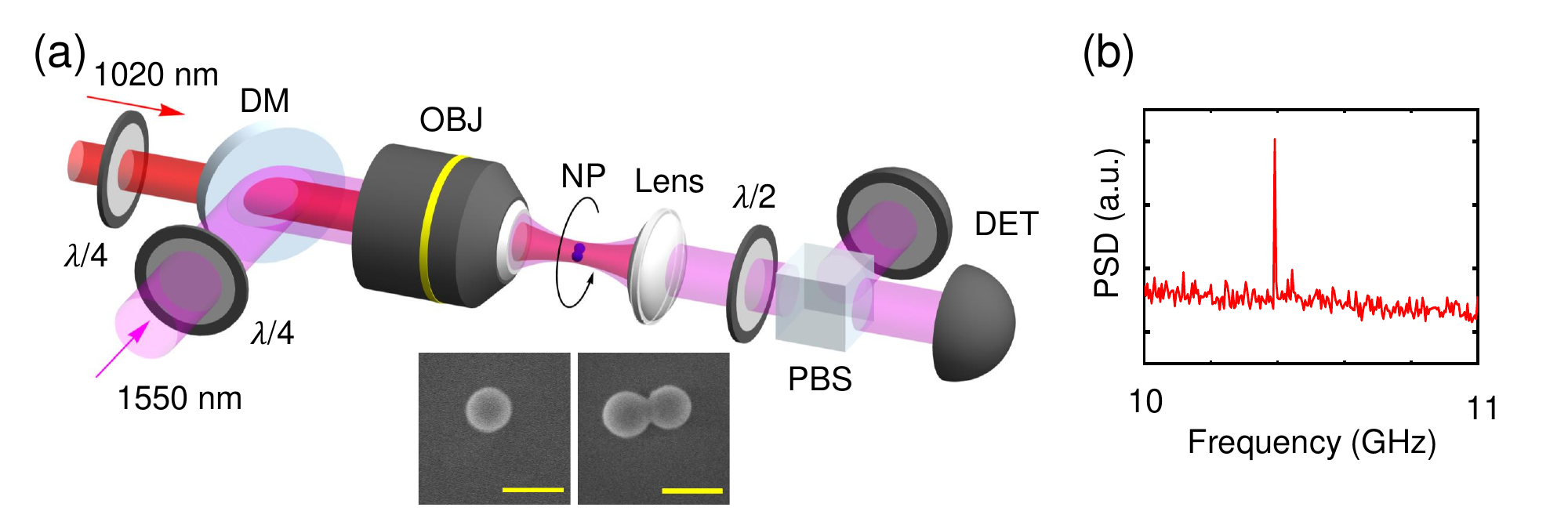}
	\caption{ (a) A silica nanoparticle (NP) is levitated by a tightly focused 1550 nm laser in vacuum. The polarization of the 1550 nm laser is controlled by a quarter waveplate. It is split by a polarizing beam splitter (PBS) and sent to detectors for monitoring the rotation of the levitated nanoparticle. A 1020 nm laser applies an external torque for testing the sensitivity of torque detection. Inset: SEM images of a silica nanosphere (left) and a silica nanodumbbell (right). The scale bar is 200 nm for both images. 
	(b) A measured power spectral density (PSD) of the rotation of an optically levitated nanoparticle . The corresponding rotation frequency is 5.2 GHz. Figure adapted from Ref. \cite{ahn2020ultrasensitive}.
	}
	\label{FigRotation}
\end{figure}

In 1998, a calcite microparticle trapped by optical tweezers in water was driven to rotate up to 357 Hz by a circularly polarized laser \cite{friese1998optical}. 
In 2013, an optically levitated vaterite microsphere (diameter: 4.4 $\mu$m) was driven to rotate up to 5 MHz in vacuum \cite{arita2013laser}. In 2018, two groups reported ultrafast rotation of silica nanoparticles  at about 1 GHz, which was mainly limited by the bandwidth of the detectors \cite{PhysRevLett.121.033603,Novotny2018}. By using faster detectors and optimizing other conditions, the rotation speed of a levitated silica nanoparticle was further increased to more than 5 GHz in vacuum \cite{ahn2020ultrasensitive,jin20216}. Figure \ref{FigRotation} shows an experimental schematic of driving a levitated nanoparticle in vacuum to rotate with a circularly polarized laser. More recently, an optically levitated silica nanodumbbell was driven to rotate at high speed near a flat surface and nanostructures \cite{ju2023near}, which have potential applications in measuring Casimir torque and investigating non-Newtonian gravity.

\subsection{Torque}

If the system is limited by thermal noise, the minimum torque that can be measured is $M_{\mathrm{min}}=\sqrt{4 k_B T I \gamma}/\sqrt{\Delta t}$ \cite{ahn2020ultrasensitive,Li2016}. Here $k_B$ is the Boltzmann constant, $T$ is the environmental temperature, $I$ is the moment of inertia of the levitated nanoparticle, $\gamma$ is the rotational damping rate, and $\Delta t$ is the measurement time. For a 150 nm-diameter silica nanosphere, a measured torque detection sensitivity of $(4.2 \pm 1.2) \times 10^{-27}$ N m Hz$^{-1/2}$ at room temperature has been reported \cite{ahn2020ultrasensitive}. This sensitivity is several orders better than the best-reported nanofabricated torque sensor at milikelvin temperatures.

\subsection{Electric and magnetic fields}

Electric fields have been used to calibrate the force sensitivity of levitated objects \cite{ranjit2016zeptonewton} \cite{hempston2017force}, to couple the Cartesian degrees of freedom of the levitated potential \cite{timberlake2019static}, to drive motion in non Cartesian degrees of freedom \cite{blakemore2022librational}, and has also been used as as a field for actuating feedback cooling with lower noise properties and greater linearity that optical AOM's \cite{tebbenjohanns2021quantum}. They also allow for quantification and control of the charge present on a levitated object \cite{frimmer2017controlling}.

For electric field sensing it is important to note that particle launching mechanisms often charge the particle \cite{moore2014search}, that levitated silica spheres frequently possess a semi-permanent dipole moment \cite{priel2021background}, and a pressure dependant complex internal and surface structure that can modify the charge state of the particle during pumpdown \cite{ricci2022chemical}. As such, calibration of any levitated sensor under its final operating conditions (especially with regards to pressure and optical power) is essential \cite{hebestreit2018calibration}.

\section{Fundamental Physics}
\subsection{Gravitational force}

Experimentally witnessing the quantum nature of gravity has been a long elusive goal in modern physics.  In part due to its extremely feeble nature, gravity is arguably the least understood fundamental force in nature. The gravitational attraction between two protons in the nucleus of an atom is approximately $10^{36}$ times weaker than their electromagnetic repulsion, reflecting the vast difference in the energy scale of quantum gravity versus the other Standard model interactions. 
An explanation for this mysterious ``hierarchy problem'' could be obtained by performing precise tests of the gravitational inverse square law of attraction at sub-millimeter distances. Deviations from the gravitational inverse square law are typically parametrized in terms of a Yukawa correction to the Newtonian potential, \cite{Murata_2015}:
\begin{equation}
    V(r) = \frac{G_N M m}{r}(1 + \alpha e^{-r/\lambda})
\end{equation}
where $G_N$ is Newton's constant, $m$ and $M$ are the sphere and attractor masses (if point-like), $r$ is their separation, $\alpha$ parameterizes the strength of the new force, and $\lambda$ parameterizes its characteristic range. For a realistic geometry an integral needs to be performed over the test and attractor mass volumes.  If the deviation from the ISL is mediated by a new force carrier with mass $m_\phi$, then typically $\lambda = \hbar /(m_\phi c$). 
Precision tests of gravitation require not only extreme sensitivity, but also the ability to mitigate or subtract out background forces of non-gravitational origin, including electromagnetic effects from patch charges or the Casimir effect. Previous experiments have employed sensitive torsion balances \cite{Kapner2007} testing the gravitational inverse square law down to length scales below $50$ microns. Cryogenic microcantilevers \cite{Geraci2008} and torsional oscillators \cite{Chen2016} have been used to search for inverse square law violations at even shorter distances. With an optically-trapped nanosphere as a test mass, several orders of magnitude of improvement is possible in the search for new gravity-like forces at the micron distance scale, in a regime where several theories of physics beyond the Standard Model predict possible new signatures \cite{geraci2010,moore2021searching}. In the quantum regime, operating a nanosphere matter-wave interferometer near a surface, as described further in Sec. \ref{sec:future}, may enable future tests of Newtonian gravity down to the 5-$\mu$m length scale or below, yielding important clues about the interface of gravity with the other three known forces in nature.

\begin{figure}[t]
    \centering
    \includegraphics[width=0.99\columnwidth]{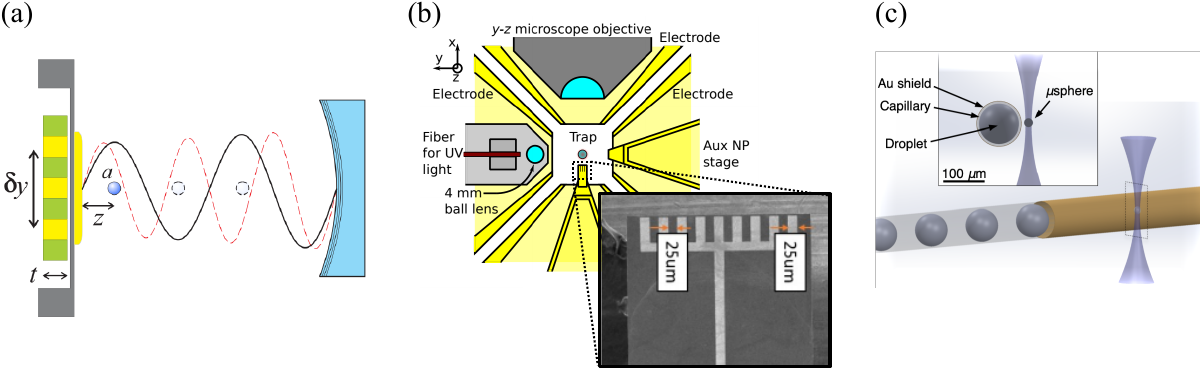}
    \caption{Adapted from Ref. \cite{moore2021searching}. Attractor designs for short-range gravity measurements using levitated systems proposed to-date.  (a) A subwavelength nanosphere is trapped at the antinode of a standing wave potential, several hundred nm from a gold coated membrane shield.  A patterned mass with different density materials (green and yellow) is oscillated behind the shield. Figure reproduced from Ref. \cite{geraci2010}. (b) A linear array of Au and Si fingers, each 25~$\mu$m wide, is oscillated at several $\mu$m distance from a 5~$\mu$m diameter sphere trapped in the center of the surrounding shielding electrodes~\cite{Wang:2017}.  An additional microfabricated shield between the moving fingers and sphere can be positioned on an independent stage~\cite{Kawasaki:2020oyl}.  Figure reproduced from~\cite{Wang:2017,Kawasaki:2020oyl}.  (c) A Au-coated microfluidic capillary is positioned at several $\mu$m distance from a $\sim$10-20~$\mu$m diameter sphere, and alternating droplets of a dense fluid (e.g. a polytungstate salt solution, $\rho \approx 3$~g/cm$^3$) and carrier fluid (e.g. mineral oil, $\rho \approx 0.8$~g/cm$^3$) flow through the capillary at $\sim$50~Hz~\cite{Moore:2018spie}.}
    \label{fig:attractors}
\end{figure}

Optically trapped dielectric spheres can function as a test mass  when held using optical radiation pressure near a gravitational source mass. Possible trapping geometries include confining the nanoparticle in an optical standing wave formed near the surface of a mirror, either in an optical cavity \cite{geraci2010} or in free-space \cite{montoya2022scanning}. Non-Newtonian Gravity-like forces and Casimir forces can be tested by monitoring the motion of the sphere as a gravitational source mass is brought behind the mirror. Other approaches involving an optical levitation trap are also being investigated \cite{Monteiro:2020wcb, Wang:2017, Kawasaki:2020oyl}. Several orders of magnitude of improvement is possible in the search for new gravity-like forces at the micron distance scale due to the sensitivity of the technique. Setups for test masses and attractor masses are shown in Fig. \ref{fig:attractors}.  Ref. \cite{moore2021searching} shows the potential reach along with theoretical predictions for new fifth forces that are Yukawa-type corrections to gravity at short distance scales using spheres of sizes 300 nm and 20 $\mu$m, currently being experimemtally investigated \cite{ranjit2016zeptonewton,Monteiro:2020wcb, montoya2022scanning,Kawasaki:2020oyl}.

Advances in sensitivity made possible by pushing the sensitivity of these sensors into the quantum regime along with improved understanding and mitigation of systematic effects due to background electromagnetic interactions such as the Casimir effect and patch potentials will enable several orders of magnitude of improvement in the search for new physics beyond the Standard model.

\subsection{High frequency gravitational waves} The extreme force sensitivity made possible by optical levitation lends itself to the search for weak astrophysical signals, including feeble strain signals from Gravitational waves or impulses from passing Dark Matter.
One of the most interesting sources of Gravitational waves in the high-frequency regime arises from physics Beyond the Standard Model. The QCD axion is a well-motivated dark matter candidate that naturally solves the strong CP problem in strong interactions and explains the smallness of the neutron's electric dipole moment \cite{axion1,axion2,PTViolation,Moody:1984ba}. The Compton wavelength of the QCD axion with axion decay constant $f_a \sim 10^{16}$ GeV (at the Grand-Unified-Theory [GUT] energy scale) matches the size of stellar mass BHs and allows for the axion to bind with the BH “nucleus,” forming a gravitational atom in the sky. A cloud of axions grows exponentially around the BH, extracting energy and angular momentum from the BH \cite{stringaxiverse,stringaxiverse2}. Axions in this cloud produce gravitational radiation through annihilations of axions into gravitons. For annihilations, the frequency of the produced GWs is given by twice the mass of the axion: $f=145\ \mathrm{kHz}$, which lies in the optimal sensitivity range for optically leviated sensors when $f_a$ is around the GUT scale. The signal is coherent, monochromatic, long-lived, and thus completely different from all ordinary astrophysical sources. The fraction of the BH mass the axion cloud carries can be as high as ${10}^{-3}$ \cite{stringaxiverse2}, leading to strain signals detectable within the sensitivity band of optically levitated sensors \cite{GWprl,aggarwal_2022}.  

A Michelson interferometer configuration with Fabry-P\'erot arms can be employed as in Ref. \cite{aggarwal_2022}, where a dielectric object is suspended at an anti-node of the standing wave inside each Fabry-P\'erot arm. A second laser can be used to read out the position of the object as well as cool it along the cavity axes, as described for a similar setup in Ref. \cite{GWprl}. The optical potential for this trap is 
$
U=\frac{1}{c}\int{I(\vec{r})(\epsilon(\vec{r})-1)d^3\vec{r}}
$
where \(I\) is the laser intensity, \(\epsilon\) is the relative dielectric constant, and the integration is performed over the extent of the dielectric particle. The trapping frequency along the axis of the cavity is determined by $\omega_0^2=\frac{1}{M}\frac{d^2U}{dx^2} {|}_{x=x_s}$ for a sensor of mass $M$ trapped 
at equilibrium position $x_s$.

A passing GW with frequency $\Omega_{\text{gw}}$ imparts a force on the trapped particle \cite{GWprl}, which is resonantly excited when  \(\omega_0\) $=\Omega_{\text{gw}}.$ Unlike a resonant-bar detector, \(\omega_0\) is widely tunable with laser intensity. 
The second cavity arm permits rejection of common mode noise, for example from technical laser noise or vibration.

The minimum detectable strain $h_{\rm{limit}}$ for a particle with center-of-mass temperature $T_\mathrm{CM}$ is approximately \cite{GWprl}
\begin{equation}
h_{\mathrm{limit}}=\frac{4}{\omega_0^2L}\sqrt{\frac{k_BT_{\mathrm{CM}}\gamma_gb}{M}\left[1+\frac{\gamma_{\mathrm{sc}}}{N_i\gamma_g}\right]}H\left(\omega_0\right), \label{eq:strain}
\end{equation}
where the cavity response function $H\left(\omega\right) \approx \sqrt{1+4\omega^2/\kappa^2}$ 
for a cavity of linewidth $\kappa$. Here $N_i=k_BT_\mathrm{CM}/\hbar\omega_0$ is the mean initial phonon occupation number of the center-of-mass motion. $\gamma_g = \frac{32P}{\pi \bar{v} \rho t}$ is the gas damping rate at pressure $P$ with mean gas speed $\bar{v}$ for a disc of thickness $t$ and density $\rho$, and $b$ is the bandwidth.

The photon recoil heating rate \cite{GWprl,jain2016direct} 
$\gamma_{sc}=\frac{V_c\lambda\omega_0}{4L}\frac{1}{\int{dV(\epsilon-1)}}\frac{1}{\mathcal{F}_{\rm{disc}}}
$ is inversely proportional to the disc-limited finesse $\mathcal{F}_{{\rm{disc}}}$, i.e. ($2\pi$) divided by the fraction of photons scattered by the disc outside the cavity mode. The integral is performed over the extent of the suspended particle. Here $V_c$ is the cavity mode volume \cite{GWprl}.  While for a nanosphere the scattering and recoil is nearly isotropic \cite{jain2016direct}, for a disc, if the beam size is smaller than the radius of the object and the wavefront curvature at the surface is small, the scattered photons acquire a stronger directional dependence and tend to be recaptured into the cavity mode. 
This reduces the variance of the recoil direction of the levitated object caused by the scattered photons.

Both of the damping rates that contribute to sensitivity in Eq. \ref{eq:strain} scale inversely with the thickness of the levitated disc, for thickness smaller than radius. 
In the gas-dominated regime, $\gamma_{\rm{sc}} \ll N_i\gamma_g$, the sensitivity scales as $\sqrt{1/Mt}$ at fixed frequency. 
For sufficiently low vacuum, the sensitivity becomes photon-recoil-limited, and 
the strain sensitivity goes as \(1/M\sqrt{\mathcal{F_\mathrm{disc}}}\).

For the application of GW detection, using high-mass, high-trap-frequency, disc- or plate-like microparticles is ideal for achieving maximal sensitivity with this technique. Recent experimental work has demonstrated optical trapping of Yb-doped $\beta-$NaYF sub-wavelength-thickness high-aspect-ratio hexagonal prisms with a micron-scale radius \cite{winstone2022optical}. The prisms are trapped in vacuum using an optical standing wave, with the normal vector to their face oriented along the beam propagation direction, yielding much higher trapping frequencies than those typically achieved with microspheres of similar mass. 
This plate-like geometry is planned to be used in high frequency gravitational wave searches in the Levitated Sensor Detector, currently under construction \cite{aggarwal_2022}.
Furthermore, the Yb-doped NaYF has previously been shown to exhibit internal cooling via laser refrigeration, which may enable higher trapping intensity and thus higher trap frequencies for gravitational wave searches approaching the several hundred kHz range \cite{winstone2022optical}.

\subsection{Casimir force and torque}
According to quantum electrodynamics (QED), vacuum is not empty, but full of virtual photons due to the zero-point energy. The quantum vacuum fluctuations will lead to an attractive force between neutral metal plates in vacuum, which is known as the Casimir force. The Casimir force can dominate the interactions at nanoscales, and must accounted in searching for new force (such as the non-Newtonian gravity) at small separations. The Casimir force has been measured between two objects with different geometries and between three objects. Besides linear momentum, the virtual photons also have angular momentum, which can lead to a Casimir torque. So far, there has been only one experimental report on the Casimir torque, which was measured with a liquid crystal \cite{somers2018measurement}. A levitated nanoparticle in vacuum will an ultrasensitive force and torque detector. Thus it will be ideal for study the Casimir force and Casimir torque \cite{Xu2017Detecting}.

\subsection{Dark Matter}

Dark matter can also be detected by observing the interaction of passing massive particles with the levitated nano-objects. For example, a recent search has been performed for composite dark matter particles scattering from an optically levitated nanogram mass, cooled to an effective temperature $\sim$200~$\mu$K~\cite{Monteiro:2020wcb}. Similar techniques may allow detection of sufficiently low momentum transfers that sub-MeV dark matter scattering coherently from 10~nm diameter spheres can be detected\cite{Afek_2022, Carney_2021}. Large arrays of such trapped objects are possible, and can enable lower cross-sections to be reached~\cite{Afek_2022,moore2021searching}. Such detectors are intrinsically sensitive to the direction of the dark matter scatter, allowing an unambiguous determination of the astrophysical origin of a signal if detected~\cite{Monteiro:2020wcb,Afek_2022,moore2021searching}.

\section{Summary and Future Perspectives: Experiment and Theory \label{sec:future}}

\subsection{Biological/chemistry applications}

Optical levitation has been used in wide range of biophysical experiments since the first demonstration that bacteria and viruses can be optically trapped by Ashkin and Dziedzic\cite{ashkin1987}.  One early application of optical levitation was to select and transfer seed crystals used in structural biology to prepare samples for single-crystal protein x-ray diffraction\cite{Bancel98}.  Single-beam laser tweezers have also enabled several seminal advances in the measurement of biologically relevant forces with pico-Newton sensitivity\cite{svoboda94}.  Chemically tethering single-molecule motor proteins to optically trapped microspheres has enabled fundamental insights into the mechanical properties of single DNA molecules\cite{bustamante2003}, the equilibrium free energy difference between distinct molecular conformational states\cite{liphardt2002}, and the asymmetric hand-over-hand mechanism underlying the motion of kinesin motor proteins\cite{asbury2003}.  Gold nanorods have also been used in place of transparent microspheres as molecular handle, leading to photothermal heating that increases the translational stepping speed of myosin-V motor proteins\cite{iwaki2015}.

Optically trapped polystyrene microspheres have been used for the sensing of pathogenic viruses, including the in situ detection of H7N9 influenza virus DNA based on a sandwich assay using fluorescence from biotinylated quantum dots\cite{Cao2016}.  Detection limits of 1 pM have been observed which are two orders of magnitude lower than conventional methods.  More recently, individual PhiX174 viruses with a diameter of just 25 nanometers and have been trapped using plasmonic double nanohole (DNH) apertures in a gold film\cite{Burkhartsmeyer2020}.  Optomechanical measurement of the virus’ acoustic radial breathing mode frequency (32 GHz) were also made using extraordinary acoustic Raman spectrosocpy.  

Beyond the characteirzation of biomolecules and viruses in liquid buffers, both single- and dual-beam optical trapping have been used to open up new possibilities in the spectroscopic characterization of biological cells and other bioaerosols\cite{santarpia2020} in air.  For example, chemical information about optically levitated biological specimens can be acquired through the analysis of inelastic Raman light scattering\cite{gessner2004}. Optical levitation has also been used to hold biological samples for background-free x-ray fluorescence spectroscopy using synchrotron radiation\cite{ vergucht2015}.  More recently, photophoretic forces have also been used to deliver aerosolized viruses into the focus of an x-ray free-electron laser for femtosecond x-ray diffractive imaging\cite{Awel2022}.

\subsection{Optical binding and Levitated particle arrays}

\begin{figure}[hb]
	\includegraphics[scale=0.38]{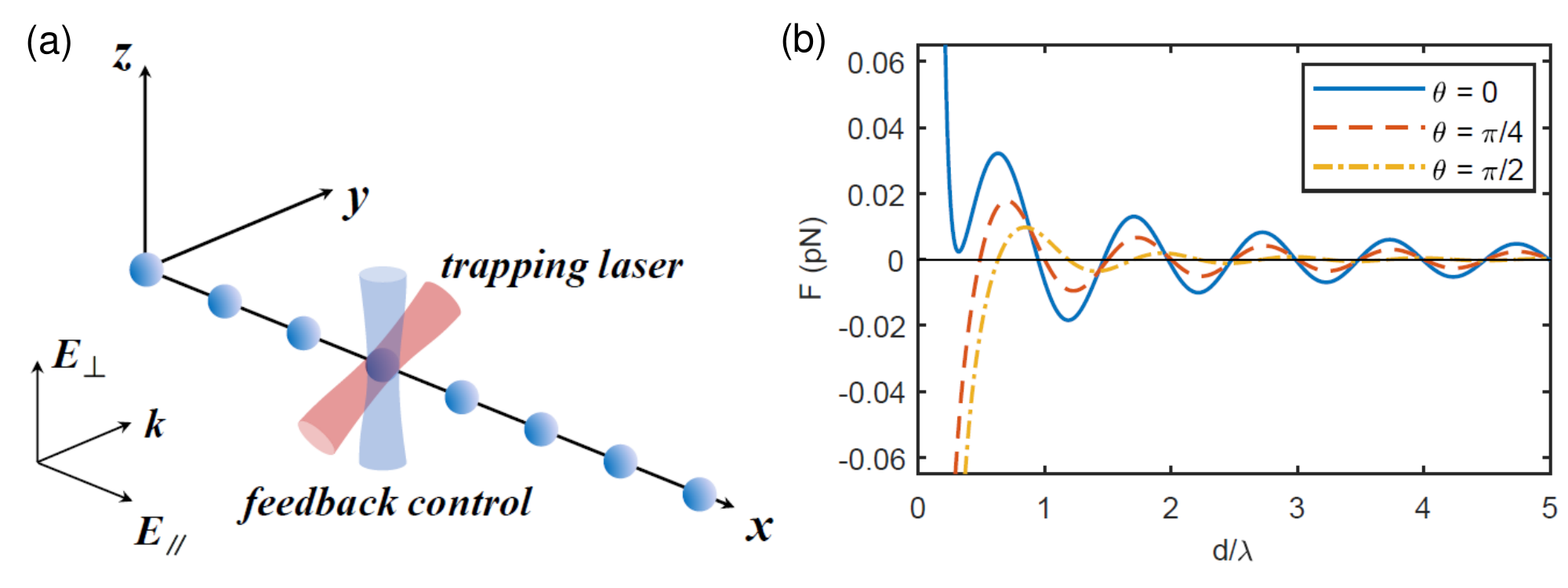}
	\caption{ (a) A one-dimensional array of optically levitated nanospheres. (b) The optical binding force between two optically levitated nanospheres as a function of their separation. $\theta$ is the angle between the polarization of the trapping laser and the $x$ axis. Figure adapted from Ref. \cite{liu2020prethermalization}.
	}
	\label{FigBinding}
\end{figure}

While most  studies in levitated optomechanics in vacuum have been focused on single particles, there is a growing number of works on multiple particles. An array of optically trapped ultracold atoms have been used for quantum simulation and quantum computing. An array of optically levitated nanoparticles in vacuum will  offer new opportunities that do we can not do with a single levitated nanoparticle \cite{liu2020prethermalization,yan2023demand}. Fig. \ref{FigBinding} illustrates the effect of the laser polarization on optical binding forces for two optically levitated nanospheres separated by varying distances.

\subsection{Near field interactions}
Several experiments have realized methods for trapping nanoparticles near surfaces, including silicon and SiN membranes \cite{Winstone2018, Novotnysurface} and near photonic crystal structures \cite{Magrini2018}, and near a metallic mirror \cite{montoya2022scanning}. Extending such a setup to allow systematically scanning the position of the particle with respect to the surface creates opportunities for three dimensional surface force microscopy. 
To date no 2-D image of a surface has been rendered using a levitated nanoparticle as a probe of the surface potential. Due to the high degree of force sensitivity and ability to constrain the particles motion to the zero point level this is a potentially useful future goal for the field. Furthermore, a greater understanding of the distribution of surface charges and permanent and induced dipoles over the levitated object itself is an another near-term goal that can be realized by studying nanoparticles trapped nearby objects.
Another interesting application enabled by near surface trapping is to study radiative heat transfer at the nanoscale \cite{Rousseau2009}.

\subsection{High mass matter wave interferometry}

Wave-particle duality is at the core of quantum mechanics. Wave-like behavior of massive particles including electrons, neutrons, and atoms has been studied over the past several decades, and recently even molecules as large as 25,000 atomic mass units have been shown to exhibit wave-like behavior \cite{arndt2019}. It is natural to ask how massive of an object can behave like a wave, i.e. be placed into a quantum superposition state after undergoing diffraction. Optically levitated nanoparticles offer a promising route to push the limits of quantum theory in this manner and to test foundational aspects of quantum theory.
In particular, optically-levitated dielectric objects in ultra-high vacuum exhibit an excellent decoupling from their environment, making them highly promising systems for precision sensing. An exciting prospect is to harness uniquely quantum aspects of sensing in these levitated nanoparticle systems, taking advantage of wave-particle duality in nanoparticles of unprecedented size \cite{Ulbricht:2014,andyhart2015}.  Given the recent experimental demonstration of quantum ground state cooling of optically trapped nanoparticles, experiments become increasingly feasible to demonstrate the wave-like nature of the states of ultracold nano-objects.  Matter wave interference of nanoparticles cooled to act as a point like source can be realized even for thermal states significantly above the quantum ground state \cite{andyhart2015}.

\begin{figure}
\centering
\includegraphics[width=0.9\textwidth]{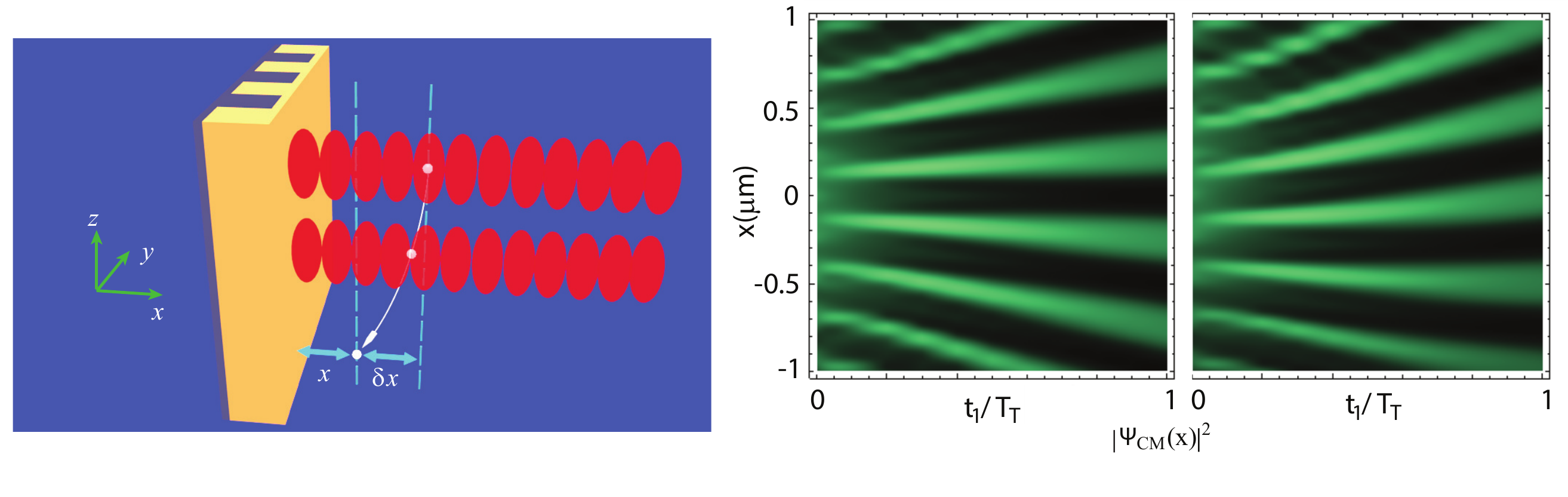}
\caption{ Adapted from Ref.~\cite{andyhart2015}. An example of a nanoparticle matter wave interference protocol. (left) Schematic of setup. A nanosphere is trapped and cooled near a surface. The optical trap is turned off and the wave packet expands as the particle falls. The particle is then diffracted from a
light grating, and its position is measured after an additional free-fall time. The acceleration of the nanosphere towards the surface from Casimir and gravitational forces shifts the interference pattern that is generated when the nanosphere diffracts from the grating. (middle) Density plot of $|\Psi(x)|^2$ following the grating for zero acceleration for releasing the trap at $\omega_0=2\pi \times 100$ Hz from its ground state.  (right) As in middle panel, with $a_\pi=4 \times 10^{-7} g$ constant acceleration. Parameters chosen as in Ref. \cite{andyhart2015} with mass of $10^6$ amu and interrogation time $T_T$ of 0.25s.}
\label{fig:matterwavesetup}
\vspace{-5mm}
\end{figure}

A possible accelerometer protocol utilizing nanosphere interference is shown in Fig. \ref{fig:matterwavesetup}. Here the center of mass temperature of a nanoparticle is sufficiently cooled so that the nanoparticle acts as a coherent source. In one possible scheme, the particle is allowed to free fall for a duration equal to the Talbot time. At this point a light-pulse grating is applied to diffract the wave packet of the falling particle, and after another duration of the Talbot time the location of the particle is measured \cite{Bateman2014}.  After repeated measurements, an interference pattern appears one particle at a time, with the locations of the fringe maxima and minima being sensitive to the acceleration experienced parallel to the grating direction during the free fall time. As the locations of the interference maxima do not depend on the initial transverse velocity, such approaches may have inherent sensitivity advantages when compared to simple free fall schemes, depending on backgrounds and noise sources \cite{andyhart2015}. The greatly reduced wave packet expansion of nanoparticles when compared with atoms used in atom interferometry provide a route to measure more highly localized forces and interactions, for example in close proximity to surfaces \cite{andyhart2015}. Technical challenges that must be overcome to realize interference protocols which generate an interference pattern one-particle-at-a-time include a fast on-demand loading mechanism compatible with ultra high vacuum and potentially cryogenic operation. 

In massive quantum-superposition experiments, background gas collisions arising from imperfect vacuum are a serious limitation to the masses of the particles that can be placed in quantum superposition and to the associated coherence times. Furthermore if optically-trapped nanoparticles are diffracted from a light grating and placed into a superposition state, blackbody emission from the surface of a room temperature sphere can decohere the quantum superposition.  To overcome these limitations, interferometry with levitated nanoparticles in a cryogenic, extreme high vacuum chamber may be a route forwards. In addition cryogenic laser refrigeration of levitated nanoparticles could assist to mitigate this decoherence \cite{rahman2017laser}. 

Additionally, quantum point defects such as the negatively-charged nitrogen vacancy center have been proposed to enable ultra-sensitive tests of quantum mechanics focused on novel demonstrations of matter wave interferometry using macroscopic test masses. When optically trapping individual nanodiamonds that contain intrinsic NV centers, laser heating ultimately limits the achievable base pressure since residual absorption leads to crystal heating that cannot be dissipated via thermal conduction. This presents an obstacle to preparing COM quantum states.  Nanoparticle state control, when coupled with the ability to laser cool the nanodiamond material, will present an opportunity for pursuing matter wave interference.


\subsection{Detecting high energy radiation}

Levitated optomechanical experiments are naturally highly sensitive to momentum impulses \cite{moore2021searching}. Experiments have therefore been designed such that study of nuclear decays is possible by directly observing the motion of a levitated optomechanical sensor embedded with a decaying radioactive species \cite{wang2024mechanical}. As the species decays, the momentum from the outbound high energy ejecta is offset by an equal and opposite momentum impulse on the levitated sensor.

This allows for the study of nuclear processes without having to transform the outbound nuclear ejecta directly into electrons and may therefore be sensitive to parameter spaces not covered by existing detector technologies. For example, heavy sterile neutrinos \cite{carney2023searches}.

Another proposal would be to use a levitated silica sphere as a platform/substrate to actually modify nuclear decays via the nuclear Purcell effect changing weak decay rates \cite{tkalya2018decay}. In this case a radioactive species is also embedded with the lattice of the levitated object (silica in this proposal), which provides a set of cavity like boundary conditions to modify the low lying energy state of thorium 229.

\subsection{Squeezing and non classical state preparation} Correlated phase spaces have been reported as "classical squeezing" by several groups, both in real time \cite{penny2023sympathetic} and by reconstructed phase spaces sampled over several hours of experiment time in a pulsed operation\cite{rashid2016experimental}. It should be noted that these examples are completely classical effects that do not compress an axis of the phase space below any limit defined by an uncertainty principle. However genuine squeezing of the quantum ground state state of a mechanical oscillator may well be achieved in the near future due to the proliferation of ground state capable levitated quantum systems. 

\subsubsection{ Levitated optomechanical state preparation}
A major catalyst for the interest in levitated optomechanical oscillators was the expectation that it should be possible to create an environment where the levitated oscillator can be prepared into its quantum mechanical ground state \cite{Rom10a}.  Tantalizing was the suggestion that it could be done at room temperature.   Nearly a decade after such predictions, the first experiments demonstrating ground state preparation have been achieved.  Of course, mechanical number states are not the only possible states that can be engineered and in this section we briefly survey some of the non-thermal levitated mechanical oscillator states that have been reported.  

\begin{figure}
\centering
\includegraphics[width=\linewidth]{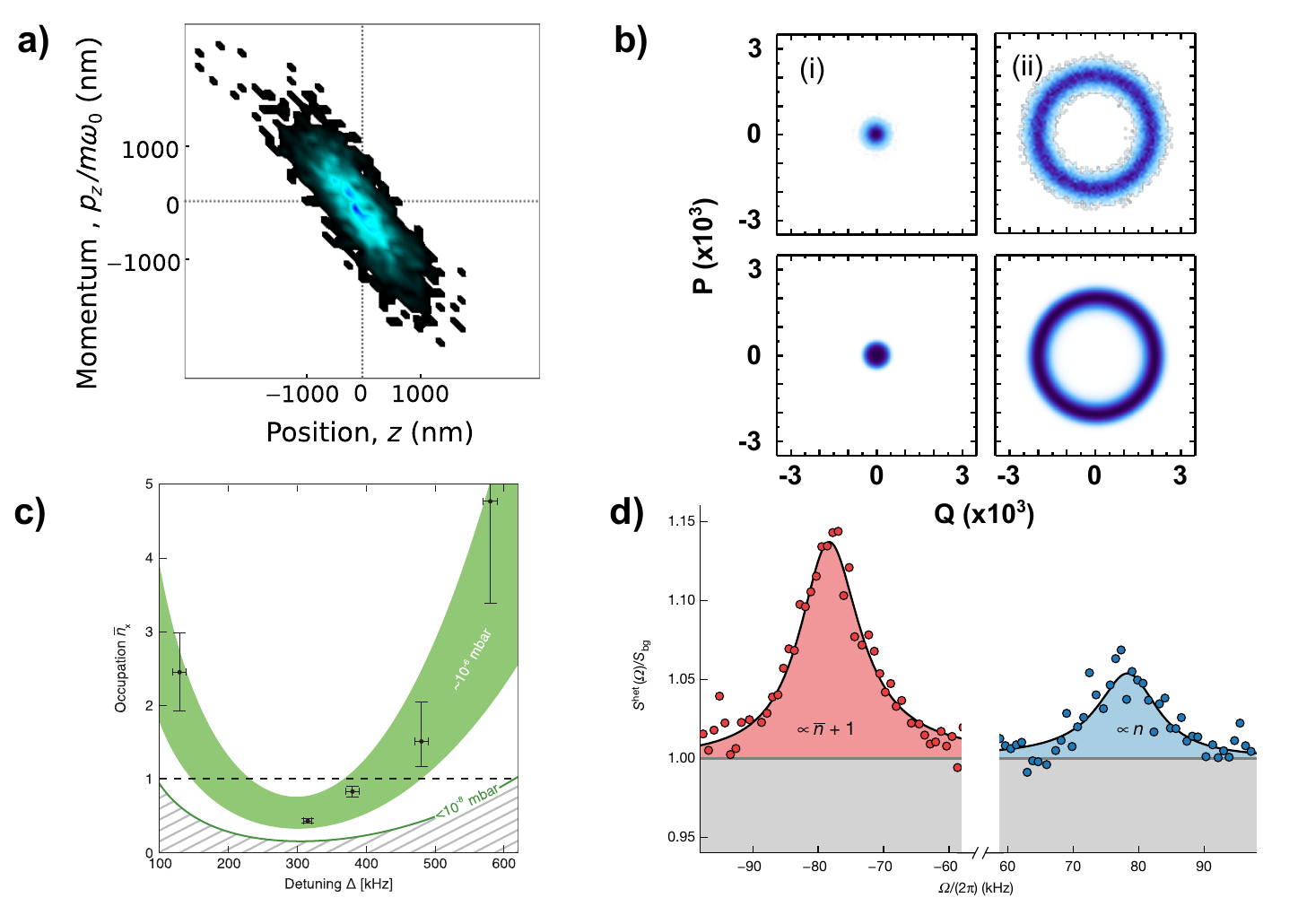}
\caption{a). Classical squeezing in  a levitated optomechanical oscillators's phase space distribution \cite{Ulb16a}. b) Phase space distribution of an optical tweezer phonon laser. The top row is the experiment and bottom row theory.  The first column is below and the second column is above threshold \cite{Pet19a}. c) The average phonon number as a function of laser tweezer detuning from the optical cavity.  The green band is the theoretical expectation taking into account system parameters\cite{Del20a}. d)  The Stokes (left) and anti-Stokes (right) sidebands measured via heterodyne detection. The black lines are theory fits from which the sideband power ratio gives the average phonon number\cite{Teb21a}.  }
\label{figX}
\end{figure}

 Passive (or free-running) levitated optomechanical oscillators, meaning no feedback mechanism has been introduced to modulate the oscillator's dynamics, naturally exhibit thermal statistics.   It is only through engineered feedback, either passive or active, that the oscillators mechanical state can be controlled.  The first experiments evidencing deviation from thermal statistics reported correlated phase spaces that exhibit  ``classical squeezing''. The phase spaces were reconstructed over several hours and relied on a pulsed experiment \cite{Ulb16a}. One example of the deviation from the phase space of a thermal oscillator is shown in Fig. \ref{figX}a). A second experiment reconstructed the phase space in real time to demonstrate classical squeezing of two coupled levitated oscillators \cite{Bar23a}. These experiments are classical in the sense that they do not compress an axis of the phase space below a limit defined by an uncertainty principle. However genuine squeezing of the quantum ground state state of a mechanical oscillator may well be achieved in the near future due to the proliferation of ground state capable levitated quantum systems. 

A second modulation of the free running levitated optomechanical oscillators thermal statistics has been realized. It was informed by the discovery that the center-of-mass oscillator dynamics in one direction resembled that of a single mode phonon laser \cite{Pet19a}. By carefully tuning the applied nonlinear feedback cooling and linear feedback gain for one mode of oscillation the thermal statistics of the oscillator were converted to the statistics of a single mode laser. The change in phonon statistics were confirmed by measuring the phonon number distribution, seeing the change in the zero-delay energy correlation function of the oscillator dynamics from that of a thermal state (bunching) to that of a laser (unity) and observing the change in the oscillator's phase space distribution, see Fig. \ref{figX}b). 

As mentioned in the introductory paragraph of this section, most recently a number of groups have reported cooling the oscillator's center-of-mass into its quantum ground state \cite{Del20a,Teb21a,Kam22a,Pio23a}.  In one experiment, a levitated oscillator (approximately 140 nm diameter) situated at the node in an optical cavity was cooled to its ground state by a coherent scattering. The cavity suppressed both elastic scattering, via the particle spatial position, and cooled the center-of-mass motion since spectrally the cavity suppressed Stokes scattering and enhanced anti-Stokes scattering. The experiments achieved a minimum average phonon number of $\bar{n} = 0.43\pm0.003$ when the optical tweezer is detuned from the cavity resonance by the levitated oscillator's oscillation frequency, see Fig. \ref{figX}c) \cite{Del20a}.  A second experiment was conducted without a cavity in a cryogenic environment.  The reduced gas temperature of 60K leads to cryogenic pumping the gas to a pressure on the order of $10^{(-9)}$ mbar.  A combination of nonlinear feedback cooling and cold damping via the Coulomb force result in average occupation numbers of $\bar{n}=0.66\pm0.08$.  Stokes and anti-Stokes thermometry, see Fig. \ref{figX}d) was used to determine the average phonon number \cite{Teb21a}.  A third approach to ground state cooling took advantage of a standing wave trap along one oscillator direction and employed optical cold damping to achieve the center-of-mass ground state with $\bar{n} = 0.85\pm0.20$ \cite{Kam22a}.  Finally, in a recent experiment  \cite{Pio23a}, again using coherent scattering into an optical cavity mode, two center-of-mass oscillation directions were cooled to their motional ground state with average phonon occupation numbers of  $\bar{n}_1 = 0.83\pm0.10$ and  $\bar{n}_2 = 0.81\pm0.12$.   

\subsection{Tests of quantum foundations - collapse models, entanglement}

\begin{figure}[hb]
\centering
	\includegraphics[scale=0.3]{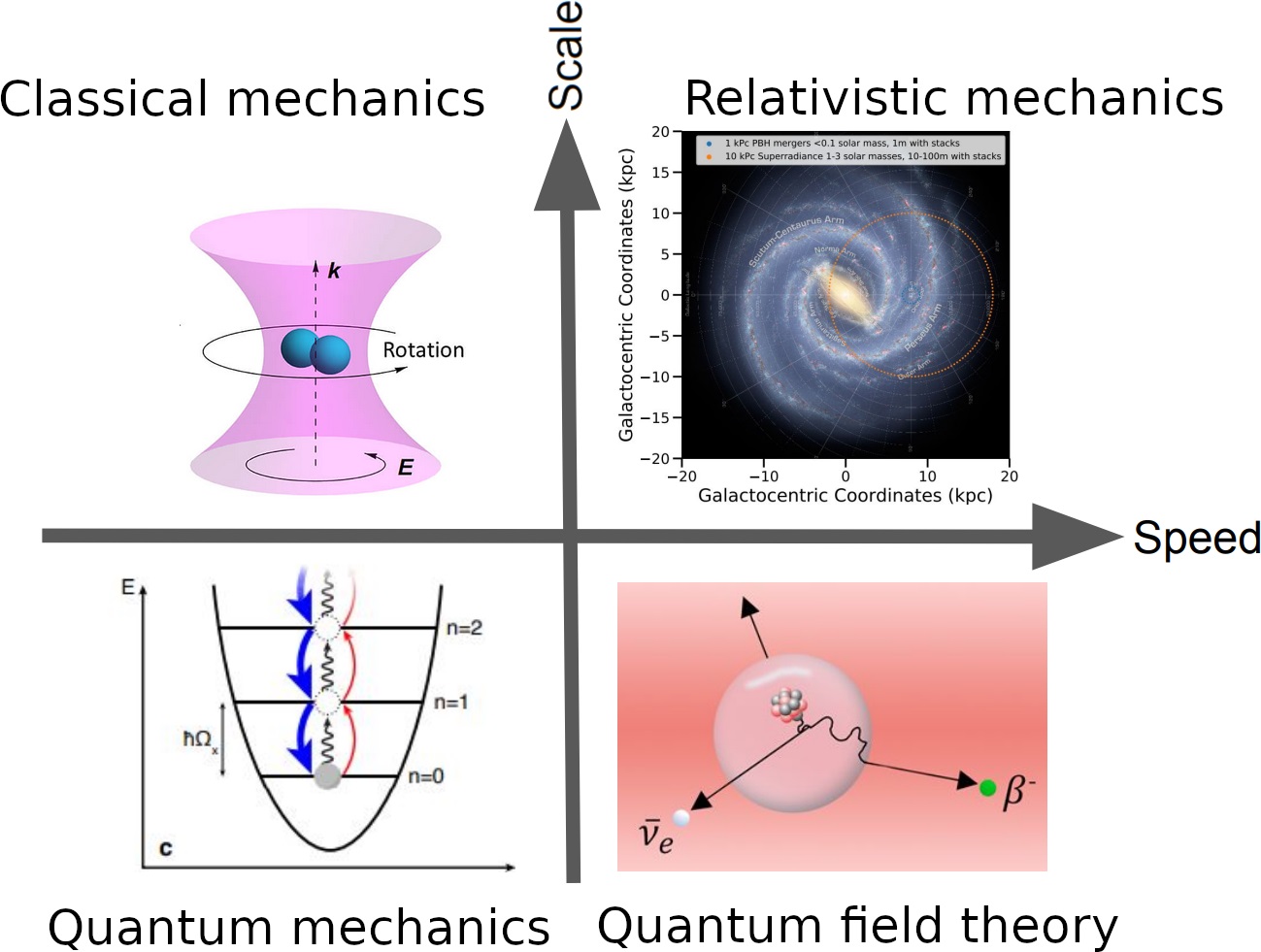}
	\caption{Levitated optomechanics contributes to the study of a wide parameter space of modern physics. A) Classic mechanics B) Astronomy and astrophysics (adapted from \cite{winstone2022optical}), C) Quantum physics and the regime of quantum to classic switchover (adapted from \cite{delic2019motional}), D) High energy particle physics (adapted from \cite{carney2023searches}). The upper zone of the "scale" axis denotes sizes far larger than $10^{-9} m$ and the lower part sizes close to or smaller than $10^{-9}m$. The right hand side of the "speed" axis denotes speeds near the speed of light while its left hand side denotes speeds much less than the speed of light.
	}
	\label{fig:domains_of_physics}
\end{figure}

Quantum mechanics has held up to every experimental test performed thus far, although most quantum mechanical phenomena have been observed in microscopic systems. It is postulated and generally believed that quantum mechanics should apply equally well to the macroscopic and microscopic worlds. In particular, apart from decoherence resulting from interactions with their environment, it contains no fundamental mechanism for preventing macroscopic objects from existing in a coherent linear superposition of different quantum states.  Since the early days of quantum mechanics, this fact has been somewhat unsettling, as it conflicts with the absence of macroscopic quantum superpositions in our everyday life.  Moreover, the lack of a fundamental mechanism to prevent macroscopic superpositions is closely related to the measurement problem in quantum mechanics.  To reconcile this conflict, some theorists have formulated an \textit{ad hoc} postulate of ``measurement induced collapse'' to describe the outcomes of measurements of a microscopic system in a superposition of two quantum states, using a macroscopic device \cite{BassiRMP}.  By contrast, at all other times, the evolution a quantum system is described by unitary evolution under the Schr$\ddot{{\rm{o}}}$dinger equation.  
As a potential resolution of these issues, models have been proposed that explain the absence of observations of macroscopic quantum superpositions and eliminate the need for ``measurement induced collapse'' by providing a mechanism for the collapse of superpositions of sufficiently massive objects.  In several of these models, the collapse is gravity-induced \cite{Penrose1996,Bassi:2017,BassiRMP,Diosi1989}.  Experimentally testing such collapse models probes the interface between quantum mechanics and gravity, and a confirmation of gravity-induced collapse would be a revolutionary advance in our understanding of quantum theory and reveal an interplay between gravity and quantum mechanics that would need to be considered in developing any theory of quantum gravity. Improving tests of collapse models beyond previous work can be achieved by creating quantum superpositions involving more massive objects and longer coherence times.  A promising approach to achieve this is interferometry with levitated nanoparticles \cite{oriol2011}. A figure of merit describing how macroscopic a quantum superposition is or its  ``macroscopicity'' has been defined for objects ranging from individual atoms to larger scale mechanical oscillators \cite{BassiRMP}.   Optically levitated systems represent a method to test spontaneous localization models at intermediate scales \cite{BassiRMP,Bassi:2017}. Bounds on such models are shown in Ref. \cite{Carlesso:19}. \textcolor{black}{Limits from interferometric tests with large molecules are reported in Ref. \cite{arndt2019}.}
We also note that collapse models also can lead to anomalous heating effects in both clamped and levitated mechanical oscillators, and can also be searched for with non-interferometric methods \cite{BassiRMP, Carlesso:19, Zheng2020,Vinante2020,VinantePRL2020,Donadi2020,Pontin_collapse}. For a recent review of limits from non-interferometric tests, see Ref. \cite{Carlesso:2022}.

Any fundamental decoherence mechanism whether gravitational related or not would need to be distinguished from technical sources of decoherence. One challenging aspect of optically levitated systems stems from the fact that for optically trapped particles in vacuum, the center of mass temperature can be quite cold while the internal  temperature  can  be  significantly  above  room  temperature.   As  such  a  warm  object  is  diffracted from  a  light  grating and  placed  into  a  superposition state, blackbody emission from  the  surface  of  the  spheres  can  decohere  the  quantum  superposition \cite{Ulbricht:2014,andyhart2015}.  Optical refrigeration techniques or operation in a cryogenic environment with minimal laser heating may be routes for improving the blackbody emission limited lifetime of such macroscopic quantum superpositions \cite{Luntz-Martin:21,rahman2017laser}. Magnetic levitation of superconducting particles has also been proposed as a method for obtaining larger quantum superpositions without the effects of laser heating \cite{Oriolracetrack}.

While probing physics at the Planck scale directly ($\sim 10^{19}$ GeV) may be hopelessly impossible, examining the role that gravity plays in uniquely quantum phenomena such as entanglement can provide insight into the quantum nature of gravity. Proposals have recently been presented for using macroscopic superpositions of massive nanoparticles to test whether the gravitational field can entangle the states of two masses \cite{Marletto:2017,Bose:2017,Carlesso_2019}.  
 
By developing new methods based on interferometry with levitated nanoparticles, despite the weakness of gravity, the phase evolution induced by the gravitational interaction of two levitated neutral test masses in adjacent matter-wave interferometers could detectably entangle them via graviton mediation even when they are placed far enough apart to keep other interactions at bay. 
The state of embedded spins in the masses can be used as a witness to probe the entanglement as described in Ref. \cite{Bose:2017}.  Such experiments require an ultra-high-vacuum cryogenic environment to minimize spurious environmental perturbations and technical noise. Although requiring much larger than has been achieved thus far in terms of massive quantum superpositions, as levitated quantum optomechanics techologies continue to improve, such tests represent an interesting future possibility.

\section{Conclusion}
Using optical methods to measure, trap, cool, and manipulate mechanical systems in the quantum regime yields a rich array of possible applications, ranging from precision measurement to quantum information science. In this tutorial, we have presented the basic principles of levitated optomechanics, discussed the interface of levitated optomechanical systems with other quantum systems, and provided a perspective on many exciting prospects for this rapidly evolving field. Fig. \ref{fig:domains_of_physics} illustrates the wide range of regimes accessible to or able to be probed by levitated optomechanical systems, ranging from quantum to classical, and non-relativistic to highly relativistic.




\begin{backmatter}
\bmsection{Funding}
AG is supported in part by NSF grants PHY-2110524 and PHY-2111544, the Heising-Simons Foundation, and the John Templeton Foundation. AG, MB, NV, PP, and GW are supported in part by and ONR Grant N00014-18-1-2370.  AG and GW are supported by the W.M. Keck Foundation. PP is supported in part by AFOSR grant FA9550-16-1-0362, NSF grant DMR-155507, and the U.S. Department of Energy (DOE), Office of Science (SC), Basic Energy Sciences (BES), Division of Materials Science and Engineering, Synthesis and Processing Sciences Program.  TL is supported by NSF grant PHY-2110591 and ONR Grant N00014-18-1-2371.

\bmsection{Acknowledgments}
We thank Andrew Dana for useful discussions and for assistance with Visualization 1.

\bmsection{Disclosures}
The authors declare no conflicts of interest.

\bmsection{Data availability} Data underlying the results presented in this paper are not publicly available at this time but may be obtained from the authors upon reasonable request.


\end{backmatter}

\bibliography{sample}

\end{document}